\documentclass[11pt,draftcls,onecolumn]{IEEEtran}

\IEEEoverridecommandlockouts

\usepackage[cmex10]{amsmath}
\usepackage{textcomp}
\usepackage{amssymb}

\usepackage{cite}

\usepackage{graphicx}
\usepackage{color}

\usepackage{mathtools}

\newtheorem{theorem}{Theorem}
\newtheorem{lemma}{Lemma}

\begin{document}

\title{Degrees of Freedom of the $K$-User Interference Channel in the Presence of Intelligent Reflecting Surfaces}

\author{\normalsize Ali H. Abdollahi Bafghi, Vahid Jamali, Masoumeh Nasiri-Kenari, and Robert Schober\\

\thanks{
Ali H. Abdollahi Bafghi and Masoumeh Nasiri-Kenari  are with the Department of Electrical Engineering, Sharif University of Technology, Tehran, Iran (email: aliabdolahi@ee.sharif.edu, mnasiri@sharif.edu).  {Vahid Jamali is with the Department of Electrical and computer Engineering, Princeton University, Princeton, NJ 08544 USA (email: jamali@princeton.edu). Robert Schober is with the Institute for Digital Communications, Friedrich-Alexander-University Erlangen-N$\rm {\ddot u}$rnberg (FAU), Germany (email: robert.schober@fau.de).}

}

}

\maketitle

\begin{abstract}
In this paper, we study the degrees of freedom (DoF) region and sum  DoF of the time-selective $K$-user interference channel in the presence of intelligent reflecting surfaces (IRSs). We consider four types of IRSs, namely 1) active IRSs, which are able to amplify,  attenuate, and add a phase shift to the received signal, 2) passive IRSs, which are able to attenuate and add a phase shift to the received signal, 3)  passive lossless IRSs, which are only able to add a phase shift to the received signal, and 4) $\varepsilon$-relaxed passive lossless IRSs, which are able to scale the received signal by a number between $1-\varepsilon$ and $1$ in addition to adding a phase shift. The performance of ideal active IRSs serves as an upper bound for the performance of passive and  passive lossless IRSs. We derive inner and outer bounds for the  DoF region and   lower and upper bounds for the sum  DoF of the $K$-user interference channel in the presence of an active IRS and  prove that the maximum value $K$ for the sum   DoF can be achieved if the number of IRS elements exceeds a certain  finite value. Then, we introduce   probabilistic inner and outer bounds for the  DoF region and  probabilistic lower and upper bounds for the sum  DoF of  the $K$-user interference channel in the presence of a passive  IRS  and prove that the lower bound for the  sum  DoF asymptotically approaches $K$ as the number of IRS elements grows large. For the DoF analysis of passive lossless IRSs, first we approximate it by the $\varepsilon$-relaxed passive lossless IRS and introduce a probabilistic lower bound for the corresponding sum  DoF. We prove that this bound asymptotically tends to $K$. In addition, we define a relaxed type of DoF called $\rho$-limited DoF. We introduce a lower bound for the $\rho$-limited sum DoF of the passive lossless IRS-assisted $K$-user interference channel and prove that this lower bound asymptotically also tends to $K$.
\end{abstract}

\begin{IEEEkeywords}
Time-selective $K$-user interference channel,  DoF region, sum  DoF, active and passive intelligent reflecting surface.

\end{IEEEkeywords}

\section{Introduction}

\label{section 1}
Smart radio environments are an emerging topic in  wireless communication research as they promise a significant improvement in   the capacity and performance of wireless communication channels. A smart radio environment is a radio environment which contains intelligent and reconfigurable elements. These elements can alter the communication channel by properly manipulating the impinging electromagnetic waves while reflecting them, e.g., changing the phase or attenuating the amplitude of the reflected wave. In conventional radio environments, physical objects may impair the transmission of information signals by causing phenomena like blockage and multipath. In contrast, a smart radio environment  can shape the wireless channel in  a manner that improves the capacity of the wireless network \cite{Renzo}.

Recently, intelligent reflecting surfaces (IRSs) have been proposed to facilitate smart radio environments \cite{Liaskos}.  An IRS is an array of electromagnetic elements, which are designed to have  controllable electromagnetic properties. An  IRS has a  control unit, which  allows the controlled reflection of the impinging electromagnetic wave. In particular, by controlling the phase of the IRS elements, the reflected wave can be amplified or attenuated in a desired direction. Therefore, IRSs facilitate the realization of a smart radio environment,  in particular, they offer the following advantages: 1)  two-dimensional IRSs facilitate easy deployment on  building facades for outdoor applications or on  room walls for indoor applications, 2)  IRSs can be flexibly reconfigured to jointly optimize the phase shifts of their elements to focus the reflected beam in a specific direction, and 3)  IRSs can improve channel capacity using techniques such as interference suppression \cite{Gong}.

IRS-assisted networks have been studied from various perspectives including channel modeling \cite{Najafi}, IRS optimization  \cite{Rui2}, and system analysis \cite{Najafi},  see \cite{Gong} for a recent survey. Maximizing the achievable rate of IRS-assisted systems is an important design objective, thus, we review some related works in this area.
In \cite{Rui2}, the authors studied the fundamental capacity limit of IRS-assisted  multiple-input multiple-output (MIMO) communications  by jointly optimizing the IRS phase shift matrix and the MIMO transmit covariance matrix. 
In \cite{ Karasik}, the authors studied   IRS-assisted communication systems, where the transmitter controls  the IRS using a finite-rate link. They derived information theoretic limits, and proved that the capacity can be achieved by  jointly encoding the information in the transmitted signal and the IRS phase shifts.
In \cite{Perovic},  the authors studied the optimization of the channel capacity of  millimeter wave channels in the presence of an IRS, where the line-of-sight path was not available.
In \cite{Huang},  the authors studied multi-user downlink communication in the presence of an IRS. They maximized the sum-rate subject to individual quality-of-service  guarantees by optimizing the transmit powers and the IRS phase shifts.
In \cite{Guo},  the authors studied  IRS-assisted downlink communication in a multi-user multiple-input single-output (MISO) system. They maximized the weighted sum-rate of all users  by jointly optimizing the beamforming vector at the base station (BS) and discrete phase shifts  at the IRS.
In \cite{Di},  the authors  studied transmission from a multi-antenna BS to multiple users via an IRS with discrete phase levels in a downlink system. They proposed a hybrid beamforming scheme assuming a reflection-dominated one-hop propagation model between the BS and the users and optimized the IRS phase shifts for maximization of the sum rate. 
In \cite{Mu}, the  authors studied a downlink  non-orthogonal multiple-access (NOMA) IRS-assisted  system, where a BS  served multiple users. They maximized the sum-rate of the users by jointly optimizing the beamforming vector at the BS and the phase shift matrix at the IRS, subject to successive interference cancellation decoding rate constraints and IRS scattering element constraints.
In \cite{Ozdogan}, the  authors  studied the usage of an IRS for  rank improvement of  MIMO communication channels.

The exact capacity expressions for  multiuser communication channels are exceedingly complicated for most scenarios. In contrast, DoF is an  analytically more tractable performance metric which indicates  how the capacity scales in the high signal-to-noise ratio (SNR) regime. 
Although the achievable rate and capacity of various IRS-assisted communication channels have been studied, multiuser channels have not been studied from a  DoF perspective.
In this paper, we study the  DoF of the time-selective $K$-user interference channel in the presence of an IRS. 
 It is well known that the sum DoF of the time-selective $K$-user interference channel without IRS is $\frac{K}{2}$ \cite{Cadambe1}.
However, we show that an IRS-assisted $K$-user interference channel is able to enlarge the DoF region and provide the full $K$ sum DoF, when the number of IRS elements is sufficiently large. More specifically, we consider four different types of IRSs. First, as an idealized system, we introduce active IRSs (whose elements can amplify and attenuate the signal in addition to adding a phase shift).
Then, we introduce passive IRSs  (whose elements can attenuate the signal in addition to adding a phase shift), passive lossless IRSs (whose elements can only apply a phase shift to the signal), and $\varepsilon$-relaxed passive lossless IRSs  (whose elements are capable of scaling the signal  by a number between $1-\varepsilon$ and $1$ in addition to adding a phase shift).
{We note that although active IRS has been proposed as a  viable option in the literature, see e.g., \cite{Alexandropoulos}-\cite{Dai}, in this paper, we consider active IRS to obtain an upper bound for the DoF achievable with the other types of IRSs and  use the framework developed for the analysis of active IRSs as the basis for the analysis of other kinds of IRSs.}  In addition, we use $\varepsilon$-relaxed passive lossless IRSs to approximate passive lossless IRSs.
In the following, we summarize the main results of this paper for each type of IRS:

\begin{itemize}
\item
 Using a modified interference alignment scheme,  we derive  the  DoF region of the $K$-user interference channel in the presence of  active IRSs for the case where the network matrix $\bf N$, which is a $K\times K$ matrix whose elements characterize the connectivity of the network, is  fixed for all channel uses. We show that this region only depends on the network matrix $\bf N$ and is also applicable for a partially connected network (without IRS). Subsequently, we derive inner and outer bounds for the DoF region for the general case where the network matrix can change from one channel use to the next one and provide lower and upper bounds for the sum  DoF. We show that a sum  DoF of $K$ can be achieved by choosing the number of IRS elements larger than a certain  finite value.
\item
{Due to the reduced capabilities of passive IRSs,
the analysis of active IRSs is not applicable to passive IRSs and a new analysis is needed, which explicitly accounts for the additional passivity condition. In particular, for passive IRSs, a deterministic DoF improvement cannot be guaranteed.}
Thus, we   derive probabilistic inner and outer bounds for the  DoF region and probabilistic lower and upper bounds for the sum  DoF of the $K$-user interference channel in the presence of  passive IRSs {\color{black}based on the mathematical framework developed for  active IRSs.} We also prove that the lower bound for the sum  DoF  asymptotically approaches $K$ as  the number of IRS elements tends to infinity. 
\item

We introduce the $\varepsilon$-relaxed passive lossless IRS as an approximation of passive lossless IRSs and develop a new lower bound for the sum  DoF of  the $\varepsilon$-relaxed passive lossless IRS-assisted $K$-user interference channel. We prove that this lower bound  asymptotically approaches $K$ as the number of IRS elements tends to infinity. In addition, to analyze the passive lossless IRS, we define a relaxed type of DoF called $\rho$-limited DoF. We show that by equipping the passive lossless IRS with a sufficient number of elements,  the $\rho$-limited sum DoF approaches $K$.

\end{itemize}

{We note that the DoFs of the interference channel have been extensively studied in the literature,} including   the DoF of the MIMO interference channel \cite{Gou2},  the DoF region of the interference channel \cite{Ramamoorthy},\cite{Khalil},  and  the DoF of the partially connected interference channel \cite{Ruan}-\cite{Sheng2}.  Interference alignment techniques play an essential role in proving DoF achievability theorems for interference channels. An overview of  available results on interference alignment is provided in \cite{Richard}. {In this paper, we will show that an IRS can be used to convert a fully connected wireless network into a partially connected wireless network with correlated channel coefficients by eliminating some cross links, which improves the DoF. To prove this, a different kind of stochastic analysis is needed  compared to that for conventional networks, and is presented in this paper.}

{ Furthermore, we note that, at first glance, IRS-assisted and relay-assisted interference channels seem to be similar. However, there are fundamental differences. {In particular, for conventional relays, the output in the $t$-th time slot is a function of the signals received at the relay in time slots $t'\in\{1,...,t-1\}$, i.e., the relay has memory. In contrast, the IRS output in the $t$-th time slot is a function of the received signal in the $t$-th time slot only, i.e., an IRS only changes the electromagnetic properties of the incident wave and reflects it instantaneously.}
In fact, it was shown in \cite{Jafar6} that conventional relays cannot increase the DoFs of the $K$-user interference channel. However, in this paper, we prove that an IRS can improve the DoF due to its instantaneous reflection property, i.e., the IRS provides an output signal for the signal received in the same time slot, which a conventional relay is not capable of doing.
{Moreover, in the literature, also an idealized type of relay, referred to as instantaneous relay (IR), has been studied \cite{El Gamal}, whose output in the $t$-th time slot is a function of its received signals in time slots $t'\in\{1,...,t\}$.} 
For IRs, it is implicitly assumed that the processing of the received signal comprising RF chains, digital processing, and re-transmission can occur within a small fraction of a channel use, which is an idealized assumption and may not be realizable in practice. Furthermore, for IRs, the signals received at  different antennas are jointly processed.
 On the other hand, IRSs assume only the instantaneous reflection of impinging waves, which does not require RF chains nor complex signal processing.
Also, the signals received at different IRS elements are reflected individually.
 While IRSs can be considered as special cases of IRs with limited functionalities,  the literature on IRs is limited to the general idealized model of IRs and the corresponding DoF results are not applicable to IRSs \cite{Lee}-\!\!\cite{Azari}.

The DoF of the IR-assisted two-user interference channel, when all nodes have the same number of antennas, was studied in \cite{Lee}. It was shown that a DoF of $\frac{3}{2}$ can be achieved.
The DoF of the $M$ antenna three-user interference channel in the presence of an IR was studied in \cite{Qiang}. It was proved that $2M$ DoFs are achievable.
The DoF of the IR-assisted two-way $K$-user interference channel, when the IR has $2K$ antennas, was studied in \cite{Cheng} and it was shown that $K$ DoFs can be achieved.
The DoF of the IR-assisted two-user interference channel for arbitrary numbers of IR transmit and receive antennas was studied in \cite{Liu}, and an inner and two outer bounds were derived.
For the IR-assisted $K$-user interference channel, when all nodes have the same number of antennas, an achievable scheme and an outer bound were proposed in \cite{Azari}. Though the DoFs in some special cases were obtained in \cite{Azari}, a general achievable DoF was not derived. In these studies \cite{Lee}-\cite{Azari},  a general framework that could be used for the analysis of the time-selective $K$-user interference channel assisted by an IRS was not provided.}

The remainder of this paper is organized as follows. The considered system model is presented in Section \ref{sec2}. In Sections \ref{sec3} and \ref{sec4}, we provide the main DoF results for the $K$-user interference channel in the presence of  active and passive IRSs, respectively. The DoF results for the   passive lossless and $\varepsilon$-relaxed passive lossless IRSs are given in  Section \ref{sec5}. In Section \ref{sec6}, we provide  numerical results to illustrate our DoF analysis.  Finally, Section \ref{sec7} concludes the paper.

\textbf{Notations:}
We denote sets and vector spaces by calligraphic upper case letters. Bold letters denote matrices and $x_{i,j}$ denotes the entry in the $i$-th row and $j$-th column of matrix ${\bf X}$. $\mathbb{R}$ is the set of real numbers. For a set $\cal A$, $|\cal A|$ denotes the cardinality of $\cal A$. ${\bf V}^{T}$ and ${\bf V}^{H}$ are the transpose and Hermitian  of matrix $\bf V$, respectively. ${\textrm{diag} (a_1,...,a_m)}$ refers to a diagonal matrix with main diagonal elements $a_1,...,a_m$. $\det({\bf H})$ denotes the determinant of  square matrix $\bf H$. Sequence $a(Q)$ converges to its limit $a^*$ with an order of at least $O(g(Q))$, if
\begin{equation*}
\mathop {\lim }\limits_{Q \to \infty } \frac{{\left| {a(Q) - {a^*}} \right|}}{{|g(Q)|}} < \infty   .
\end{equation*}
We say  sequence $a(n)$ goes to infinity with an order of $O(n^l)$, if $0 < \mathop {\lim }\limits_{n \to \infty } \frac{{\left| {a(n)} \right|}}{{{n^l}}} < \infty $. Whereas  a function $f(\rho)$ is $o(\log(\rho))$, if
\begin{equation*}
\mathop {\lim }\limits_{\rho  \to \infty } \frac{{|f(\rho )|}}{{\log (\rho )}} = 0.
\end{equation*}
  $\Pr\{{\cal A}\}$ denotes the probability measure of event $\cal A$. $\mu(\cal A)$ refers to the Lebesgue measure of  set $\cal A$.
 We also define $\mathbb{N}=\{1,2,3,...\},   \mathbb{W}=\{0,1,2,3,...\},\mathbb{Z}=\{...,-1,0,1,...\}$, and $\Phi  = \{ \}$. For random variables $X$ and $Y$,  the differential entropy of $X$ and the mutual information between $X$ and $Y$, denoted by $H(X)$ and $I(X;Y)$, respectively, are defined as follows:
\begin{equation*}
H(X) = \int\limits_x { - \log ({f_X}(x)){f_X}(x)dx} ,
\end{equation*} 
\begin{equation*}
I(X;Y)=H(X)+H(Y)-H(X,Y),
\end{equation*}
where $f_{X}(x)$ is the probability distribution of $X$.

\section{System Model and Preliminaries}

\label{sec2}

\subsection{System Model}

\begin{figure}
\centering
\includegraphics[width=10cm]{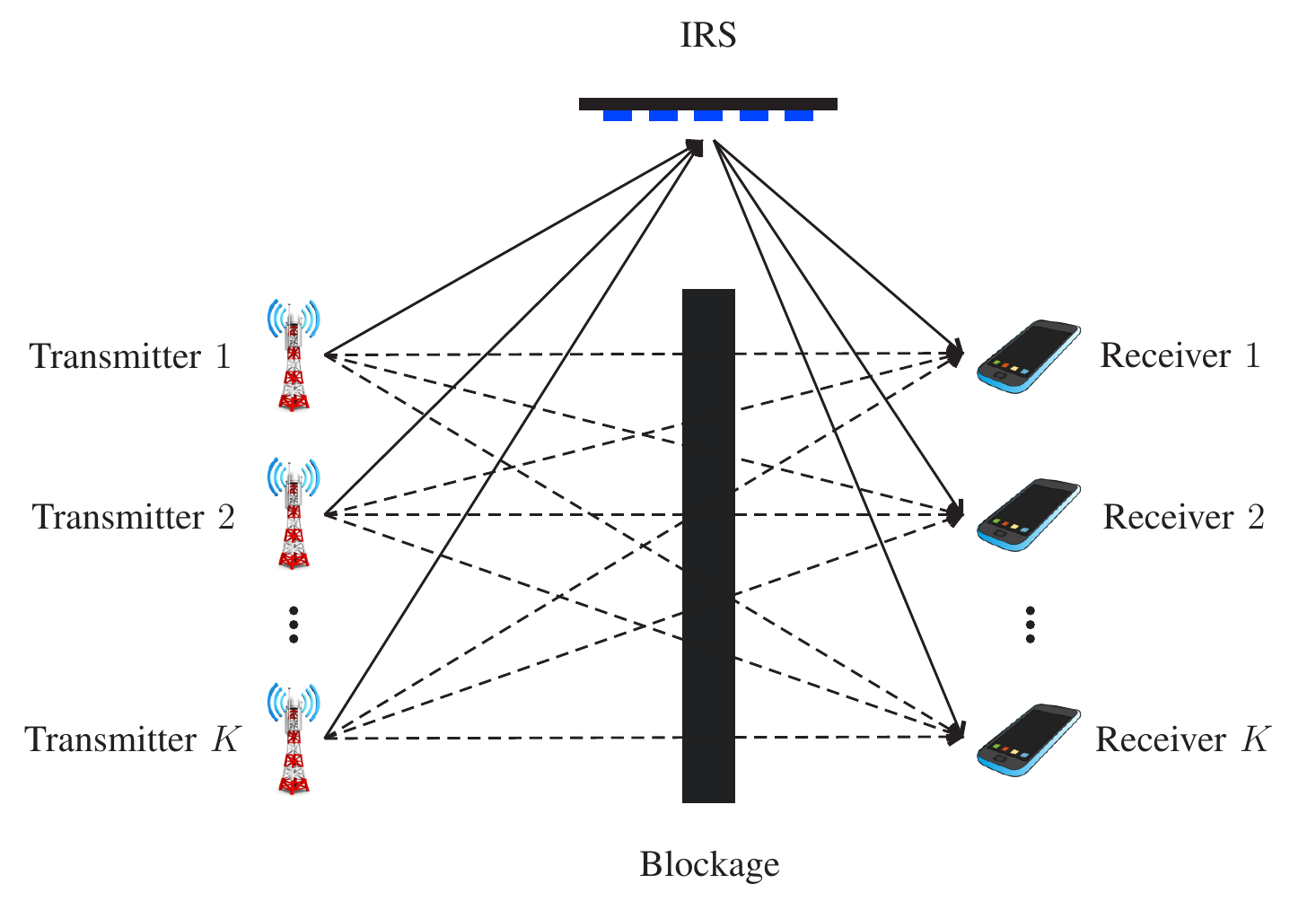}
\caption{Illustration of the IRS-assisted $K$-user interference channel. Direct links are shown by dashed black arrows, and links between the IRS and transmitters or receivers are shown by continuous black arrows. The black rectangle represents a blockage of the direct links.}
 \label{illustration}
\end{figure}

We consider a time-selective $K$-user interference channel in the presence of a $Q$-element IRS, where $K$ single-antenna transmitters send their messages to $K$ single-antenna receivers.  In this system, the $i$-th transmitter  sends  message $w^{[i]}$ to the $i$-th receiver. An illustration of the system model is shown in Fig. \ref{illustration}.

As is customary for DoF analysis \cite{Cadambe1}-\cite{Ramamoorthy}, we assume that the channel is time-selective.
 The received signal at the $j$-th receiver in the $t$-th time slot  is given by $Y^{[j]}({t})$ and can be written as follows:
\begin{equation}
\scalebox{.95}[1]{${Y^{[j]}}({t}) = \sum\limits_{i = 1}^K {{H^{[ji]}}({t}){X^{[i]}}({t})}  + \sum\limits_{u = 1}^Q {{H_{\rm IR}^{[j{u}]}}({t}){X_{\rm IRS}^{[{u}]}}({t})}  + {Z^{[j]}}({t}),\quad t\in\{1,...,T\}$}
\label{model receiver}
\end{equation}
where $X^{[i]}({t})$ is the output of the $i$-th transmitter, ${H^{[ji]}}({t})$ is the channel coefficient from the $i$-th transmitter to the $j$-th receiver, ${{X_{\rm IRS}^{[{u}]}}({t})}$ is the output of the $u$-th IRS element, ${{H_{\rm IR}^{[j{u}]}}({t})}$ is the channel coefficient from the $u$-th  IRS element  to the $j$-th receiver,
  ${Z^{[j]}}({t})$ is the additive white complex Gaussian noise at the $j$-th receiver having variance $N_0$, and $T$ is the number of time slots.    The received signal at the $q$-th IRS element in the $t$-th time slot, $Y_{\rm IRS}^{[{q}]}({t})$, is given as follows:
\begin{equation}
\scalebox{.95}[1]{${Y_{\rm IRS}^{[{q}]}}({t}) = \sum\limits_{i = 1}^K {{H_{\rm TI}^{[{q}i]}}({t}){X^{[i]}}({t})},$}
\label{model relay}
\end{equation}
where ${H_{\rm TI}^{[{q}i]}}({t})$ is the channel coefficient from the $i$-th transmitter to the $q$-th IRS element ($q \in \{1,...,Q\}$). 
The output of the $u$-th IRS element is given by
\begin{equation}
\scalebox{.95}[1]{${X_{\rm IRS}^{[{u}]}}({t}) =\tau^{[u]}({t}){Y_{\rm IRS}^{[{u}]}}({t})= \rho^{[u]}({t}) e^{j\phi^{[u]}({t})}{Y_{\rm IRS}^{[{u}]}}({t}),u \in \{1,...,Q\},$}
\label{func relay3}
\end{equation}
where $\tau^{[u]}(t)$, $\phi^{[u]}({t})\in[0,2\pi]$, and $\rho^{[u]}({t})$ are the coefficient, phase shift, and amplitude  added to the received signal by the $u$-th IRS element, respectively.
The amplitude factor $\rho^{[u]}(t)$ depends on the type of IRS and we consider the following cases: 1) active IRS, 2) passive IRS, 3) passive lossless IRS, and 4) $\varepsilon$-relaxed passive lossless IRS, as explained in the following. If $\rho^{[u]}({t})\in \mathbb{R}^+$, we call the IRS ``active". Active IRSs are proposed in \cite{Alexandropoulos}-\cite{Dai} and we provide a framework for the characterization of the corresponding achievable DoF, which will then be used for analyzing the DoF of the other types of IRSs. In addition, the performance of active IRSs serves as an upper bound for the performance of the other more realistic IRS models. 
From a physical point of view, an active IRS is equipped with controllable amplifier  and phase shifter\footnote{Active IRSs do not employ receive/transmit RF chains  and the incident wave is only amplified while being reflected \cite{Alexandropoulos}-\!\!\cite{Dai}. This amplification may introduce a low level of noise, which however is negligible compared to the noise introduced by the RF chains at the receivers \cite{Bousquet}. Moreover, in this paper, the communication channel is analyzed in the high signal-to-noise ratio (SNR) regime. Hence, the amount of the additive Gaussian noise does not affect the DoF results. Furthermore, we note that the signal amplification at the IRS elements does not cause a significant delay in the reflection of the incident wave \cite{Bousquet}.}. Thus, it can amplify or attenuate the received wave in addition to applying a phase shift\footnote{The amplifier integrated into the elements of the active IRS can be realized by different existing active components, such as current-inverting converters \cite{Loncar} or asymmetric current mirrors \cite{Bousquet}.}.
If $\rho^{[u]}( t)\in [0,1]$, we  call the IRS ``passive", if $\rho^{[u]}( t)=1$, we call it ``passive lossless", and if $1-\varepsilon \le\rho^{[u]}( t)\le1$, where $0<\varepsilon<1$, we call it ``$\varepsilon$-relaxed passive lossless".
Passive and $\varepsilon$-relaxed passive lossless IRSs are equipped with a controllable resistor in addition to a phase shifter to realize $\rho^{[u]}(t)<1$ and $1-\varepsilon<\rho^{[u]}(t)<1$, whereas the passive lossless IRS is  equipped only with a phase shifter.

We assume that in the $t$-th time slot, the channel coefficients of the direct links   ($H^{[ji]}(t'),\forall i,j,\forall t'\in\{1,...,t\}$) and the concatenation of the transmitter-IRS and IRS-receiver channel coefficients (i.e., $H_{\rm TI}^{[ui]}(t')$ $H_{\rm IR}^{[ju]}(t'),\forall i,j,u,\forall t'\in\{1,...,t\}$)   are  known  at  the transmitters, the receivers, and the IRS\footnote{We note that the analysis of the DoF for imperfect CSI is much more complicated than that for perfect CSI and requires more complicated analysis tools, see, e.g. \cite{Jafar7,Jafar8}. Hence, the DoF analysis for imperfect CSI is out of the scope of this paper but constitutes an important direction for future research.}.  For channel estimation schemes for IRS-assisted systems, we refer to \cite{Cui}-\cite{Al-Nahhas-n}.
We also assume that
the channel coefficients ${H^{[ji]}}({t})$, ${{H_{\rm IR}^{[j{u}]}}({t})}$, and ${H_{\rm TI}^{[{q}i]}}({t})$ are independent random variables for all $i,j,u,q,$ and $t$ with a continuous cumulative probability distribution. We note that these channel coefficients are complex in general, whereby, the real and imaginary parts  must be independent random variables with a continuous cumulative probability distribution (e.g., complex Gaussian distribution).

{The assumption of independent channel coefficients for each element of the IRS requires the elements to be sufficiently spaced, i.e., by more than half a wavelength \cite{Sanguinetti}.}  In addition, the  independence of the channel coefficients with respect to time (time-selectivity) can be realized by interleaving.
Note that, in this paper, we start with the  idealized active IRS model.
Then, we relax the idealized assumptions of the initial model step by step and study their impact in each step.

{\textit{Remark 1}}: {When the spacing between the IRS elements is more than half a wavelength, the correlation between the elements is not perfectly zero; however, the correlation is small such that modeling the corresponding channels as independent is a reasonable approximation. Hence, independent channels have been widely assumed in the IRS literature \cite{Wu4}-\!\!\cite{Huang10}. Moreover, if the element spacing is more than $\frac{\lambda}{\sqrt 2}$, where $\lambda$ denotes the wavelength, the aperture area is not efficiently exploited, and in addition, grating lobes start to appear, which deteriorate the performance \cite{Hannan}. Thus, an element spacing between $\frac{\lambda}{2}$ and $\frac{\lambda}{\sqrt 2}$ provides an appropriate trade-off between avoiding channel correlation and avoiding grating lobes \cite{Hannan}. We note that IRSs with less than $\frac{\lambda}{2}$ element spacing have been reported in the literature \cite{Najafi},\cite{Sanguinetti}; however, for simplicity of analysis, we focus on $\frac{\lambda}{2}$ element spacing, which leads to negligible channel correlation without grating lobes.}

We assume that the maximum transmit power of each transmitter is $\rho $. Denoting  the size of the message of the  $i$-th transmitter by ${\left| {{w^{[i]}}} \right|}$, we say the  rates ${r_i}(\rho ) = \frac{{\log (\left| {{w^{[i]}}} \right|)}}{T}, \forall i$, are achievable, if the decoding error probability of  all messages goes to zero as the number of time slots $T$ tends to infinity. ${\cal C}(\rho)$ denotes the closure of the set of all achievable rates ${{\bold r}}(\rho ) = ({r_1}(\rho ),...,{r_K}(\rho ))$.

\subsection{Preliminaries}

In the following, we introduce some basic definitions that are used throughout the paper. 

\textbf{Degrees of freedom (DoF)}:
Similar to \cite{Cadambe1}, for the $K$-user interference channel, we define the  DoF region ${\cal D}$ as follows:
\begin{equation}
\scalebox{.87}[1]{${{\cal D}} = \left\{ {({d_1},...,{d_K}) \in {\mathbb{R}}_ + ^K:\forall ({\alpha_1},...,{\alpha_K}) \in {\mathbb{R}}_ + ^K,} {{\alpha_1}{d_1} + ... + {\alpha_K}{d_K} \le \mathop {\lim \sup }\limits_{\rho  \to \infty } \left( {\mathop {\sup }\limits_{{{\bold r}}(\rho ) \in {{\cal C}}(\rho )} \left( {{\alpha_1}{r_1} + ... + {\alpha_K}{r_K}} \right)} \right)\frac{1}{\log(\rho)}} \right\},$}
\label{exact DoF}
\end{equation}
where $d_j$ is the DoF for the $j$-th receiver. { $\alpha_j$ in (\ref{exact DoF}) is an auxiliary variable needed for the definition of the DoF region. Since optimizing $r_1,...,r_K$ leads to a multi-objective optimization problem, for given weights $\alpha_1,...,\alpha_K$, one point on the boundary of the rate region is obtained. The entire rate region is obtained by varying the weights.}

\textbf{$\rho$-limited DoF}:
We define the $\rho$-limited DoF as a relaxed version of the original DoF.  In this definition for the DoF, the parameter $\rho$ has a given value and does not go to infinity. For a $K$-user interference channel, we define the $\rho$-limited DoF region ${{\cal D}^{[\rho]}}$ as follows:
\begin{equation}
{{\cal D}^{[\rho ]}} = \bigcup\limits_{({r_1}(\rho ),...,{r_K}(\rho )) \in {\cal C}(\rho )} {\hat {\cal D}({r_1}(\rho ),...,{r_K}(\rho ))} .
\label{relaxed rho DoF}
\end{equation}
where:
\begin{equation*}
\hat {\cal D}({r_1}(\rho ),...,{r_K}(\rho )) = \left\{ {(d_1^{[\rho ]},...,d_K^{[\rho ]}) \in \mathbb{R}_ + ^K:\forall i,{r_i}(\rho ) \ge d_i^{[\rho ]}\log \left( \rho  \right)} \right\},
\end{equation*}
We will use $\rho$-limited DoF for analyzing passive lossless IRSs, for which perfect interference cancellation is not possible.

\textbf{Span}:
We denote the space spanned by the column vectors of  matrix $\bold V$ by $\textrm{span}({\bold V})$.

\textbf{Dimension}: We refer to the number of dimensions of $\textrm{span}({\bold V})$ as the dimension of $\bold V$  and  denote it by $d({\bold V})$. In other words, $d({\bold V})$ is equal to the rank of $\bf V$.

We also use this notation for  vector spaces $\cal{A}$, i.e., $d({\cal A})$ denotes the dimension of $\cal A$.

\textbf{Network matrix}: The network matrix is a $K\times K$ matrix that characterizes the topology of the network. We denote the network matrix by ${\bf N}$ where the element in the $i$-th row and $j$-th column, denoted by $n_{i,j}$, is $1$, if there is a link between the $i$-th transmitter and the $j$-th receiver, otherwise, it is $0$.
For the IRS-assisted interference channel, $\bf N$ denotes the effective network matrix that accounts for the impact of the IRS.
Note that unlike for conventional wireless channels, where the network matrix is given and fixed, in IRS-assisted systems, the IRS can be designed to change the network matrix, e.g., by canceling the signal received at some of the receivers, which effectively creates zero entries in the network matrix.

\section{$K$-user Interference Channel in the Presence of an Active IRS}

\label{sec3}

\subsection{DoF Region}

Before introducing our DoF results, we show that a $Q$-element active IRS is able to change the topology of the network  by realizing a network matrix $\bf N$, which has $Q\le K^2-K$ zero entries with probability  $1$. This can be achieved by choosing the IRS coefficients such that the following equations are satisfied:
 \begin{equation}
\scalebox{.94}[1]{$\sum\limits_{u \in \{ 1,...,Q\} } {{H_{\rm TI}^{[{u}i]}}({t}){H_{\rm IR}^{[j{u}]}}({t}){\tau ^{[u]}}({t})}  =  - {H^{[ji]}}({t}), \quad  i \ne j,(i,j) \in {\cal A},\left| {\cal A} \right| = Q,$}
\label{found cancel}
\end{equation}
where ${\cal A}=\{(i,j)|n_{i,j}=0\}$ and   we impose $i\ne j$ as the direct links should not be removed, of course. Note that the system of  equations in (\ref{found cancel}) is solvable almost surely because if we write (\ref{found cancel}) in matrix form, ${\bf H}\tau={\bf h}$, where ${\bold{H}}$ is a matrix with elements  ${H_{\rm TI}^{[{u}i]}}({t}){H_{\rm IR}^{[j{u}]}}({t}),(i,j)\in {\cal A}$, ${\bold{\tau}}$ is a column vector with elements  ${\tau ^{[u]}}({t})$, and ${\bold h}$ is a column vector with elements  $- {H^{[ji]}}({t}),(i,j)\in {\cal A}$, then $\det ({\bf H})$ will be a nonzero polynomial in terms of $H_{\rm TI}^{[{u}i]}$ and $H_{\rm IR}^{[j{u}]}$ and by Lemma \ref{lemma1}, we have $\Pr\{\det({\bf H})=0\}=0$.
In addition, we introduce some notations  for simplicity of presentation. We rewrite  (\ref{model receiver}) and (\ref{model relay}) in vector form as follows:
\begin{equation}
\scalebox{.94}[1]{${\bold{y}^{[j]}} = \sum\limits_{i = 1}^K {{\bold{H}^{[ji]}}{\bold{x}^{[i]}}}  + \sum\limits_{u = 1}^Q {{\bold{H}_{\rm IR}^{[j{u}]}}{\bold{x}_{\rm IRS}^{[{u}]}}}  + {\bold{z}^{[j]}},\quad{\bold{y}_{\rm IRS}^{[q]}} = \sum\limits_{i = 1}^K {{\bold{H}_{\rm TI}^{[{q}i]}}{\bold{x}^{[i]}}}  ,$}
\label{output}
\end{equation}
where $\bold{x}^{[i]}$ is a $T \times 1$ column vector containing the channel inputs ${X}^{[i]}( t)$, i.e.,
\begin{equation}
\scalebox{.94}[1]{${{\bf{x}}^{[i]}} = {\left[ {\begin{array}{*{20}{c}}
{{X^{[i]}}({1})}&{{X^{[i]}}({2})}& \cdots &{{X^{[i]}}({T})}
\end{array}} \right]^T}.$}
\end{equation}
Vectors $\bold{y}^{[i]}$, $\bold{y}_{\rm IRS}^{[q]}$, $\bold{x}_{\rm IRS}^{[u]}$, and $\bold{z}^{[j]}$  are defined  similarly. ${\bold{H}^{[ji]}}$ is a diagonal matrix defined as follows:
\begin{equation*}
\scalebox{.94}[1]{${{\bf{H}}^{[ji]}} = \textrm{diag}\left( {{H^{[ji]}}({1}),{H^{[ji]}}({2}), \ldots ,{H^{[ji]}}({T})} \right).$}
\end{equation*}
Matrices ${\bold{H}_{\rm IR}^{[j{u}]}}$ and ${\bold{H}_{\rm TI}^{[{q}i]}}$ are defined similarly. Now, we are ready to provide a theorem for the DoF region of the  $K$-user interference channel in the presence of an active IRS, where we assume  the network matrix is fixed across all time slots. This theorem will be the basis of our subsequent theorems for active and passive IRSs.
\begin{theorem}
{Consider a $K$-user interference channel  assisted by a $Q$-element active IRS. Based on (\ref{found cancel}), if the IRS elements are set such that network matrix $\bf N$ is fixed and at most {$Q$ of its non-diagonal elements are $0$ across all $T$ time slots,} then, the   DoF region is given as follows:}
\begin{equation}
\scalebox{.94}[1]{${{\cal D}_{\bf N}} = \left\{ {({d_1},...,{d_K}) \in \mathbb{R}_ + ^K\left| {\begin{array}{*{20}{c}}
{\forall i,j \in \{ 1,...,K\} ,i \ne j:({d_i} + {d_j}) \le 2 - \max \{ {n_{i,j}},{n_{j,i}}\} ,}\\
{\forall i \in \{ 1,...,K\} :0 \le {d_i} \le 1,}
\end{array}} \right.} \right\},$}
\label{DoF region const active}
\end{equation}
 where $n_{i,j}$ is  the  element in the $i$-th row and $j$-th column of network matrix $\bf N$.
\label{thm  DoF reg const active IRS}
\end{theorem}
\begin{IEEEproof}
The proof is provided in Appendix \ref{appendix1}.
\end{IEEEproof}

\textit{Remark 2}: We note that Theorem \ref{thm  DoF reg const active IRS} is valid for any interference channel with network matrix $\bf N$ regardless of whether or not it is equipped with an IRS. Nevertheless, Theorem \ref{thm  DoF reg const active IRS} provides insight into which network matrices are more beneficial for DoF region improvement and serves as the basis for the design of the IRS in this paper.

Now, we give inner and outer bounds for the  DoF region of the active IRS-assisted $K$-user interference channel, when the network matrix is allowed to change across different time slots.  First, we introduce the inner bound which is based on time sharing of the  DoF region provided in Theorem \ref{thm  DoF reg const active IRS}.
\begin{theorem}
Consider a $Q$-element active IRS-assisted $K$-user interference channel and denote ${\cal N}_Q$ as  the set of all possible network matrices with at most $Q$ zero non-diagonal elements facilitated by the IRS. 
Then, the following region is achievable:
\begin{equation}
{{\cal D}_{active - in}} = \bigcup\limits_{{\bf{a}} \in {\cal A}} {\left\{ {\sum\limits_{i = 1}^{\left| {{{\cal N}_Q}} \right|} {{a_i}{{\bf{d}}_i}} \left| {{{\bf{d}}_i} \in {{\cal D}_{{{\bf{N}}_i}}},{{\bf{N}}_i} \in {{\cal N}_Q}} \right.} \right\}} ,
\label{DoF region active inner2}
\end{equation}
where set ${\cal D}_{{\bf N}_i}$ is given by (\ref{DoF region const active}) and
\begin{equation}
{\cal A} = \left\{ {{\bf{a}}\left| {0 \le {a_i} \le 1,\sum\limits_{i = 1}^{\left| {{{\cal N}_Q}} \right|} {{a_i} = 1} } \right.} \right\}.
\label{A_eq}
\end{equation}
\label{active IRS inner  DoF reg}
\end{theorem}
\begin{IEEEproof}
The proof is provided in Appendix \ref{appendix2}.
\end{IEEEproof}

\textit{Remark 3}: {Since the cardinality of set ${\cal N}_Q$ grows with an order of $2^{K(K-1)}$, the complexity of the evaluation of the DoF increases rapidly with $K$. In contrast, the computational complexity of symbol decoding in the proposed achievability scheme grows with an order of $n^{3(K^2-K-Q)}$,
where $n$ is an integer auxiliary variable.
This complexity decreases rapidly with $Q$, i.e., if we set $Q=K^2-mK$, the order of complexity will be $n^{3K(m-1)}$.}

As the DoF gives  a capacity approximation, which is accurate within $o(\log(\rho))$, we have:
\begin{equation*}
 \bigcup\limits_{{\bf{a}} \in {\cal A}} {\left\{ {\left(\sum\limits_{i = 1}^{\left| {{{\cal N}_Q}} \right|} {{a_i}{{\bf{d}}_i}}\right)\log(1+\rho)+o(\log(\rho)) \left| {{{\bf{d}}_i} \in {{\cal D}_{{{\bf{N}}_i}}},{{\bf{N}}_i} \in {{\cal N}_Q}} \right.} \right\}}\subseteq {{\cal C}(\rho)} .
\end{equation*}

Next, we introduce an outer bound for the  DoF region of the $K$-user interference channel with an active IRS. The corresponding theorem shows that similar to the inner bound introduced in Theorem \ref{active IRS inner  DoF reg}, the outer bound depends on the network matrix of each time slot and the percentage of their occurrence.

\begin{theorem}
Consider a $Q$-element active IRS-assisted $K$-user interference channel and denote ${\cal N}_Q$ as  the set of all possible network matrices with at most $Q$ zero non-diagonal elements facilitated by the IRS. For each network matrix ${\bf N}\in {\cal N}_Q$, we define the  following parameters:
\begin{equation}
\forall i,j \in \{ 1,...,K\} :{{\tilde d}^{[ij]}}_{{\bf N}} = 2-\frac{{{n_{i,j}} + {n_{j,i}}}}{2}.
\label{DoF region active outer1}
\end{equation}
In addition, for each ${\bf a}\in {\cal A}$ with $\cal A$  given in (\ref{A_eq}), we define the following set:
\begin{equation}
{{\cal D}'({\bf a})} =  {\left\{ {{\bf d}\left| \begin{array}{l}
\forall i,j \in \{ 1,...,K\} ,i \ne j:{d_i} + {d_j} \le \sum\limits_{m = 1}^{\left| {{{\cal N}_Q}} \right|} {{a_m}{{\tilde d}^{[ij]}}_{{\bf N}_m}} \\
{{\bf N}_m} \in {{\cal N}_Q}: {{\bf N}_m} \ne {{\bf N}_{m'}},m \ne m' 
\end{array} \right.} \right\}} .
\label{equation_a_m}
\end{equation}
Then, the  DoF region is a subset of the following region:
\begin{equation}
{{\cal D}_{active - out}} = \bigcup\limits_{{\bf a} \in {\cal A}} {{\cal D}'({\bf a})}.
\label{DoF region active outer2}
\end{equation}

\label{active IRS outer  DoF reg}
\end{theorem}
\begin{IEEEproof}
The proof is provided in Appendix \ref{appendix3}.
\end{IEEEproof}

The inner bound given by ${{\cal D}_{active-in}}$ and the outer bound given by ${{\cal D}_{active-out}}$ have two differences which cause the outer bound to not necessarily coincide with the inner bound: 1) the expression $\max\{n_{i,j},n_{j,i}\}$ in  ${{\cal D}_{active-in}}$ is replaced by  $\frac{n_{i,j}+n_{j,i}}{2}$ in  ${{\cal D}_{active-out}}$, 2) ${{\cal D}_{active-out}}$ is not necessarily achievable with time sharing.
For the  approximation of the capacity region, from Theorem \ref{active IRS outer  DoF reg} we obtain:
\begin{equation*}
{{\cal C}}(\rho) \subseteq \bigcup\limits_{{\bf a} \in {\cal A}} {{\cal C}'({\bf a})},
\end{equation*}
where:
\begin{equation*}
{{\cal C}'({\bf a})} =  {\left\{ {{\bf d}(\log(1+\rho)+o(\log(\rho)))\left| \begin{array}{l}
\forall i,j \in \{ 1,...,K\} ,i \ne j:{d_i} + {d_j} \le \sum\limits_{m = 1}^{\left| {{{\cal N}_Q}} \right|} {{a_m}{{\tilde d}^{[ij]}}_{{\bf N}_m}} \\
{{\bf N}_m} \in {{\cal N}_Q}: {{\bf N}_m} \ne {{\bf N}_{m'}},m \ne m' 
\end{array} \right.} \right\}} .
\end{equation*}

\textit{Remark 4}:
In this section, we have not explicitly included a power constraint for  the active IRS. However, if a power constraint is considered, e.g., $|\tau^{[u]}(t)|<P,\forall u$, then, in $\Pr\{|\tau^{[u]}(t)|<P,\forall u\}$ fraction of time slots, in which the power constraint is satisfied, the results are applicable and  by setting $P$ sufficiently large, $\Pr\{|\tau^{[u]}(t)|<P,\forall u\}$ tends to $1$. The analysis of an IRS with a strict power constraint, i.e., $|\tau^{[u]}(t)|<1,\forall u$, will be presented in Section \ref{sec4}.

\subsection{Sum  DoF}

Next, we use the inner and outer bounds given in Theorems \ref{active IRS inner  DoF reg} and \ref{active IRS outer  DoF reg} to provide lower and upper bounds for the sum  DoF of an active IRS-assisted $K$-user interference channel in Theorems \ref{active IRS sum in} and \ref{active IRS sum up}, respectively.

\begin{theorem}
Assume $W\in \{0,1,...,K\}$. Then, with an active IRS with $Q=W(K-1)+W(K-W)$ elements, the following sum  DoF is achievable\footnote{When  an integer $W$ that satisfies $Q= W(K-1)+W(K-W)$ does not exist, we adopt the largest $Q'\le Q$, for which there exists an integer $W$, such that $Q'= W(K-1)+W(K-W)$ holds. This implies that we use only $Q'$ elements of the IRS instead of  $Q$ elements.}:
\begin{equation}
{DoF}_{active-low}=\frac{K+W}{2}.
\end{equation}
\label{active IRS sum in}
\end{theorem} 
\begin{IEEEproof}
The proof is provided in Appendix \ref{appendix4}.
\end{IEEEproof}
\begin{theorem}
For an active IRS with $Q\le K(K-1)$ elements, the  sum  DoF is upper bounded by:
\begin{equation}
{DoF}_{active-up}=\frac{K}{2}+\frac{Q}{2(K-1)}.
\end{equation}
\label{active IRS sum up}
\end{theorem} 
\begin{IEEEproof}
The proof is provided in Appendix \ref{appendix5}.
\end{IEEEproof}

\textit{Corollary 1}: If we set $W=K$ and $Q=K(K-1)$ in Theorem \ref{active IRS sum in}, the maximum sum DoF of $K$ is achieved. {If $Q>K(K-1)$, there is no further DoF gain. Nevertheless, the additional IRS elements are still beneficial as they can be exploited to improve the link budget, i.e., to increase the receive SNR.}

Corollary 1 reveals that the maximum possible sum DoF of the $K$-user interference channel {\color{black}assisted by an active IRS is $K$}, whereas without an IRS, the maximum sum DoF is $\frac{K}{2}$ \cite{Cadambe1}. From Theorem \ref{active IRS sum in}, we can see that with an active IRS, we require a quadratically large number of elements, $Q=K(K-1)$, to achieve the maximum sum DoF of $K$. { This achievability scheme can be interpreted as follows. An active IRS with $Q=W(K-1)+W(K-W)$ elements can decompose the $K$-user interference channel into a channel with $W$ interference-free users, which can achieve $W$ sum DoF, and $(K-W)$ interfering users, which can achieve $\frac{K-W}{2}$ sum DoF. Thus, an overall sum DoF of $\frac{K+W}{2}$ is achievable.}

Since the  DoF provides a capacity approximation which is accurate within $o(\log(\rho))$, we can assert from Theorem \ref{active IRS sum in} that the sum capacity of the $K$-user interference channel assisted by an active IRS with $Q=W(K-1)+W(K-W)$ elements, where $0\le W \le K$, exceeds $\left( {\frac{{K + W}}{2}}-\epsilon \right)\log (1 + \rho ) + o(\log (\rho )),\forall \epsilon>0$. In addition, Theorem \ref{active IRS sum up} implies that the sum capacity of the $K$-user interference channel assisted by an active IRS with $Q$ elements, where $0\le Q\le K(K-1)$, is less than 
 $\left( {\frac{K}{2} + \frac{Q}{{2(K - 1)}}} \right)\log (1 + \rho ) + o(\log (\rho ))$. Furthermore, with a $K(K-1)$-element active IRS, the sum capacity can be approximated by $K\log(1+\rho)+ o(\log (\rho ))$.  In the next section, we use Theorems \ref{active IRS inner  DoF reg}-\ref{active IRS sum up} to derive  bounds for  the  DoF, and  consequently for the approximate capacity of the $K$-user interference channel in the presence of a passive IRS.

\section{$K$-user Interference Channel in the  Presence of a Passive IRS}

\label{sec4}

After analyzing the $K$-user interference channel in the presence of an active IRS, we study the class of passive IRSs. Due to the random channel realizations and the inability of passive IRSs to amplify signals, the DoF improvement introduced by the IRS is not deterministic. Thus, we derive probabilistic bounds.
The main difference between passive and active IRSs is that the set of possible network matrices in a time slot depends on the realization of the channel coefficients in that time slot, which introduces randomness.
Although the set of network matrices that can be realized by a passive IRS depends on the channel realization, for sufficiently large $T$, all possible sets occur with high probability in a particular fraction of time slots, which is proportional to the probability of realization of a particular set.
In other words, for $T$ realizations of discrete random variable $X$, the event $x$ will occur approximately $T\Pr\{x\}$ times with high probability for large enough $T$. Thus, for a sufficiently large number of time slots, we can ensure that a specific set of network matrices occurs in a fraction of $T$. Therefore, if we replace the set ${\cal N}_Q$ by the set of realizable network matrices, we can use the framework introduced in the previous section for these time slots. In the following, this procedure is described in detail.

\subsection{DoF Region}

Let us  define some new sets and random variables.
Define sets ${\tilde {\cal N}}_{Q_i},i\in\{1,...,2^{\left| {{{\cal N}_Q}} \right|-1}\}$, as all  subsets of ${\cal N}_Q$ including the  full matrix ${\bf N}_1$  with:
\begin{equation}
{{n_{{1_{i,j}}}} = 1,\forall i,j \in \{ 1,...,K\} }.
\label{all one}
\end{equation}
 In addition, we define event ${\cal E}_{Q_i}$ as the event that the network matrices ${\bf N}\in {\tilde {\cal N}}_{Q_i}$ can be realized and the network matrices ${\bf N}\in {\tilde {\cal N}}^c_{Q_i}$ cannot be realized  with a $Q$-element passive IRS, where $ {\tilde {\cal N}}^c_{Q_i}={\cal N}_Q-{\tilde {\cal N}}_{Q_i}$. Note that the events ${\cal E}_{Q_i}$ are distinct for $\forall i$. Now, we introduce a probabilistic outer bound for the  DoF region of the $K$-user interference channel assisted by a $Q$-element passive IRS.
\begin{theorem}
Define  the sets ${\cal D}''({{\bf{a}}^{[1]},...,{\bf{a}}^{[{2^{\left| {{{\cal N}_Q}} \right| - 1}}]}},\delta )$ as follows:
\begin{equation}
 \scalebox{.85}[1]{${\cal D}''({{\bf{a}}^{[1]}},...,{{\bf{a}}^{[{2^{\left| {{{\cal N}_Q}} \right| - 1}}]}},\delta ) = \left\{ {{\bf{d}}\left| {\begin{array}{*{20}{l}}
{\forall i,j \in \{ 1,...,K\} ,i \ne j:{d_i} + {d_j} \le \sum\limits_{l = 1}^{{2^{\left| {{{\cal N}_Q}} \right| - 1}}} {\left( {\Pr \left\{ {{{\cal E}_{{Q_l}}}} \right\} + \delta } \right)\sum\limits_{m = 1}^{\left| {{{\tilde {\cal N}}_{{Q_l}}}} \right|} {a_m^{[l]}{{\tilde d}^{[ij]}}_{{\bf{N}}_m^{[l]}}} } }\\
{{{\bf{a}}^{[l]}} \in {{\cal A}_l},{\bf{N}}_m^{[l]} \in {{\tilde {\cal N}}_{{Q_l}}}: {\bf{N}}_m^{[l]} \ne {\bf{N}}_{m'}^{[l]},m \ne m'}
\end{array}} \right.} \right\},$}
\label{DoF region passive outer1}
\end{equation}
where ${{\tilde d}^{[ij]}}_{{\bf{N}}_m^{[l]}}$ is given by (\ref{DoF region active outer1}) and  set ${\cal A}_l$ is given as follows:
\begin{equation}
{{\cal A}_l} = \left\{ {{{\bf{a}}^{[l]}}\left| {0 \le a_m^{[l]} \le 1,\sum\limits_{m = 1}^{\left| {{{\tilde {\cal N}}_{{Q_l}}}} \right|} {a_m^{[l]}}  = 1} \right.} \right\}, l \in \{1,...,{2^{\left| {{{\cal N}_Q}} \right| - 1}}\}.
\label{A_i_eq}
\end{equation}
 In addition, we define the set ${\cal D}_{out}(\delta)$ as follows:
\begin{equation}
{{\cal D}_{out}}(\delta ) = \bigcup\limits_{\forall l:{{\bf{a}}^{[l]}} \in {{\cal A}_l}} {{\cal D}''({{\bf{a}}^{[1]}},...,{{\bf{a}}^{{[2^{\left| {{{\cal N}_Q}} \right| - 1}]}}},\delta )} .
\label{DoF region passive outer2}
\end{equation}
Then, if the channel coefficients for all $T$ time slots are drawn independently and identically distributed  (i.i.d.) from a  continuous cumulative probability distribution, for $\forall \epsilon , \delta >0$, there exists a number $T'$ such that for $\forall T>T'$, we have: 
\begin{equation}
\Pr \left\{ {{\cal D} \subseteq {{\cal D}_{out}(\delta) }} \right\} > 1 - \epsilon,
\label{DoF region passive outer3}
\end{equation}
where $\cal D$ is the  DoF region for $T$ time slots.

\label{passive outer  DoF reg}
\end{theorem}
\begin{IEEEproof}
The proof is provided in Appendix \ref{appendix6}.
\end{IEEEproof}

{Theorem \ref{passive outer  DoF reg} suggests that, by letting the number of time slots $T$ approach infinity, with a probability close to $1$, we have ${{\cal D} \subseteq {{\cal D}_{out}(0) }}$.}
The difference of the outer bound for passive IRS in Theorem \ref{passive outer  DoF reg} and outer bound for the active IRS in Theorem \ref{active IRS outer  DoF reg} is that the coefficients $a_m^{[l]}$ are more restricted. In particular, the coefficients corresponding to network matrices which are not realizable in ${\cal E}_{Q_l}$ are zero. In other words, in $\Pr \{{\cal E}_{Q_l}\}$ fraction of time slots, in which ${\cal E}_{Q_l}$ occurs, the set of possible network matrices is ${\tilde {\cal N}}_{Q_l}$, whereas for  active IRSs,  the set of possible network matrices was ${\cal N}_Q$ for all time slots. This difference  causes the region in (\ref{DoF region passive outer2}) to be smaller than the region in (\ref{DoF region active outer2}). For the approximate capacity region, Theorem \ref{passive outer  DoF reg} indicates that:
\begin{equation*}
\Pr\left\{{{\cal C}(\rho)}\subseteq \bigcup\limits_{\forall l:{{\bf{a}}^{[l]}} \in {{\cal A}_l}} {{\cal C}''({{\bf{a}}^{[1]}},...,{{\bf{a}}^{{[2^{\left| {{{\cal N}_Q}} \right| - 1}]}}},\delta )}\right\}\ge 1-\epsilon ,
\end{equation*}
where:
\begin{equation*}
 \scalebox{.78}[1]{${\cal C}''({{\bf{a}}^{[1]}},...,{{\bf{a}}^{[{2^{\left| {{{\cal N}_Q}} \right| - 1}}]}},\delta ) = \left\{ {{{\bf{d}}(\log(1+\rho)+o(\log(\rho)))}\left| {\begin{array}{*{20}{l}}
{\forall i,j \in \{ 1,...,K\} ,i \ne j:{d_i} + {d_j} \le \sum\limits_{l = 1}^{{2^{\left| {{{\cal N}_Q}} \right| - 1}}} {\left( {\Pr \left\{ {{{\cal E}_{{Q_l}}}} \right\} + \delta } \right)\sum\limits_{m = 1}^{\left| {{{\tilde {\cal N}}_{{Q_l}}}} \right|} {a_m^{[l]}{{\tilde d}^{[ij]}}_{{\bf{N}}_m^{[l]}}} } }\\
{{{\bf{a}}^{[l]}} \in {{\cal A}_l},{\bf{N}}_m^{[l]} \in {{\tilde {\cal N}}_{{Q_l}}}: {\bf{N}}_m^{[l]} \ne {\bf{N}}_{m'}^{[l]},m \ne m'}
\end{array}} \right.} \right\}.$}
\end{equation*}

In the next step, we introduce a probabilistic inner bound for the   DoF region of  a $K$-user interference channel assisted by a $Q$-element passive  IRS. To this end, we  introduce some additional notations. Consider a network matrix ${\bf N}\in {\tilde {\cal N}}_{Q_i}$,  then we define set ${\cal M}_{{\bf N}}$ as follows:
\begin{equation}
{\cal M}_{{\bf N}} = \left\{ {(i,j)\left| {i,j \in \{ 1,...,K\} ,{n_{i,j}} = 0} \right.} \right\}.
\label{M_N}
\end{equation}
Thus, to realize network matrix $\bf N$,  the IRS elements' coefficients must satisfy the following equations for each $t\in \{1,...,T\}$:
\begin{equation}
\sum\limits_{u \in \{ 1,...,Q\} } {{H_{\rm TI}^{[{u}i]}}({t}){H_{\rm IR}^{[j{u}]}}({t}){\tau ^{[u]}}({t})}  =  - {H^{[ji]}}({t}),(i,j) \in {\cal M}_{{\bf N}}.
\label{Interference cancelation passive IRS}
\end{equation}
We can rewrite (\ref{Interference cancelation passive IRS}) in matrix form,  ${\bold{H}}_{\bf N}{\bold{\tau}}_{\bf N}={\bold h}_{\bf N}$, where ${\bold{H}}_{\bf N}$ is a matrix with elements ${H_{\rm TI}^{[{u}i]}}({t}){H_{\rm IR}^{[j{u}]}}({t})$, $(i,j)\in {\cal M}_{{\bf N}}$, ${\bold{\tau}}_{\bf N}$ is a column vector with elements ${\tau ^{[u]}}({t})$, and ${\bold h}_{\bf N}$ is a column vector with elements $- {H^{[ji]}}({t}),(i,j)\in {\cal M}_{{\bf N}}$. Since the number of variables in (\ref{Interference cancelation passive IRS}) can exceed the number of equations, i.e., $Q\ge |{\cal M}_{{\bf N}}|$,  we use the pseudo-inverse\footnote{The pseudo-inverse is one of the solutions of (\ref{Interference cancelation passive IRS}). We use this particular solution to derive the achievable DoFs as it leads to a tractable interference alignment scheme and asymptotic analysis. For the outer bound, as is evident from the definition of event ${\cal E}_{Q_l}$, all solutions of (\ref{Interference cancelation passive IRS}) are considered.} to calculate $\tau_{\bf N}$, i.e.,
\begin{equation}
\tau^*_{\bf N}={\bold{H}}^H_{\bf N}{({\bold{H}}_{\bf N}{\bold{H}}^H_{\bf N})}^{-1}{\bold h}_{\bf N}.
\label{pseudo}
\end{equation}
 Note that if we choose $|{\cal M}_{\bf N}|$  of columns of   matrix ${\bf H}_{\bf N}$ to construct a square matrix ${\tilde{\bf H}}_{\bf N}$, then  $\det({{\tilde {\bold H}}_{\bf N}})$ is a non-zero polynomial in terms of $H_{\rm TI}^{[{u}i]}( t)$ and ${H_{\rm IR}^{[j{u}]}}( t)$ and by Lemma \ref{lemma1}, $\Pr \{\det({{\tilde {\bold H}}_{\bf N}})=0\}=0$, so ${\tilde {\bold{H}}}_{\bf N}$ is full rank almost surely. On the other hand,  ${rank}({\bold{H}}_{\bf N})={rank}({\tilde {\bold{H}}}_{\bf N})={rank}({\bold{H}}_{\bf N}{\bold{H}}^H_{\bf N})$ holds, so ${\bold{H}}_{\bf N}{\bold{H}}^H_{\bf N}$ is full rank and invertible almost surely. Note that the main motivation for increasing the number of IRS elements is
that by increasing $Q$, the probability of the event, in which the coefficients $\tau^*_{\bf N}$ are realizable by a passive IRS, increases.
Now, we define event ${\cal F}_{{Q_i}}$ in the $t$-th time slot as follows:
\begin{equation*}
{{\cal F}_{{Q_i}}} = \left\{ {\forall u \in \{ 1,...,Q\} ,\forall {\bf N} \in {{\tilde {\cal N}}_{{Q_i}}}:\left| {{{\tau^{[u]}_{\bf N}}^* ( t)}} \right| \le 1} \right\}
\end{equation*}
\begin{equation}
\bigcap {\left\{ {\forall {\bf N} \in {\tilde {\cal N}}_{{Q_i}}^c:\exists u \in \{ 1,...,Q\}  \to \left| {{{\tau^{[u]}_{\bf N}}^* ( t)}} \right|> 1} \right\}}.
\end{equation}
Note that similar to ${\cal E}_{Q_i}$,  events ${\cal F}_{{Q_i}}$ are distinct $\forall i$. Now, we introduce a probabilistic inner bound for the DoF region of the $K$-user interference channel assisted by a passive IRS.
\begin{theorem}
Define set ${\cal D}_{{\tilde {\cal N}}_{Q_i}}$ as follows:
\begin{equation}
{{\cal D}_{{{\tilde {\cal N}}_{{Q_i}}}}} = \bigcup\limits_{{\bf{a}} \in {{\cal A}_i}} {\left\{ {\sum\limits_{j = 1}^{\left| {{{\tilde {\cal N}}_{{Q_i}}}} \right|} {{a_j}{{\bf{d}}_j}} \left| {{{\bf{d}}_j} \in {{\cal D}_{{{\bf{N}}_j}}},{{\bf{N}}_j} \in {{\tilde {\cal N}}_{{Q_i}}}} \right.} \right\}} ,
\label{DoF region passive inner1}
\end{equation}
where  set ${\cal D}_{{\bf N}_j}$ is given by (\ref{DoF region const active}) and set ${\cal A}_i$ is given by (\ref{A_i_eq}). In addition, we define  set ${\cal D}_{in}(\delta)$ as follows:
\begin{equation}
{{\cal D}_{in}(\delta)} = \left\{ {\sum\limits_{i = 1}^{{2^{\left| {{{\cal N}_Q}} \right|-1}} } {\left( {\Pr \{ {{\cal F}_{{Q_i}}}\}  - \delta } \right){{\bf d}_i}} \left| {{{\bf d}_i} \in {{\cal D}_{{{\tilde {\cal N}}_{{Q_i}}}}},i \in \{ 1,...,{2^{\left| {{{\cal N}_Q}} \right|-1}} \} } \right.} \right\}.
\label{DoF region passive inner2}
\end{equation}
Then, if the channel coefficients for all $T$ time slots are drawn i.i.d. from a  continuous cumulative probability distribution, for $\forall \epsilon , \delta >0$, there exists a number $T'$ such that for $\forall T>T'$, we have: 
\begin{equation}
\Pr \left\{ {{{\cal D}_{in}(\delta) } \subseteq {\cal D}} \right\} > 1 - \epsilon,
\label{DoF region passive inner3}
\end{equation}
where $\cal D$ is the  DoF region for $T$ time slots.

\label{passive inner  DoF reg}
\end{theorem}
\begin{IEEEproof}
The proof is provided in Appendix \ref{appendix7}.
\end{IEEEproof}

{Similar to Theorem \ref{passive outer  DoF reg}, Theorem \ref{passive inner  DoF reg} indicates that, as the number of time slots $T$ approaches infinity, with  probability  $1$, we have ${{{\cal D}_{in}(0) } \subseteq {\cal D}}$.}
Similar to the outer bound, the main difference between the inner bound for the passive IRS in Theorem \ref{passive inner  DoF reg} and the inner bound for the active IRS in Theorem \ref{active IRS inner  DoF reg} is that the coefficients $a_m$ are more restricted. In particular, the coefficients corresponding to the network matrices, which are not achievable in ${\cal F}_{Q_l}$ are zero. In other words, in $\Pr \{{\cal F}_{{Q_l}}\}$ fraction of time slots, in which ${\cal F}_{{Q_l}}$ has occurred, the set of achievable network matrices is ${\tilde {\cal N}}_{Q_l}$, whereas for the active IRS,  the set of achievable network matrices was ${\cal N}_Q$ in all time slots. 
 This difference will cause  region (\ref{DoF region passive inner2}) to be smaller than region (\ref{DoF region active inner2}). For the approximate capacity region,  Theorem \ref{passive inner  DoF reg} leads to:
\begin{equation*}
\Pr\left\{ \bigcup\limits_{{\bf{a}} \in {{\cal A}_i}} {\left\{ {\left(\sum\limits_{j = 1}^{\left| {{{\tilde {\cal N}}_{{Q_i}}}} \right|} {{a_j}{{\bf{d}}_j}}\right)(\log(1+\rho)+o(\log(\rho))) \left| {{{\bf{d}}_j} \in {{\cal D}_{{{\bf{N}}_j}}},{{\bf{N}}_j} \in {{\tilde {\cal N}}_{{Q_i}}}} \right.} \right\}}\subseteq {{\cal C}(\rho)} \right\} \ge 1-\epsilon.
\end{equation*}

Theorems \ref{passive outer  DoF reg} and \ref{passive inner  DoF reg} represent the outer and inner bounds for the DoF region as a function of  $\Pr\{{\cal E}_{Q_l}\}$ and $\Pr\{{\cal F}_{Q_l}\}$, respectively.
In the following theorem, we characterize the behavior of $\Pr\{{\cal F}_{{Q_l}}\}$ for large values of $Q$.

\begin{theorem}
Assume that the imaginary and real parts of all channel coefficients are zero mean and their probability distributions  have the following properties:
\begin{equation}
{f_{H_{{\rm{T}}{{\rm{I}}_r}}^{[ui']}(t)}}(h) = {f_{H_{{\rm{T}}{{\rm{I}}_r}}^{[i']}(t)}}(h),\forall i' \in \{ 1,...,K\} ,\forall u \in \{ 1,...,Q\} ,
\label{prop1}
\end{equation}
\begin{equation}
{f_{H_{{\rm{I}}{{\rm{R}}_r}}^{[ju]}(t)}}(h) = {f_{H_{{\rm{I}}{{\rm{R}}_r}}^{[j]}(t)}}(h),\forall j \in \{ 1,...,K\} ,\forall u \in \{ 1,...,Q\},
\label{prop2}
\end{equation}
\begin{equation}
{f_{H_{{\rm{T}}{{\rm{I}}_i}}^{[ui']}(t)}}(h) = {f_{H_{{\rm{T}}{{\rm{I}}_i}}^{[i']}(t)}}(h),\forall i' \in \{ 1,...,K\} ,\forall u \in \{ 1,...,Q\} ,
\label{prop3}
\end{equation}
\begin{equation}
{f_{H_{{\rm{I}}{{\rm{R}}_i}}^{[ju]}(t)}}(h) = {f_{H_{{\rm{I}}{{\rm{R}}_i}}^{[j]}(t)}}(h),\forall j \in \{ 1,...,K\} ,\forall u \in \{ 1,...,Q\},
\label{prop4}
\end{equation}
\begin{equation}
\int {|h{|^n}{f_{H_{{\rm{T}}{{\rm{I}}_r}}^{[i']}(t)}}(h)dh < \infty } ,\int {|h{|^n}{f_{H_{{\rm{I}}{{\rm{R}}_r}}^{[j]}(t)}}(h)dh < \infty ,\forall i',j \in \{ 1,...,K\} } ,0 \le n \le 4,
\label{prop5}
\end{equation}
\begin{equation}
\int {|h{|^n}{f_{H_{{\rm{T}}{{\rm{I}}_i}}^{[i']}(t)}}(h)dh < \infty } ,\int {|h{|^n}{f_{H_{{\rm{I}}{{\rm{R}}_i}}^{[j]}(t)}}(h)dh < \infty ,\forall i',j \in \{ 1,...,K\} } ,0 \le n \le 4,
\label{prop6}
\end{equation}
\begin{equation}
\int {|h{|^n}{f_{H_r^{[ji']}(t)}}(h)dh < \infty } ,\int {|h{|^n}{f_{H_i^{[ji']}(t)}}(h)dh < \infty ,\forall i',j \in \{ 1,...,K\} } ,0 \le n \le 2.
\label{prop7}
\end{equation}
where  index $r$ indicates  the real  part and  index $i$ indicates  the imaginary  part of the channel coefficients.
In addition, without loss of generality, assume that ${\tilde {\cal N}}_{Q_1}={\cal N}_{Q}$. Then, we have:
\begin{equation}
\mathop {\lim }\limits_{Q \to \infty } \Pr \{ {{\cal F}_{{Q_1}}}\}  = 1,
\end{equation}
\begin{equation}
\mathop {\lim }\limits_{Q \to \infty } \Pr \{ {{\cal F}_{{Q_i}}}\}  = 0,i\ne 1,
\end{equation}
\label{sum  DoF passive asymptotic}
where the order of convergence is at least $O(\frac{1}{Q})$.
\end{theorem}
\begin{IEEEproof}
The proof is provided in Appendix \ref{appendix10}.
\end{IEEEproof}

We note that (\ref{prop1})-(\ref{prop4}) imply that the distributions of the channel coefficients from the transmitters to the IRS and from the IRS to the receivers are identical for all IRS elements, i.e., they do not depend on index $u\in\{1,...,Q\}$. 

Considering the definitions of ${\cal E}_{Q_l}$ and ${\cal F}_{{Q_l}}$, we can see that $\Pr\{{\cal E}_{Q_1}\}\ge\Pr\{{\cal F}_{{Q_1}}\}$, because if ${\cal F}_{{Q_1}}$ occurs and we can realize all network matrices using the pseudo inverse (see (\ref{pseudo})), then  ${\cal E}_{Q_1}$ occurs as well, i.e., ${{\cal F}_{{Q_1}}} \subseteq {{\cal E}_{Q_1}}$. So, based on  Theorem
\ref{sum  DoF passive asymptotic}, we obtain $1 - O(\frac{1}{Q}) \le \Pr \{ {{\cal E}_{Q_1}}\}  \le 1$, which leads to the following relations:
\begin{equation}
\mathop {\lim }\limits_{Q \to \infty } \Pr \{ {{\cal E}_{Q_1}}\}  = 1,
\end{equation}
\begin{equation}
\mathop {\lim }\limits_{Q \to \infty } \Pr \{ {{\cal E}_{{Q_i}}}\}  = 0,i\ne 1,
\end{equation}
where the order of convergence is at least $O(\frac{1}{Q})$.

\textit{Remark 5}: Although we proved that for a sufficiently large  number of elements of the passive IRS, $\Pr\{{\cal F}_{{Q_1}}\}$ tends to $1$, we note that for very large numbers of elements, the assumption of i.i.d. channel coefficients may break down. 
Therefore, considering a maximum number of IRS elements, for which the i.i.d. assumption holds,  the slope of the decay of the probability measure $\Pr\{{\cal F}_{{Q_i}}\},i\ne 1$, determines how closely   $\Pr\{{\cal F}_{{Q_1}}\}$ can approach $1$, i.e., the faster $\Pr\{{\cal F}_{{Q_i}}\},i\ne 1$, vanishes as $Q$ increases, the more DoF can be achieved in a practical setting. Note that if we make more specific assumptions on the distribution of the channel coefficients, faster decays than $O(\frac{1}{Q})$ can be achieved.

\subsection{Sum DoF}

Now, based on Theorems \ref{passive outer  DoF reg} and \ref{passive inner  DoF reg}, we derive probabilistic  upper and lower bounds for the sum  DoF of the $K$-user interference channel in the presence of a passive IRS.
\begin{theorem}
Consider a  $K$-user interference channel with a $Q$-element passive IRS. Then, for $\forall {\bf d}\in {\cal D}$, if the channel coefficients for all $T$ time slots are drawn i.i.d. from a  continuous cumulative probability distribution, for $\forall \epsilon , \delta >0$, there exists a number $T'$ such that for $\forall T>T'$, we have: 
\begin{equation}
\Pr \left\{ {\sum\limits_{k = 1}^K {{d_k}}  \le \sum\limits_{i = 1}^{{2^{\left| {{{\cal N}_Q}} \right| - 1}}} {\left( {\Pr \{ {{\cal E}_{{Q_i}}}\}  + \delta } \right)\mathop {\max }\limits_{{\bf N} \in {\tilde {\cal N}}_{Q_i}} \left( {\frac{K}{2} + \frac{{\left| {{{\cal M}_{{\bf N}}}} \right|}}{{2(K - 1)}}} \right)} } \right\} > 1 - \epsilon ,
\end{equation}
where set ${\cal M}_{{\bf N}}$ is given by (\ref{M_N}).
\label{theorem passive sum  DoF}
\end{theorem}
\begin{IEEEproof}
The proof is provided in Appendix \ref{appendix8}.
\end{IEEEproof}
Theorem \ref{theorem passive sum  DoF} demonstrates that, if we let the number of time slots $T$ approach infinity, with probability close to $1$, we obtain ${\sum\limits_{k = 1}^K {{d_k}}  \le \sum\limits_{i = 1}^{{2^{\left| {{{\cal N}_Q}} \right| - 1}}} {\left( {\Pr \{ {{\cal E}_{{Q_i}}}\}  } \right)\mathop {\max }\limits_{{\bf N} \in {\tilde {\cal N}}_{Q_i}} \left( {\frac{K}{2} + \frac{{\left| {{{\cal M}_{{\bf N}}}} \right|}}{{2(K - 1)}}} \right)} }$.
This theorem also reveals that the approximate sum capacity of the $K$-user interference channel assisted by a $Q$-element passive IRS is upper bounded by $\left( {\sum\limits_{i = 1}^{{2^{\left| {{{\cal N}_Q}} \right| - 1}}} {\left( {\Pr \{ {{\cal E}_{{Q_i}}}\}  + \delta } \right)\mathop {\max }\limits_{{\bf{N}} \in {{\tilde {\cal N}}_{{Q_i}}}} \left( {\frac{K}{2} + \frac{{\left| {{{\cal M}_{\bf{N}}}} \right|}}{{2(K - 1)}}} \right)} } \right)\log (1 + \rho ) + o(\log (\rho ))$. The upper bound depends on the distribution of the channel coefficients via the probability measure $\Pr\{{\cal E}_{Q_l}\}$. 

To establish a lower bound, we have to introduce a new definition. For each set ${\tilde {\cal N}}_{Q_i}$, we define the subsets ${\tilde {\cal N}}^W_{Q_i},W\in\{0,...,K\}$, as follows:
\begin{equation}
\scalebox{.82}[1]{${\tilde {\cal N}}_{{Q_i}}^W = \left\{ {{\bf{N}}\left| {{\bf{N}} \in {{\tilde {\cal N}}_{{Q_i}}},\exists {\cal B}\subseteq\{1,...,K\}:\left| {\cal B} \right| = W,i' \in \{ 1,...,K\} ,j' \in {\cal B},i' \ne j' \to {n_{i',j'}} = 0} \right.,}  {i' \in {\cal B},j' \in {{\cal B}^c} \to {n_{i',j'}} = 0} \right\}$},
\end{equation}
where
\begin{equation*}
{\cal B}^{c}=\{1,...,K\}-{\cal B},
\end{equation*}
\begin{equation}
{\tilde {\cal N}}_{{Q_i}}^0 = \left\{ {{\bf{N}}\left| {{\bf{N}} \in {{\tilde {\cal N}}_{{Q_i}}},\forall {\cal B}:\exists(i',j') \in {{\cal C}_{\cal B}} \to {n_{i',j'}} = 1} \right.} \right\},
\end{equation}
\begin{equation*}
{{\cal C}_{\cal B}} = \left\{ {(i',j')\left| {i' \in \{ 1,...,K\} ,j' \in {\cal B},i' \ne j'} \right.} \right\}\bigcup {\left\{ {(i',j')\left| {i' \in {\cal B},j' \in {{\cal B}^c}} \right.} \right\}} .
\end{equation*}
Now, we present a probabilistic lower bound for the  sum  DoF.
\begin{theorem}
Consider the $Q$-element passive IRS-assisted $K$-user interference channel. Then, for  $\forall {\bf d}\in {\cal D}$, if the channel coefficients for all $T$ time slots are drawn i.i.d. from a  continuous cumulative probability distribution, for $\forall \epsilon , \delta >0$, there exists a number $T'$ such that for $\forall T>T'$, we have: 
\begin{equation}
\Pr \left\{ {\mathop {\max }\limits_{{\bf d} \in {\cal D}}\sum\limits_{k = 1}^K {{d_k}}  \ge \sum\limits_{i = 1}^{{2^{\left| {{{\cal N}_Q}} \right| - 1}}} {\left( {\Pr \{ {{\cal F}_{{Q_i}}}\}  - \delta } \right)\mathop {\max }\limits_{{\tilde {\cal N}}_{{Q_i}}^W \ne \Phi } \left( {\frac{{K + W}}{2}} \right)} } \right\} > 1 - \epsilon .
\end{equation}
\label{sum  DoF lower passive}
\end{theorem}
\begin{IEEEproof}
The proof is provided in Appendix \ref{appendix9}.
\end{IEEEproof}
The achievability scheme for the passive IRS in Theorem \ref{sum  DoF lower passive} is based on the proportion of channel realizations, for which the achievability scheme developed for active IRSs can be realized with passive IRS elements.
 This theorem shows that, when the number of time slots $T$ approaches infinity, with probability  $1$, we achieve $ {\mathop {\max }\limits_{{\bf d} \in {\cal D}} \sum\limits_{k = 1}^K {{d_k}}  \ge \sum\limits_{i = 1}^{{2^{\left| {{{\cal N}_Q}} \right| - 1}}} {\left( {\Pr \{ {{\cal F}_{{Q_i}}}\} } \right)\mathop {\max }\limits_{{\tilde {\cal N}}_{{Q_i}}^W \ne \Phi } \left( {\frac{{K + W}}{2}} \right)} }$. 
Theorem \ref{sum  DoF lower passive} also indicates that the approximate sum capacity of the $K$-user interference channel assisted by a $Q$-element passive IRS is lower bounded by $\left( {\sum\limits_{i = 1}^{{2^{\left| {{{\cal N}_Q}} \right| - 1}}} {\left( {\Pr \{ {{\cal F}_{{Q_i}}}\}  - \delta } \right)\mathop {\max }\limits_{{\tilde {\cal N}}_{{Q_i}}^W \ne \Phi } \left( {\frac{{K + W}}{2}} \right)} } \right)\log (1 + \rho ) + o(\log (\rho ))$. 
In addition,  we note that the sum  DoF tends to $K$ in probability, because ${\tilde {\cal N}}_{Q_1}$ includes the identity matrix  (i.e., ${\tilde {\cal N}}^K_{Q_1}\ne \Phi$) and by Theorem \ref{sum  DoF passive asymptotic}, we have $\mathop {\lim }\limits_{Q \to \infty } \Pr \{ {{\cal F}_{{Q_1}}}\}  = 1$, i.e., the DoF is lower bounded by $K(1-O(\frac{1}{Q}))$ for sufficiently large $Q$.
Thus,  by choosing a large  $Q$, the approximate sum capacity of the $K$-user interference channel in the presence of a passive IRS is lower bounded by $(K-\epsilon)\log(1+\rho)+o(\log(\rho)),\forall\epsilon>0$, which reveals that the passive IRS can asymptotically achieve the same performance as an active IRS.

\section{$K$-user Interference Channel in the Presence of a Passive Lossless and an $\varepsilon$-Relaxed Passive Lossless IRS}

\label{sec5}

In this section, we  study the  $K$-user interference channel in the presence of a passive lossless and an $\varepsilon$-relaxed passive lossless IRS. Note that we introduce the $\varepsilon$-relaxed passive lossless IRS as an approximation of the passive lossless IRS, i.e., when $\varepsilon \to 0$. The outer bound introduced in Theorem \ref{passive outer  DoF reg} is valid for the  $\varepsilon$-relaxed passive lossless IRS, if we define  the event ${\cal E}_{Q_i}$ such that the network matrices ${\bf N}\in {\tilde {\cal N}}_{Q_i}$ can be realized and the network matrices ${\bf N}\in {\tilde {\cal N}}^c_{Q_i}$ cannot be realized  by the  $Q$-element $\varepsilon$-relaxed passive lossless IRS. The inner bound introduced in Theorem \ref{passive inner  DoF reg} is applicable to the $\varepsilon$-relaxed passive lossless IRS but, in this case, its asymptotic sum  DoF does not tend to $K$, i.e., the bound is not tight asymptotically. Therefore, we introduce a new lower bound for the sum  DoF which is tighter and asymptotically tends to $K$.
For the passive lossless IRS, the interference cancellation scheme, which was introduced in Sections \ref{sec3} and \ref{sec4}, cannot be applied because  the passive lossless IRS is unable to control the amplitude of the received signal, i.e., only a phase shift can be applied.
Therefore, we provide theorems in term of the $\rho$-limited sum DoF and prove that  the value $K$ can be achieved for the $\rho$-limited sum DoF of  the $K$-user interference channel assisted by a passive lossless  IRS for each value of $\rho\in \mathbb{R}_{+}$.

\subsection{$\varepsilon$-Relaxed Passive Lossless IRS}
The idea of the new lower bound for  the  sum  DoF of the $K$-user interference channel in the presence of an $\varepsilon$-relaxed passive lossless IRS is that we  use a properly chosen subset of IRS elements for complete interference cancellation. This subset contains $K^2$ elements.
Assume   $Q\ge K^2$ for the number of  IRS elements.
We define the sets  ${{\cal N}_{k'}},k' \in \left\{ {1,...,{Q\choose K^2}} \right\}$, as  all subsets of $\{1,...,Q\}$ with $K^2$ members. Then, we define random variables $\tau_{k'}^{[u]}$ as the solutions to the following set of equations:
\begin{equation}
\left\{ {\begin{array}{*{20}{c}}
{\sum\limits_{u \in {{\cal N}_{k'}}} {{H_{\rm TI}^{[{u}i]}}({t}){H_{\rm IR}^{[j{u}]}}({t})\tau _{k'}^{[u]}({t})}  =  - {H^{[ji]}}({t}),i,j \in \{ 1,...,K\} ,i \ne j}\\
{\sum\limits_{u \in {{\cal N}_{k'}}} {{H_{\rm TI}^{[{u}i]}}({t}){H_{\rm IR}^{[j{u}]}}({t})\tau _{k'}^{[u]}({t})}  = 0,i,j \in \{ 1,...,K\} ,i = j}
\end{array}} \right..
\label{intercancel11}
\end{equation}
Note that $\tau_{k'}^{[u]}( t)$ will be the coefficient of the $u$-th IRS element, if the elements $u\in {\cal N}_{k'}$ are the operating elements of the $\varepsilon$-relaxed passive lossless IRS (other elements do not cooperate in the IRS transmission and their coefficients are set to zero). We construct a new random variable  $\Lambda ( t)$  such that:
\begin{equation}
{\Lambda }({t})=
\begin{cases}
1,  & \mathrm{if}\,\, \exists k' \in \{ 1,...,{Q\choose K^2}\} :\forall u \in {{\cal N}_{k'}},1-\varepsilon \le \left| {\tau _{k'}^{[u]}( t)} \right| \le 1 \\
0, & \mathrm{if}\,\, \forall k' \in \{ 1,...,{Q\choose K^2}\} :\exists u { \in {{\cal N}_{k'}}},1 < \left| {\tau _{k'}^{[u]}( t)} \right| or  \left| {\tau _{k'}^{[u]}( t)} \right|<1-\varepsilon
\end{cases}.
\label{Lambda construction}
\end{equation}
Note that if $\Lambda ( t)=1$, interference cancellation is possible at all receivers  using at least one subset of $\varepsilon$-relaxed passive lossless IRS elements ${\cal N}_{k'}$.
\begin{theorem}
Let $Q\ge K^2$ be the number of $\varepsilon$-relaxed passive lossless IRS  elements. If the channel coefficients for all time slots are drawn i.i.d., then for  $\forall \epsilon,\delta>0$, there exists a number $T'$ such that for $T>T'$, we have:
\begin{equation}
\Pr \left\{ {\mathop {\max }\limits_{{\bf d} \in {\cal D}}\sum\limits_{i = 1}^K {{d_i}}  \ge \frac{K}{2}\left( {\Pr \left\{ {\Lambda ({t}) = 0} \right\} - \delta } \right) + K\left( {\Pr \left\{ {\Lambda ({t}) = 1} \right\} - \delta } \right)} \right\} > 1 - \epsilon .
\label{DoFpassive}
\end{equation}
\label{theorem5}
\end{theorem} 
\begin{IEEEproof}
The proof is provided in Appendix \ref{appendix11}.
\end{IEEEproof}

Theorem \ref{theorem5} indicates that if $T$ is sufficiently large, then, in at least $T({\Pr \left\{ {\Lambda ({t}) = 0} \right\} - \delta })$ time slots, interference cancellation is not possible by utilizing $K^2$ member subsets of IRS elements, and in at least $T({\Pr \left\{ {\Lambda ({t}) = 1} \right\} - \delta })$ time slots, full interference cancellation is possible. 
In the next theorem, we provide a condition, which ensures $\mathop {\lim }\limits_{Q \to \infty } \Pr \left\{ {\Lambda ({t}) = 0} \right\} = 0$, such that  the $K$-user interference channel can achieve the asymptotic $K$ sum DoF in the presence of an $\varepsilon$-relaxed passive lossless IRS. 
\begin{theorem}
Assume that the channel coefficients meet the following conditions:
\begin{itemize}
\item
The channel coefficients for all time slots are drawn i.i.d. from a  continuous cumulative probability distribution. In addition,
 the probability distribution of the channel coefficients  satisfies (\ref{prop1})-(\ref{prop4}).
\item
For all measurable sets ${\cal A}\subseteq \mathbb{R}$ with nonzero  measure, $\mu({\cal A})>0$, and for $\forall i',j\in\{1,...,K\}$, we have:
\begin{equation}
\int\limits_{\cal A} {{f_{H_{{\rm{T}}{{\rm{I}}_r}}^{[i']}(t)}}(h)dh > 0} ,\int\limits_{\cal A} {{f_{H_{{\rm{I}}{{\rm{R}}_r}}^{[j]}(t)}}(h)dh > 0} ,\int\limits_{\cal A} {{f_{H_{{\rm{T}}{{\rm{I}}_i}}^{[i']}(t)}}(h)dh > 0} ,\int\limits_{\cal A} {{f_{H_{{\rm{I}}{{\rm{R}}_i}}^{[j]}(t)}}(h)dh > 0} .
\label{axiom1}
\end{equation}
\begin{equation}
\int\limits_{\cal A} {{f_{H^{[ji']}(t)}}(h)dh > 0} ,\int\limits_{\cal A} {{f_{H^{[ji']}(t)}}(h)dh > 0}  .
\label{axiom11}
\end{equation}
\end{itemize}
Then, we have:
\begin{equation}
\mathop {\lim }\limits_{Q \to \infty } \Pr \left\{ {\Lambda ({t}) = 0} \right\} = 0.
\label{lossless convergence1}
\end{equation}
Therefore, by Theorem \ref{theorem5}, the asymptotic $K$ sum DoF are achievable in probability for the $K$-user interference channel in the presence of $\varepsilon$-relaxed passive lossless IRS.
\label{theorem6}
\end{theorem}
\begin{IEEEproof}
The proof is provided in Appendix \ref{appendix12}.
\end{IEEEproof}
Theorem \ref{theorem6} shows that by choosing a sufficiently large  $Q$, the approximate sum capacity of the $K$-user interference channel in the presence of an $\varepsilon$-relaxed passive lossless IRS is lower bounded by $(K-\epsilon)\log(1+\rho)+o(\log(\rho)),\forall\epsilon>0$, which reveals that the $\varepsilon$-relaxed passive lossless IRS  asymptotically performs as well as an active IRS. 

Although Theorem \ref{theorem6} shows that the maximum $K$ sum DoFs are achievable for asymptotically large $Q$, similar to Remark 2, we emphasize that for very large $Q$,  the assumption of i.i.d. channel coefficients may not hold true.

\subsection{Passive Lossless IRS}

First, we introduce the following definitions:
\begin{equation}
\scalebox{.95}[1]{${\text{SIN}}{{\text{R}}_k} = \frac{{{{\left| {{\text{Re}}\left\{ {{H^{[kk]}}(t) + \sum\limits_{u = 1}^Q {H_{{\text{IR}}}^{[ku]}(t){\tau ^{[u]}}(t)H_{{\text{TI}}}^{[uk]}(t)} } \right\}} \right|}^2}\rho  + {{\left| {{\text{Im}}\left\{ {{H^{[kk]}}(t) + \sum\limits_{u = 1}^Q {H_{{\text{IR}}}^{[ku]}(t){\tau ^{[u]}}(t)H_{{\text{TI}}}^{[uk]}(t)} } \right\}} \right|}^2}\rho }}{{\sum\limits_{i = 1,i \ne k}^K {{{\left| {{\text{Re}}\left\{ {{H^{[ki]}}(t) + \sum\limits_{u = 1}^Q {H_{{\text{IR}}}^{[ku]}(t){\tau ^{[u]}}(t)H_{{\text{TI}}}^{[ui]}(t)} } \right\}} \right|}^2}} \rho  + \sum\limits_{i = 1,i \ne k}^K {{{\left| {{\text{Im}}\left\{ {{H^{[ki]}}(t) + \sum\limits_{u = 1}^Q {H_{{\text{IR}}}^{[ku]}(t){\tau ^{[u]}}(t)H_{{\text{TI}}}^{[ui]}(t)} } \right\}} \right|}^2}\rho }  + {N_0}}}, $} 
\end{equation}
\begin{equation}
{\textrm{SINR}}_k^r = \frac{{{{\left| {{\mathop{\rm Re}\nolimits} \left\{ {{H^{[kk]}}(t) + \sum\limits_{u = 1}^Q {{H_{\rm IR}^{[k{u}]}}(t){\tau ^{[u]}}(t){H_{\rm TI}^{[{u}k]}}(t)} } \right\}} \right|}^2}\frac{\rho }{2}}}{{\sum\limits_{i = 1,i \ne k}^K {{{\left| {{\mathop{\rm Re}\nolimits} \left\{ {{H^{[ki]}}(t) + \sum\limits_{u = 1}^Q {{H_{\rm IR}^{[k{u}]}}(t){\tau ^{[u]}}(t){H_{\rm TI}^{[{u}i]}}(t)} } \right\}} \right|}^2}\frac{\rho }{2}}  + \frac{{{N_0}}}{2}}},
\label{SINR_k_r}
\end{equation}
\begin{equation}
{\textrm{SINR}}_k^i = \frac{{{{\left| {{\mathop{\rm Im}\nolimits} \left\{ {{H^{[kk]}}(t) + \sum\limits_{u = 1}^Q {{H_{\rm IR}^{[k{u}]}}(t){\tau ^{[u]}}(t){H_{\rm TI}^{[{u}k]}}(t)} } \right\}} \right|}^2}\frac{\rho }{2}}}{{\sum\limits_{i = 1,i \ne k}^K {{{\left| {{\mathop{\rm Im}\nolimits} \left\{ {{H^{[ki]}}(t) + \sum\limits_{u = 1}^Q {{H_{\rm IR}^{[k{u}]}}(t){\tau ^{[u]}}(t){H_{\rm TI}^{[{u}i]}}(t)} } \right\}} \right|}^2}\frac{\rho }{2}}  + \frac{{{N_0}}}{2}}}.
\end{equation}
We use $\textrm{SINR}_k^r$ and $\textrm{SINR}_k^m$ for simplicity of proof of Theorem \ref{strict SINR}.
Now, we introduce a probabilistic lower bound for the \textit{$\rho$-limited sum DoF} of a passive lossless  IRS-assisted $K$ user interference channel and show that this lower bound can approach $K$.

\begin{theorem}
Consider the  $K$-user interference channel assisted by a $Q$-element passive lossless IRS. Then, by employing a set of coefficients $\tau^{[u]}(t)$ and for $\forall {\bf d}^{[\rho]}\in {\cal D}^{[\rho]}$, if the channel coefficients for all $T$ time slots are drawn i.i.d. from a  continuous cumulative probability distribution, for $\forall \epsilon,\varepsilon , \delta >0$, there exists a number $T'$ such that for $\forall T>T'$, we have: 
\begin{equation}
\Pr \left\{ {\mathop {\max }\limits_{{\bf d}^{[\rho]} \in {\cal D}^{[\rho]}}\sum\limits_{i = 1}^K {d_i^{[\rho ]}}  \ge K(1 - \varepsilon )\left( {\Pr \left\{ {\bigcap\limits_{k = 1}^K {\left\{ {\left\{ {{\textrm{SINR}}_k^r \ge {\rho ^{1 - \varepsilon }}} \right\}\bigcap {\left\{ {{\textrm{SINR}}_k^i \ge {\rho ^{1 - \varepsilon }}} \right\}} } \right\}} } \right\} - \delta } \right)} \right\} > 1 -  \epsilon ,
\label{m sum DoF}
\end{equation}
\label{m-limited DoF thm}
where $d^{[\rho]}_i$ is the $\rho$-limited DoF for the $i$-th user.
\end{theorem}
\begin{IEEEproof}
The proof is provided in Appendix \ref{appendix14}.
\end{IEEEproof}

Next,  we prove that by employing a sufficiently large number of elements for the passive lossless IRS, we can find coefficients $\tau^{[u]}(t)$, for which ${\Pr \left\{ {\bigcap\limits_{k = 1}^K {\left\{ {\left\{ {{\textrm{SINR}}_k^r \ge {\rho ^{1 - \varepsilon }}} \right\}\bigcap {\left\{ {{\textrm{SINR}}_k^i \ge {\rho ^{1 - \varepsilon }}} \right\}} } \right\}} } \right\}}$ can approach the value $1$ and we state some conditions, for which the order of convergence is at least $O(\frac{1}{Q})$.

\begin{theorem}
Consider the  $K$-user interference channel assisted by a $Q$-element passive lossless IRS.
Similar to Theorem \ref{sum DoF passive asymptotic},
assume that the imaginary and real parts of all channel coefficients are zero mean, their probability distributions satisfy (\ref{prop1})-(\ref{prop4}) and  have the following properties:
\begin{equation}
\int {|h{|^n}{f_{H_{{\rm{T}}{{\rm{I}}_r}}^{[i']}(t)}}(h)dh < \infty } ,\int {|h{|^n}{f_{H_{{\rm{I}}{{\rm{R}}_r}}^{[j]}(t)}}(h)dh < \infty ,\forall i',j \in \{ 1,...,K\} } ,0 \le n \le 2,
\label{prop13}
\end{equation}
\begin{equation}
\int {|h{|^n}{f_{H_{{\rm{T}}{{\rm{I}}_i}}^{[i']}(t)}}(h)dh < \infty } ,\int {|h{|^n}{f_{H_{{\rm{I}}{{\rm{R}}_i}}^{[j]}(t)}}(h)dh < \infty ,\forall i',j \in \{ 1,...,K\} } ,0 \le n \le 2,
\label{prop14}
\end{equation}
\begin{equation}
\int {|h{|^n}{f_{H_r^{[ji']}(t)}}(h)dh < \infty } ,\int {|h{|^n}{f_{H_i^{[ji']}(t)}}(h)dh < \infty ,\forall i',j \in \{ 1,...,K\} } ,0 \le n \le 2,
\label{prop15}
\end{equation}
where index $r$ indicates  the real  part and  index $i$ indicates  the imaginary  part of the channel coefficients.
Then, for each $\epsilon>0$ and $0<m<\frac{\rho}{N_0}$, there exists a number $Q'$ such that for  $Q\ge Q'$, we can find phase coefficients $\tau^{[u]}( t)$, for which we have:
\begin{equation}
\Pr \left\{ {\left\{ {{\textrm{SINR}}_k^r < m} \right\}\bigcup {\left\{ {{\textrm{SINR}}_k^i < m} \right\}} } \right\} < \epsilon,\forall k\in \{1,...,K\}.
\label{prob SINR}
\end{equation}
The order of convergence is at least $O(\frac{1}{Q})$.
\label{strict SINR}
\end{theorem}

\begin{IEEEproof}
The proof is provided in Appendix \ref{appendix13}.

\end{IEEEproof}

Theorem \ref{strict SINR} indicates that for a given  value $0<m<\frac{\rho}{N_0}$, we can ensure that the probability measure ${\Pr \left\{ {\bigcap\limits_{k = 1}^K {\left\{ {\left\{ {{\textrm{SINR}}_k^r \ge {m}} \right\}\bigcap {\left\{ {{\textrm{SINR}}_k^i \ge {m}} \right\}} } \right\}} } \right\}}$ tends to $1$, by employing a sufficient number of elements for the passive lossless IRS.  Therefore, if we consider sufficiently large $\rho$ and $Q$ and a small enough $\varepsilon$, then the $\rho$-limited sum DoF of  the $K$-user interference channel assisted by a passive lossless  IRS approaches $K$.

\section{Numerical Result}

\label{sec6}

In this section, we provide numerical results to quantify the proposed bounds.
For our simulations, we assumed that all channel coefficients are generated independently and follow  zero mean complex  Gaussian distributions. The variance of the channel coefficients from the transmitters to the IRS and from the IRS to the receivers is $\sigma _1^2 = {\left( {\frac{\lambda }{{4\pi {\rho _1}}}} \right)^2}$, respectively, and the variance of the direct links is $\sigma _2^2 = {\left( {\frac{\lambda }{{4\pi {\rho _2}}}} \right)^2}\hat h$, where $\rho_1$ is the distance between the IRS and other nodes, $\rho_2$ is the distance between each transmitter and each receiver, and  parameter $\hat h$  models a potential blockage of the direct and cross  links between the transmitters and the receivers. We set $\lambda=0.06{\rm m}$, which is the wavelength for a carrier frequency of $5$ GHz, and $\rho_1=\rho_2=25\sqrt{2} {\rm m}$.

\begin{figure}[!htb]
\centering
\begin{minipage}{.45\textwidth}
\centering
\includegraphics[width=8cm]{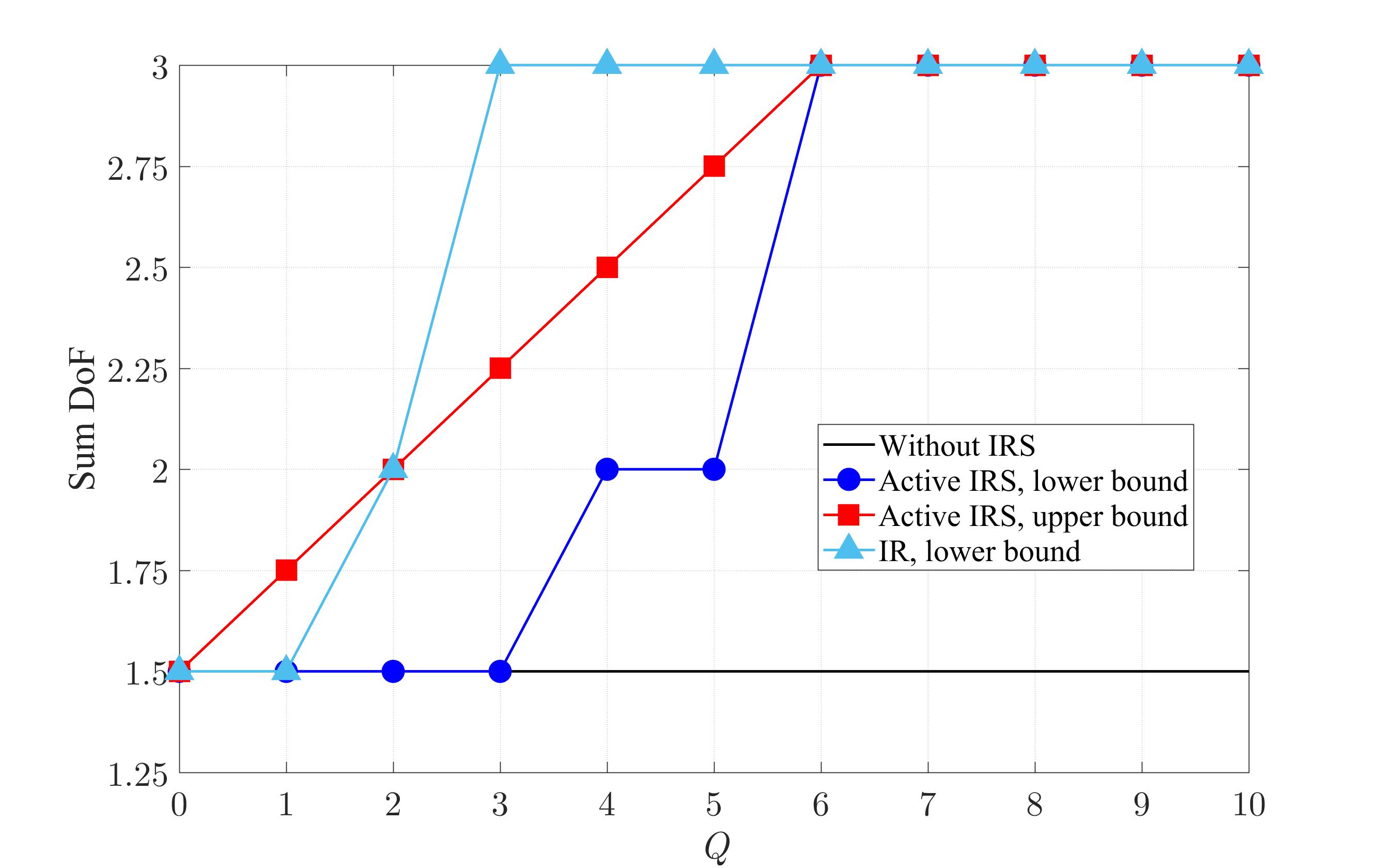}
\vspace{-30pt}
\caption{Comparison of  bounds on the sum DoF of a $3$-user interference channel in the presence of an active IRS and IR, respectively, and the sum DoF without an IRS.}
 \label{fig1}
\end{minipage}
\quad
\begin{minipage}{.45\textwidth}
\centering
\includegraphics[width=7.5cm]{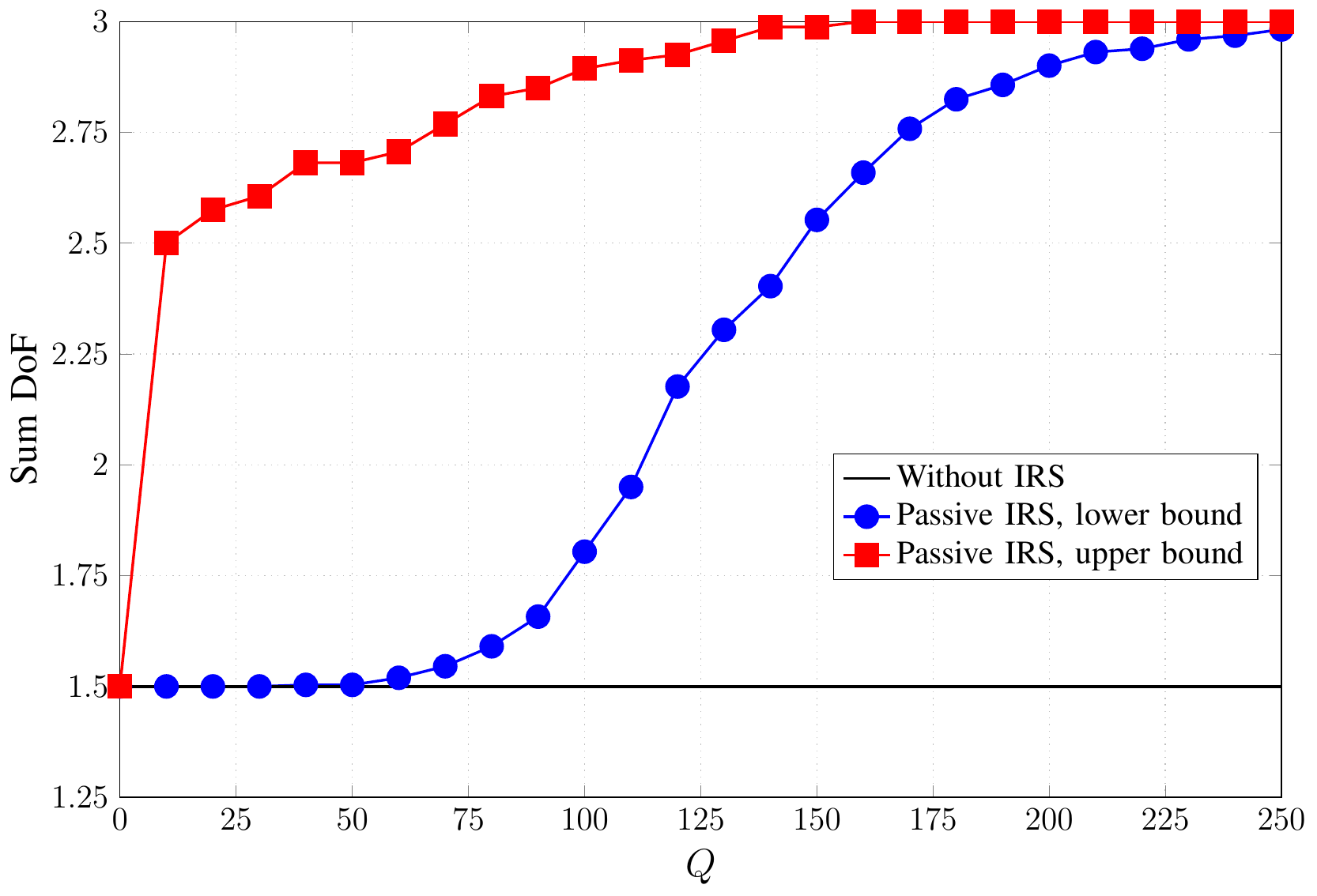}
\vspace{-30pt}
\caption{Comparison of the lower and upper bounds on the sum DoF of a $3$-user interference channel in the presence of a passive IRS and the sum DoF without an IRS.}
 \label{fig2}
\end{minipage}
\begin{minipage}{.45\textwidth}
\vspace{20pt}
\centering
\includegraphics[width=8cm]{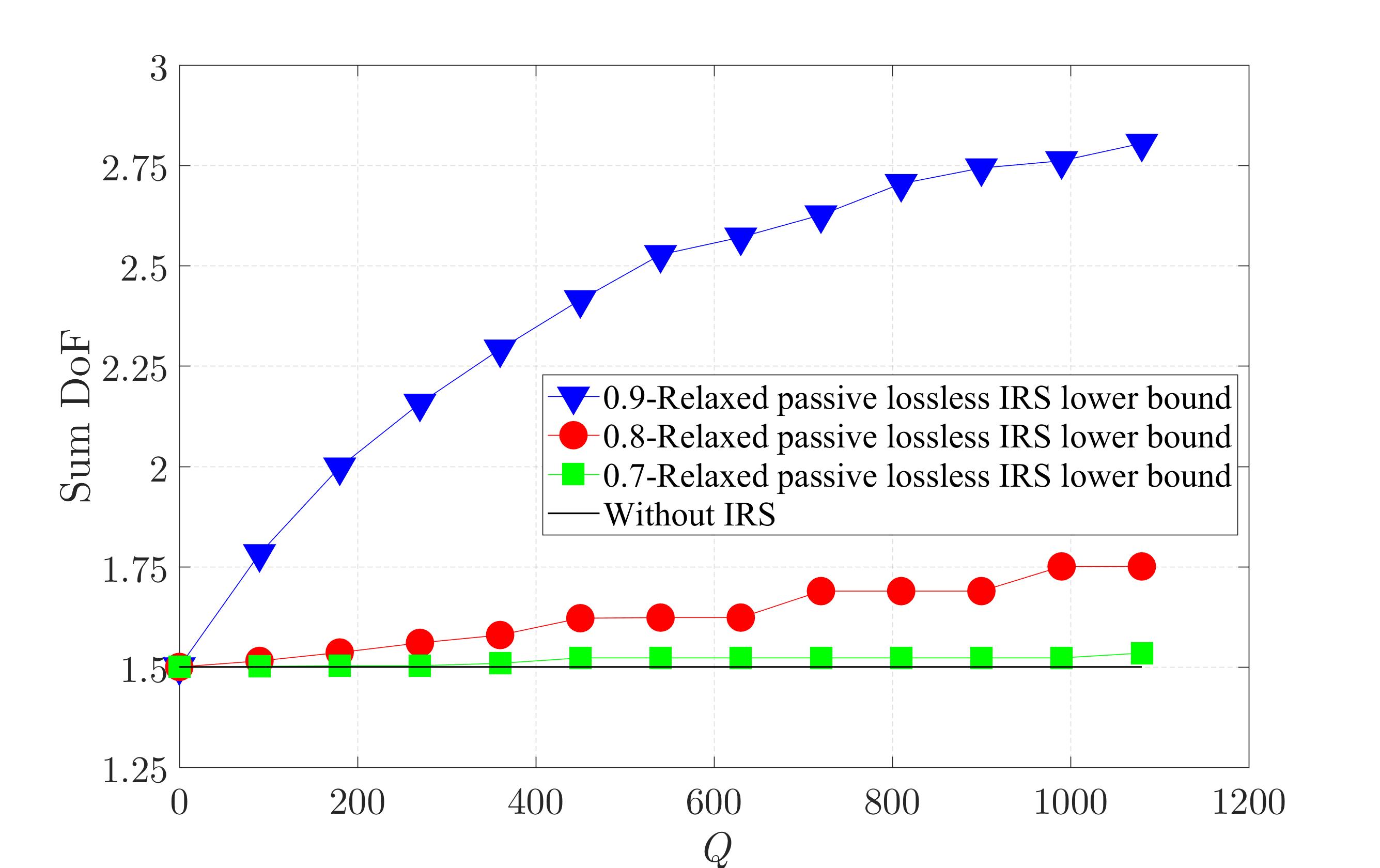}
\vspace{-30pt}
\caption{Comparison of the lower  bound on the sum DoF of a $3$-user interference channel in the presence of  $0.9,0.8$, and $0.7$-relaxed passive lossless IRSs  and the sum DoF without an IRS.}
 \label{fig3}
\end{minipage}
\quad
\begin{minipage}{.45\textwidth}
\vspace{20pt}
\centering
\includegraphics[width=8.2cm]{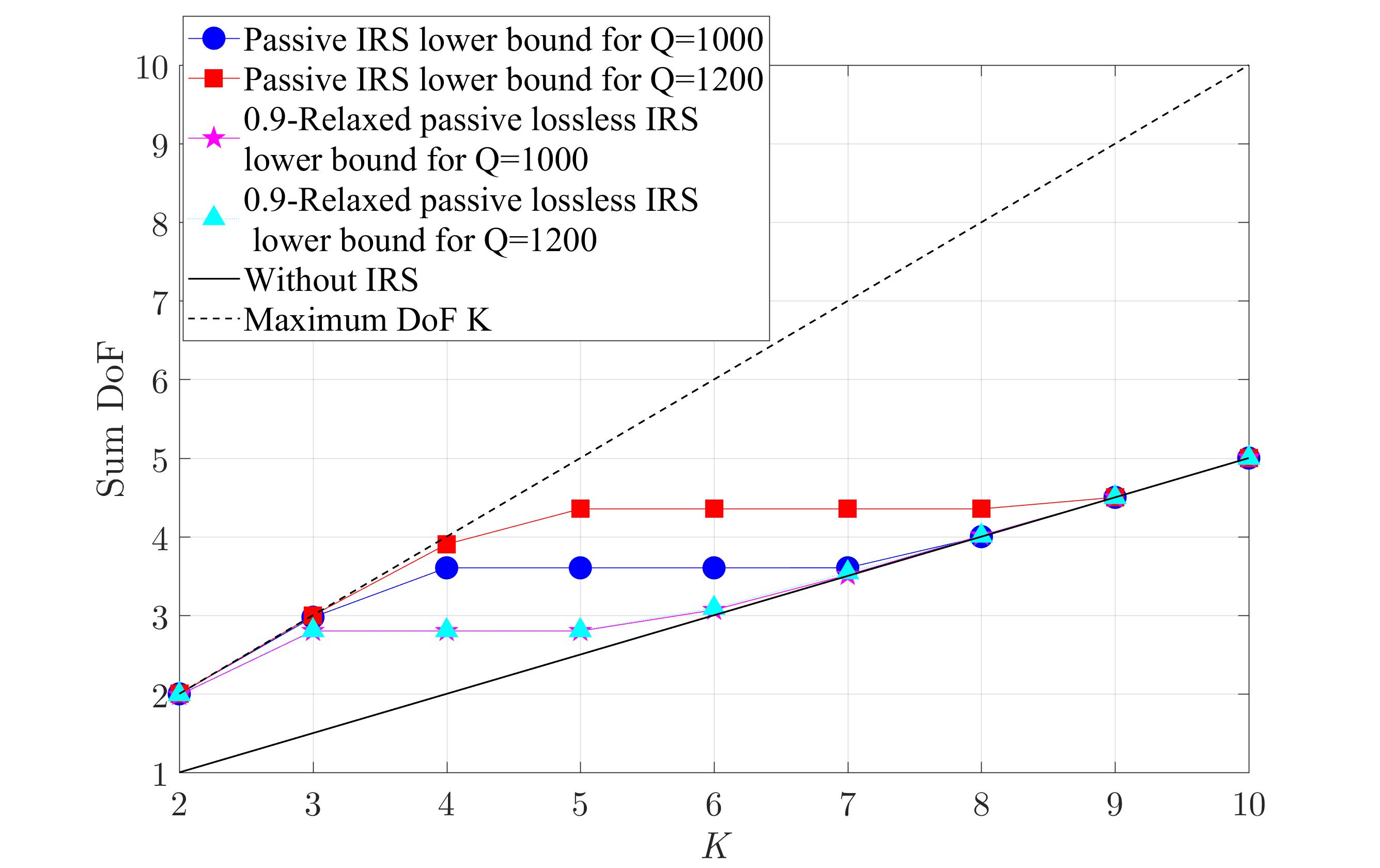}
\vspace{-30pt}
\caption{Comparison of the lower  bound on the sum DoF of a $3$-user interference channel in the presence of  passive and $0.9$-relaxed passive lossless IRS, the sum DoF without an IRS, and the maximum DoF $K$.}
 \label{fig4}
\end{minipage}
\end{figure}

{ In Fig. \ref{fig1}, we compare the lower and upper bounds on the sum DoF of a $3$-user interference channel in the presence of an active IRS, the lower bound on the sum DoF in the presence of an IR derived in \cite{me5},  and the sum DoF in the absence of an IRS. }
This figure shows that the upper bound grows linearly with $Q$, whereas the lower bound grows stepwise until both reach the maximum DoF of $3$. {\color{black}The reason for the shape of the lower bound} is that its value does not change for  $Q$ values in the interval  $[W(K-1)+W(K-W),(W+1)(K-1)+(W+1)(K-W-1))$. Also, we can see that, by  increasing the number of IRS elements, the gap between the lower bound and the upper bound decreases and becomes zero for $Q=K(K-1)\ge 6$. {Moreover, since active IRSs have more limited processing capabilities than IRs,  the achievable DoF of IRs are not less than the achievable DoF of IRSs. Nonetheless, the realization of ideal IRs is challenging, whereas active IRSs have been realized, see \cite{Loncar},\cite{Bousquet}.}
 Note that the DoF in this figure does not depend on the values of $\rho_1$, $\rho_2$, and $\hat h$.

 In Fig. \ref{fig2}, we compare the lower and upper bounds on the sum DoF for a $3$-user interference channel in the presence of a passive IRS and the sum DoF in the absence of the IRS. For this figure, we have assumed $\hat h=2.5*10^{-7}$. Similar to the active IRS, the gap between the lower bound and the upper bound decreases as $Q$ increases. Also, we can observe that for $Q\ge200$ both bounds converge and a sum DoF close to $3$ (the maximum sum DoF) can be achieved. The reason for the slower growth of the DoF with $Q$ compared to active IRS is the more limited capability of passive IRS to control the amplitude of the signal, as only attenuation is possible.

In Fig. \ref{fig3}, we compare the lower bound of the sum DoF for the $0.9,0.8$, and $0.7$-relaxed passive lossless IRS-assisted $3$-user interference channel and the sum DoF in the absence of the IRS.  For this figure, we assumed $\hat h=5*10^{-10}$.
We note that in this figure, the upper bound  
(\ref{upper bound}), which is used in the proof of convergence is employed instead of $\Pr\{\Lambda( t)=0\}$ due to its reduced computational complexity. Fig. \ref{fig3} shows that by decreasing $\varepsilon$, the sum DoF and its slope decrease.

In Fig. \ref{fig4}, we compare the achievable sum DoF of the passive and $0.9$-relaxed passive lossless IRS, the sum DoF in the absence of the IRS, and the maximum DoF $K$. For the passive IRS, we assumed $\hat h=5*10^{-7}$ and for the $0.9$-relaxed passive lossless IRS, we assumed $\hat h=5*10^{-10}$. The reason for choosing different blockage parameters  is to make the achievable DoFs comparable for convenience of illustration. We can observe from this figure that when we fix the number of IRS elements, for large values of $K$, the sum DoF becomes the same as the sum DoF without the IRS. Thus, the larger $K$ is, the more IRS elements are needed to achieve DoF improvement. 

{We note that to model the impact of blockage in Figs. \ref{fig2}-\ref{fig4}, the channel coefficients of the direct links are assumed to be much smaller than those of the channels connected to the IRS. Nevertheless, this scenario is different from two-hop communication where the direct channel coefficients are equal to zero \cite{Jeon3}-\cite{Shomorony}. In fact, if the direct channel coefficients are equal to zero, these channel coefficients are not drawn from a continuous cumulative probability distribution, and thus, the assumptions made for the derivation  of the results presented in the paper are not satisfied, and hence,  these results are not applicable. On the other hand, if the variance of channel coefficients are similar, the number of IRS elements needed for DoF improvement increases, because the numerator of (\ref{Markov1}) increases and hence, higher values of $Q$ is needed for $\Pr\{{\cal F}_{Q_l}\},l\ne 1$ to be small.
}

\section{Conclusion}

\label{sec7}

In this paper, we studied  the time-selective $K$-user interference channel in the presence of different types of IRSs from a  DoF perspective. We derived inner and outer bounds for the  DoF region and lower and upper bounds for the sum  DoF of the $K$-user interferene channel in the presence of both active and passive IRSs. We also  presented a lower bound for the sum  DoF of the $K$-user interference channel in the presence of $\varepsilon$-relaxed passive lossless IRS as an approximation of passive lossless IRSs. For all of these cases, we showed that by choosing a sufficiently large number of IRS elements,  any value less than $K$ can be achieved for the sum  DoF.
From our simulations, we observed that the active IRS requires the smallest number of elements and the $\varepsilon$-relaxed passive lossless IRS requires the largest number of elements to achieve the maximum sum DoF. Thus, the  active IRS has the best  and the $\varepsilon$-relaxed passive lossless IRS has the worst DoF performance.
 In addition, we defined a relaxed type of DoF, referred to as $\rho$-limited DoF, and proved that the maximum $K$  $\rho$-limited sum DoF is achievable if the passive lossless IRS is equipped with a sufficiently large number of elements.  Interesting directions for future research include: 
1) finding tighter bounds for passive IRSs,
2) adopting physics-based channel models for the analysis, and
3) considering imperfect channel state information.

\begin{appendices}
\section{}
\label{appendix1}

\textbf{\textit{A. Achievability}}

We prove the achievability of (\ref{DoF region const active}) in several steps.  {First, we define the normalized asymptotic dimension as follows:

\textbf{Normalized asymptotic dimension}:
In the following derivations, for a given $K$ and $Q$, the dimensions of certain matrices (beamforming matrices) will have the form  $O(n^{l}),l,n \in \mathbb{N}$, where $n$ is an auxiliary variable, which can go to infinity and $l$ is an integer, which depends on $K$ and $Q$. Hence, it is convenient to define the normalized asymptotic dimension (${D_N}$) of a  matrix $\bold V$ as follows:
\begin{equation}
\scalebox{.94}[1]{${D_N}({\bold V}) = \mathop {\lim }\limits_{n \to \infty } \frac{{d({\bold V})}}{{{n^l}}},$}
\label{DN}
\end{equation}
where $l$ is the minimum integer number, for which $ \mathop {\lim }\limits_{n \to \infty } \frac{{d(\bf{V})}}{{{n^l}}} < \infty$ holds.

We also use this definition for  vector spaces $\cal{A}$, i.e., $D_N({\cal A})$ is  the normalized asymptotic  dimension of $\cal A$.
}


\textbf{Step 1: Message stream generation}

For each transmitter $i\in \{1,...,K\}$, we introduce a vector of extended symbols  ${{\tilde {\bold{x}}}^{[i]}} \in \mathbb{C}^{{d_{{{\tilde {\bold{x}}}^{[i]}}}} \times 1}$ and a $T\times {{d_{{{\tilde {\bold{x}}}^{[i]}}}}}$  beamforming matrix ${{\tilde{\bold{V}}}^{[i]}}$, whose columns are the beamforming vectors for the elements of ${{\tilde{\bold{x}}}^{[i]}}$.
So, we can write:
\begin{equation}
{\bold{x}^{[i]}} =  {{\tilde{\bold{V}}}^{[i]}}{{\tilde{\bold{x}}}^{[i]}}.
\label{partitioning1}
\end{equation}
We will determine the dimension of ${{\tilde{\bold{V}}}^{[i]}}$ and ${{\tilde{\bold{x}}}^{[i]}}$, i.e., ${{d_{{{\tilde {\bold{x}}}^{[i]}}}}}$ in step 5 of the proof. Note that ${{\tilde{\bold{V}}}^{[i]}}$ has $T$ rows because there are $T$ time slots.

\textbf{Step 2: Interference cancellation and channel equalization}

Let us define the following set:
\begin{equation}
{\cal N} = \left\{ {(i,j)\left| {i,j \in \{ 1,...,K\} ,{n_{i,j}} = 0} \right.} \right\}.
\end{equation}
Then, we design the IRS element coefficients such that the following equations are satisfied for all $t\in \{1,...,T\}$:

\begin{equation}
\sum\limits_{u \in \{ 1,...,Q\} } {{X^{[i]}}({ t}){H_{\rm TI}^{[{u}i]}}({ t}){H_{\rm IR}^{[j{u}]}}({ t}){\tau ^{[u]}}({ t})}  = -{X^{[i]}}({ t}){H^{[ji]}}({ t}),(i,j)\in {\cal N},
\end{equation}
which is equivalent to
\begin{equation}
\sum\limits_{u \in \{ 1,...,Q\} } {{H_{\rm TI}^{[{u}i]}}({ t}){H_{\rm IR}^{[j{u}]}}({ t}){\tau ^{[u]}}({ t})}  =  - {H^{[ji]}}({ t}),(i,j) \in {\cal N}.
\label{Interference cancelation}
\end{equation}
By this procedure, the cross links from the $i$-th transmitter to the $j$-th receiver, where $(i,j)\in {\cal N}$, are eliminated.
If we write (\ref{Interference cancelation}) in matrix form  (${\bold{H}}{\bold{\tau}}=\bold h$), we can see that $\det({\bold H})$ is a nonzero polynomial in terms of $H_{\rm TI}^{[{u}i]}( t)$ and ${H_{\rm IR}^{[j{u}]}}( t)$ and by Lemma \ref{lemma1}, $\Pr \{\det({\bold H})=0\}=0$, so (\ref{Interference cancelation}) has a solution almost surely.

\begin{lemma}
Consider $k$ independent random variables $X_1,...,X_k$, each constructed from a continuous cumulative probability distribution. The probability of the event that a nonzero polynomial $P_k(X_1,...,X_k)$ constructed from $X_1,...,X_k$  with finite degree assumes the value zero is zero, i.e., $\Pr\{P_k(X_1,...,X_k)=0\}=0$.
\label{lemma1}
\end{lemma}

\begin{IEEEproof}
We use strong mathematical induction to prove this Lemma. For $k=0$, the proof of the statement is obvious because the probability of the event that the value of a nonzero number is zero is zero.  For $k=1$, a polynomial $P_1(\cdot)$ with degree $n$  has at most $n$ real roots, so we have:
\begin{equation*}
\Pr \left\{ {\bigcup\limits_{i = 1}^n {\{ {X_1} = x_i^*\} } } \right\} \le \sum\limits_{i = 1}^n {\Pr \left\{ {{X_1} = x_i^*} \right\} = 0} ,
\end{equation*}
where $ x_i^*$ are the roots of polynomial $P_1$ and ${\Pr \left\{ {{X_1} = x_i^*} \right\} = 0}$ because $X_1$ is drawn from  a continuous cumulative probability distribution.
To complete the proof, we now show that assuming the statement  of Lemma \ref{lemma1} is valid for $k'\in\{1,...,k\}$, this statement must be valid for $k+1$. Assume that the polynomial $P_{k+1}$ has the following general form:
\begin{equation}
P_{k+1}({X_1},...,{X_{k + 1}}) = \sum\limits_{({i_1},...,{i_{k + 1}}) \in {{\cal A}_{k + 1}}} {{a_{{i_1},...,{i_{k + 1}}}}X_1^{{i_1}} \cdots X_{k + 1}^{{i_{k + 1}}}} ,
\label{polynomial}
\end{equation}
\begin{equation*}
{{\cal A}_{k + 1}} \subseteq {\mathbb{W}^{k + 1}},\left| {{{\cal A}_{k + 1}}} \right| < \infty , 
\end{equation*}
where   $\exists ({{i_1},...,{i_{k + 1}}})\in {\cal A}_{k+1}$ for which ${a_{{i_1},...,{i_{k + 1}}}}\ne 0$. Without loss of generality, we assume that $X_{k+1}$ is present in $P_{k+1}(X_1,...,X_{k+1})$, otherwise $P_{k+1}(X_1,...,X_{k+1})$ can be represented as a polynomial of a lower degree.
 We can  rewrite (\ref{polynomial}) in the following form:
\begin{equation}
{P_{k + 1}}({X_1},...,{X_{k + 1}}) = \sum\limits_{i \in {\cal C}} {P_k^{[i]}({X_1},...,{X_k})X_{k + 1}^i} ,
\label{rewrite}
\end{equation}
where ${{\cal C}}\subset \mathbb{W},\left| {{\cal C}} \right| < \infty $. Then, we will have:
\begin{equation*}
\Pr \left\{ {{P_{k + 1}}({X_1},...,{X_{k + 1}}) = 0} \right\}=
\end{equation*}
\begin{equation*}
\Pr \left\{ {\left\{ {{P_{k + 1}}({X_1},...,{X_{k + 1}}) = 0} \right\}\bigcap {\left\{ {\bigcap\limits_{i \in {\cal C}} {\left\{ {P_k^{[i]}({X_1},...,{X_k}) = 0} \right\}} } \right\}} } \right\}
\end{equation*}
\begin{equation}
 + \Pr \left\{ {\left\{ {{P_{k + 1}}({X_1},...,{X_{k + 1}}) = 0} \right\}\bigcap {\left\{ {\bigcup\limits_{i \in {\cal C}} {\left\{ {P_k^{[i]}({X_1},...,{X_k}) \ne 0} \right\}} } \right\}} } \right\}.
\label{rewrite2}
\end{equation}
For the first term in (\ref{rewrite2}), we have:
\begin{equation*}
\Pr \left\{ {\left\{ {{P_{k + 1}}({X_1},...,{X_{k + 1}}) = 0} \right\}\bigcap {\left\{ {\bigcap\limits_{i \in {\cal C}} {\left\{ {P_k^{[i]}({X_1},...,{X_k}) = 0} \right\}} } \right\}} } \right\}\le 
\end{equation*}
\begin{equation}
\Pr \left\{ {\bigcap\limits_{i \in C} {\left\{ {P_k^{[i]}({X_1},...,{X_k}) = 0} \right\}} } \right\} \le \mathop {\min }\limits_{i \in {\cal C}} \Pr \left\{ {P_k^{[i]}({X_1},...,{X_k}) = 0} \right\} = 0,
\label{prob_k}
\end{equation}
where (\ref{prob_k}) is true because there exists at least one $i^*\in {\cal C}$ for which $P_{k}^{[i^*]}({X_1},...,{X_k})$ is nonzero and by the correctness of the statement of the lemma  for $k'\in\{1,...,k\}$, we have $\Pr \left\{P_{k}^{[i^*]}({X_1},...,{X_k})=0\right\}=0$. Now, define the following set:
\begin{equation*}
{\cal F} = \bigcup\limits_{i \in {\cal C}} {\left\{ {({x_1},...,{x_k})\left| {P_k^{[i]}({x_1},...,{x_k}) \ne 0} \right.} \right\}} .
\end{equation*}
Then, for the second term in (\ref{rewrite2}), we have:
\begin{equation*}
\Pr \left\{ {\left\{ {{P_{k + 1}}({X_1},...,{X_{k + 1}}) = 0} \right\}\bigcap {\left\{ {\bigcup\limits_{i \in {\cal C}} {\left\{ {P_k^{[i]}({X_1},...,{X_k}) \ne 0} \right\}} } \right\}} } \right\} =
\end{equation*}
\begin{equation}
\int\limits_{({x_1},...,{x_k}) \in {\cal F}} {\Pr \left\{ {\left\{ {{P_{k + 1}}({X_1},...,{X_{k + 1}}) = 0} \right\}\left| {{X_1} = {x_1},...,{X_k} = {x_k}} \right.} \right\}{f_{{X_1}}}({x_1}) \cdots {f_{{X_k}}}({x_k})d{x_1} \cdots d{x_k}}  = 0,
\label{fffff}
\end{equation}
where $f_{X_i}(x_i)$ is the probability distribution of $X_i$.
Eq. (\ref{fffff}) follows from the fact that conditioned on $X_1,...,X_k$, where  $({x_1},...,{x_k}) \in {\cal F}$, ${P_{k + 1}}({x_1},...,x_{k},{X_{k + 1}})$ is a nonzero polynomial constructed in terms of $X_{k+1}$, so we have $\Pr \left\{ {\left\{ {{P_{k + 1}}({X_1},...,{X_{k + 1}}) = 0} \right\}\left| {{X_1} = {x_1},...,{X_k} = {x_k}} \right.}\right\}=0$ by the correctness of the lemma for $k=1$.

\end{IEEEproof}

 Now, the resulting $\tau^{[u]}( t)$ has the following form:
\begin{equation}
{\tau ^{[u]}}({ t}) =\sum\limits_{(i',j') \in {{\cal N} }} {{H^{[j'i']}}({ t}){P^{[uj'i']}}({H_{\rm TI}^{[{{u'}}i'']}}({ t}),{H_{\rm IR}^{[j''{{u'}}]}}({ t}):u' \in \{ 1,...,Q\} ,(i'',j'') \in {{\cal N}})}  ,
\end{equation}
where ${P^{[uj'i']}}(\cdot)$ are fractional polynomials.
We refer to the ratio of polynomial $P_1(\cdot)$ to non-zero polynomial $P_2(\cdot)$ as a fractional polynomial.
Hence, the equivalent channel  has the following form:
\begin{equation*}
{Y^{[j]}}({ t}) = \sum\limits_{i \in \{  1,...,K\} } {{{\tilde H}^{[ji]}( t)}{X^{[i]}( t)}}  + {Z^{[j]}( t)},
\end{equation*}
where
\begin{equation*}
{{\tilde H}^{[ji]}}(t) = {H^{[ji]}}(t) + \sum\limits_{u \in \{ 1,...,Q\} } {H_{{\text{TI}}}^{[ui]}(t)H_{{\text{IR}}}^{[ju]}(t){\tau ^{[u]}}(t)} .
\end{equation*}
For the next steps, we define matrix $\tilde{\bf H}^{[ji]}$ as follows:
\begin{equation*}
{\tilde {\bf{H}}^{[ji]}} = \textrm{diag}\left( {{{\tilde H}^{[ji]}}({ 1}),{{\tilde H}^{[ji]}}({ 2}), \ldots ,{{\tilde H}^{[ji]}}({ T})} \right).
\end{equation*}

\textbf{Step 3: Interference alignment equations for  the $j$-th receiver}

In this step, we determine the interference alignment equations for each receiver. Our interference alignment concept is slightly different from that introduced in \cite{Cadambe1}. The authors in \cite{Cadambe1}   align the subspace of interference of interfering users with the subspace of interference of \textit{one} of the interfering users. In contrast to our approach,  we force the subspace of interference of \textit{all} interfering users to align with a larger subspace with equal normalized asymptotic dimension. Note that for matrices $\bold V$ and $\bold V'$, the following relations can hold simultaneously, $d({\bold V})>d({\bold V}')$ and $D_N({\bold V})=D_N({\bold V}')$.

Hence, for the $j$-th receiver and for the set of transmitters ${{\cal B}_j} = \left\{ {i\left| {i \in \{ 1,...,K\} ,i \ne j,{n_{i,j}} = 1} \right.} \right\}$, we must have:
\begin{equation}
\textrm{span}\left( {{\tilde{\bold{H}}^{[ji]}}{{\tilde {\bold V}}^{[i]}}} \right) \subseteq {\tilde{\cal A}_j},i\in {\cal B}_j,
\label{IA15new}
\end{equation}
where ${\tilde{\cal A}_j}$ is a subspace, for which we have:
\begin{equation}
\mathop {\max }\limits_{i \in {\cal B}_j} {D_N}\left(\textrm{span}\left({\tilde{\bold{H}}^{[ji]}}{{\tilde {\bold V}}^{[i]}}\right)\right) = {D_N}({\tilde{\cal A}_j}).
\label{IA16new}
\end{equation}
In addition, we define the message subspace as:
\begin{equation}
{\tilde {\cal C}_j} = \textrm{span}\left( {{{\tilde{\bold H}}^{[jj]}}{{\tilde {\bold V}}^{[j]}}} \right),
\end{equation}
and we design the beamforming matrices $\tilde{\bf V}^{[i]}$ such that ${\tilde {\cal C}_j}$ and ${\tilde {\cal A}_j}$ are full rank and linearly independent,  so we can ensure decodability of the message stream $\tilde {\bold x}^{[j]}$ by zero forcing at the $j$-th receiver.

\textbf{Step 4: Beamforming matrix design}

We consider the following beamforming matrices  ${{\tilde{\bold{V}}}^{[i]}}$:

\begin{equation}
{\tilde {\bf{V}} ^{[i]}} = \left\{ {\left[ {\prod\limits_{(i',j') \in \tilde {\cal S}} {{{\left( {{{\tilde{\bf{H}}}^{[j'i']}}} \right)}^{{\alpha _{j'i'}}}}} } \right]{\bf w}:{\alpha _{j'i'}} \in \{ 1,...,t_i n\} } \right\},
\label{beam4new}
\end{equation}
where:
\begin{equation*}
\tilde {\cal S} = \left\{ {(i',j')\left| {i',j' \in \{ 1,...,K\} ,i' \ne j',n_{i',j'}=1} \right.} \right\},
\end{equation*}
\begin{equation*}
{\bf w} = {\left[ {\begin{array}{*{20}{c}}
1&1& \cdots &1
\end{array}} \right]^H},
\end{equation*}
\begin{equation*}
0\le t_i\le 1,\forall i\in\{1,...,K\},
\end{equation*}
where $n\in \mathbb{N}$ and $t_i$ are auxiliary variables. In particular, $n$ is employed to compute the asymptotic dimensions by letting $n\to \infty$,  and $0\le t_i\le 1$ is used to control the asymptotic dimension of $\tilde{\bf V}^{[i]}$, i.e., $d({{\tilde{\bold{V}}}^{[i]}})$. In (\ref{beam4new}), $\alpha_{j'i'}$   can assume any value of set $\{1,...,t_in\}$, so the number of columns of  ${\tilde {\bf{V}} ^{[i]}}$ can be ${(t_in)}^{K^2-K-Q}$. Note that  each value of parameter $t_i$ can be approximated by a rational number with arbitrarily small error, and by considering a sufficiently large $n$, $t_in$ will be integer and the proposed transmission scheme will be realizable.

\textbf{Step 5: Satisfaction of the interference alignment equations, decodability of message symbols, and DoF analysis}

We characterize the message  subspace $\tilde{\cal{C}}_j$  and the interference subspace $\tilde{\cal{A}}_j$ as follows:
\begin{equation}
{\tilde {\cal C} _j} = {\rm{span}}\left\{ {\left[ {\prod\limits_{(i',j') \in \tilde {\cal S}^{\cal C}} {{{\left( {{{\tilde{\bf{H}}}^{[j'i']}}} \right)}^{{\alpha _{j'i'}}}}} } \right]{\bf w}:{\alpha _{j'i'}} \in \tilde {\cal S}_{j'i'j}^{\alpha ,{\cal C}}} \right\},
\label{c_tilde_dirty_IRS}
\end{equation}
\begin{equation}
{\tilde {\cal A} _j} = {\rm{span}}\left\{ {\left[ {\prod\limits_{(i',j') \in \tilde {\cal S}^{\cal A}} {{{\left( {{\tilde{\bf{H}}^{[j'i']}}} \right)}^{{\alpha _{j'i'}}}}} } \right]{\bf w}:{\alpha _{j'i'}} \in \tilde {\cal S}_{j'i'j}^{\alpha ,{\cal A}}} \right\},
\label{a_tilde_dirty_IRS}
\end{equation}
where sets $\tilde {\cal S}^{\cal{C}}$, $\tilde {\cal S}_{j'i'j}^{\alpha,{\cal{C}}}  $, $\tilde {\cal S}^{\cal{A}}$, and $\tilde {\cal S}_{j'i'j}^{\alpha,{\cal{A}}}  $ are defined as follows: 
\begin{equation}
\tilde {\cal S}^{\cal C} = \left\{ {(i',j')\left| {i',j' \in \{ 1,...,K\} ,{n_{i',j'}} = 1} \right.} \right\},
\end{equation}
\begin{equation}
\tilde {\cal S}_{j'i'j}^{\alpha ,{\cal C}} = \left\{ {\begin{array}{*{20}{c}}
{\{ 1,...,t_jn\} ,i' \ne j',{n_{i',j'}} = 1}\\
{\{ 0\} ,j' \ne j,i' = j'}\\
{\{ 1\} ,j' = j,i' = j}
\end{array}} \right.,
\end{equation}
\begin{equation}
\tilde {\cal S}^{\cal{A}} = \left\{ {(i',j')\left| {i',j' \in \{ 1,...,K\} ,i' \ne j',n_{i',j'}=1} \right.} \right\} ,
\end{equation}
\begin{equation}
\tilde {\cal S}_{{j'i'j}}^{\alpha,{\cal{A}}}  = \{ 1,...,(\mathop {\max }\limits_{i'' \ne j,n_{i'',j}=1}{t_{i''}})n + 1\}.
\end{equation}

Now, we prove two Lemmas to show that under some conditions, subspaces ${\tilde {\cal C} _j}$ and ${\tilde {\cal A} _j}$ are full rank and linearly independent.

\begin{lemma}
\label{lemma2}
Consider three sets of variables $\{{x_i},i \in {{\cal A}_x},\left| {{{\cal A}_x}} \right| < \infty \}$, $\{{y_i},i \in {{\cal A}_y},\left| {{{\cal A}_y}} \right| < \infty\}$, and $\{{z_i},i \in {{\cal A}_z},\left| {{{\cal A}_z}} \right| < \infty \}$. Consider  the following functions:
\begin{equation}
{f_j} = \prod\limits_{i = 1}^{\left| {{{\cal A}_x}} \right|} {{{\left( {{x_i} + \sum\limits_{i' \in {{\cal C}_j},i'' \in {{\cal D}_j}} {{x_{i'}}{y_{i''}}{P_1}^{[i'i''j]}({z_k}:k \in {{\cal A}_z}) + {y_{i''}}{P_2}^{[i'i''j]}({z_k}:k \in {{\cal A}_z})} } \right)}^{a_i^j}}},
\end{equation}
\begin{equation*}
(a_1^j,...,a_{\left| {{{\cal A}_x}} \right|}^j) \in {\mathbb{W}^{\left| {{{\cal A}_x}} \right|}},j \in \{ 1,...,J\},
\end{equation*}
where ${P_1^{[i'i''j]}}(\cdot)$ and ${P_2^{[i'i''j]}}(\cdot)$ are fractional polynomials and for $\forall j$, we have $\left| {{{\cal C}_j}} \right|,\left| {{{\cal D}_j}} \right| < \infty$. If for $\forall j,j'$ with $j\ne j'$, $(a_1^j,...,a_{\left| {{{\cal A}_x}} \right|}^j) \ne (a_1^{j'},...,a_{\left| {{{\cal A}_x}} \right|}^{j'})$, then the functions $f_j$ will be linearly independent.
\end{lemma}
\begin{IEEEproof}
We prove this lemma by contradiction, so we assume that there exist $({\lambda _1},...,{\lambda _J}) \ne (0,...,0)$ such that:
\begin{equation}
\sum\limits_{j = 1}^J {{\lambda _j}{f_j}}  = 0.
\label{contradiction}
\end{equation}
We define the following operators:
\begin{equation}
{\Omega_ j} = \prod\limits_{i = 1}^{\left| {{{\cal A}_x}} \right|} {\frac{{{\partial ^{a_i^j}}}}{{\partial {x_i}^{a_i^j}}}}, j \in \{1,...,J\}.
\label{operator}
\end{equation}
Note that by convention, $\frac{{{\partial ^0}}}{{\partial x_i^0}} = 1$.
By applying operator $\Omega_j$ to $(\ref{contradiction})$  and setting $y_i=0,x_i=0$, we have:
\begin{equation*}
{\left. {{ {\Omega_ j}}\sum\limits_{j' = 1}^J {{\lambda _{j'}}{f_{j'}}} } \right|_{{y_i} = 0,{x_i} = 0}} = \prod\limits_{i = 1}^{\left| {{{\cal A}_x}} \right|} {(a_i^j)!} {\lambda _j} = 0,
\end{equation*}
so $\lambda_j=0,\forall j \in \{1,...,J\}$, which is in contrast with the assumption $({\lambda _1},...,{\lambda _J}) \ne (0,...,0)$.

\end{IEEEproof}

\begin{lemma}
Consider the set of nonzero linearly independent fractional polynomials $\{P^{[j]}(\cdot),j\in\{1,...,J\}\}$ and consider $J$ sets of variables ${{{\cal X}}}_j=\{x_i^j:i \in {\cal I},{\cal I}\subseteq \mathbb{N},|{\cal I}|<\infty\}$, $j \in \{1,...,J\}$. The determinant of the following matrix will be a nonzero fractional polynomial:
\begin{equation}
\bold{A}=\left[ {\begin{array}{*{20}{c}}
{{P^{[1]}}({{{\cal X}}_1})}&{{P^{[2]}}({{{{\cal X}}}_1})}& \cdots &{{P^{[J]}}({{{{\cal X}}}_1})}\\
{{P^{[1]}}({{{{\cal X}}}_2})}&{{P^{[2]}}({{{{\cal X}}}_2})}& \cdots &{{P^{[J]}}({{{{\cal X}}}_2})}\\
 \vdots & \vdots & \ddots & \vdots \\
{{P^{[1]}}({{{{\cal X}}}_J})}&{{P^{[2]}}({{{{\cal X}}}_J})}& \cdots &{{P^{[J]}}({{{{\cal X}}}_J})}
\end{array}} \right].
\end{equation}

\label{lemma3}
\end{lemma}
\begin{IEEEproof}
We prove this Lemma by contradiction. Thus, we assume that:
\begin{equation}
 \det (\bold{A})=0.
\label{zero_det}
\end{equation}
By Laplace's formula, the determinant of matrix $\bold{A}$ has the following form in terms of its minors:
\begin{equation}
\det (\bold{A}) = \sum\limits_{j = 1}^J {{{( - 1)}^{(1 + j)}}{{{P^{[j]}}({{{{\cal X}}}_1})}}{M_{1,j}}},
\end{equation}
where the minor $M_{1,j}$ is the determinant of the submatrix constructed by removing the first row and the $j$-th column of matrix $\bold{A}$. The minors $M_{1,j}$ are functions of ${{{\cal X}}}_2,...,{{{\cal X}}}_J$, so all the minors  $M_{1,j}$ must be zero because of relation (\ref{zero_det}) and  linear independence of  the  polynomials $\{P^{[j]}(\cdot),j\in\{1,...,J\}\}$.
Since $M_{1,j}$ is a determinant itself, we can apply a similar argument  $J$ times  and conclude that each polynomial $P^{[j]}(\cdot),j\in\{1,...,J\}$, must be a zero polynomial, which is in contrast to the assumption of the Lemma.

\end{IEEEproof}

 By the nature of $\tilde {\cal {C}}_j$ in (\ref{c_tilde_dirty_IRS}) and $\tilde {\cal A}_j$ in (\ref{a_tilde_dirty_IRS}), we can see from the statement of  Lemma \ref{lemma2}, if we choose  variables $x_k$ as $H^{[ji]}( t),(i,j) \in {\left\{ {1,...,K} \right\}^2},(i,j) \notin {\cal N}$, $y_k$ as $H^{[ji]}( t),(i,j)\in {\cal N}$, and $z_k$ as $H_{\rm IR}^{[ju]}( t),H_{\rm TI}^{[{u'}i]}( t),i,j\in {\{1,...,K\}},u,u'\in \{1,...,Q\}$, then by Lemmas \ref{lemma1}-\ref{lemma3}, the subspaces $\tilde {\cal {A}}_j$ and $\tilde {\cal C}_j$ are full rank  and linearly independent almost surely. This is because if we put the base vectors of $\tilde {\cal {C}}_j$ and $\tilde {\cal A}_j$ into a matrix and choose a square matrix by omitting some of its rows, then by Lemmas \ref{lemma2} and \ref{lemma3}, its determinant will be a nonzero polynomial constructed from independent random variables and by Lemma \ref{lemma1} its determinant will be nonzero almost surely. In the following, we first assume that $T$ is sufficiently large, but later, we show how large $T$ has to be.

Now, we analyze the dimensions of the message and interference  subspaces. From (\ref{c_tilde_dirty_IRS}) and (\ref{a_tilde_dirty_IRS}), for the message subspace $\tilde {\cal C}_j$ and the interference subspace $\tilde {\cal A}_j$ at the $j$-th receiver, we have:
\begin{equation}
d({\tilde {\cal C}_j})={{(t_jn)}^{{{K}^2} - K-Q}},
\end{equation}
\begin{equation}
d({\tilde {\cal A}_j})\le {{((\mathop {\max }\limits_{i'' \ne j,n_{i'',j}=1}{t_{i''}})n+1)}^{{{K}^2} - K-Q}}.
\end{equation}
We can see from (\ref{DN})  that $l={{{K}^2} - K-Q}$, so the normalized asymptotic dimension of ${\tilde {\cal C}_j}$ and ${\tilde {\cal A}_j}$ is:
\begin{equation}
D_N ({\tilde {\cal C}_j})={{t_j}^{{{K}^2} - K-Q}} ,
\end{equation}
\begin{equation}
D_N({\tilde {\cal A}_j})={{(\mathop {\max }\limits_{i'' \ne j,n_{i'',j}=1}{t_{i''}})}^{{{K}^2} - K-Q}}=\mathop {\max }\limits_{i'' \ne j,n_{i'',j}=1}{({t_{i''}}^{{{K}^2} - K-Q})}.
\end{equation}
Without loss of generality, we assume following value for $T$:
\begin{equation}
T = {(n + 1)^{{K^2} - K - Q}},
\label{T1}
\end{equation}
so we have:
\begin{equation}
\mathop {\lim }\limits_{n \to \infty } \frac{T}{{{n^{{K^2} - K - Q}}}} = 1.
\label{T2}
\end{equation}
By assumption (\ref{T1}) and (\ref{T2}), we can see that for interference alignment equations (\ref{IA15new}) and (\ref{IA16new}) to be satisfied, we must have the following conditions for the $j$-th receiver:
\begin{equation}
d ({\tilde {\cal C}_j})+d({\tilde {\cal A}_j})\le{{(t_jn)}^{{{K}^2} - K-Q}}+{{((\mathop {\max }\limits_{i'' \ne j,n_{i'',j}=1}{t_{i''}})n+1)}^{{{K}^2} - K-Q}}\le T,
\label{dd_j1}
\end{equation}
\begin{equation}
d ({\tilde {\cal C}_j})={{(t_jn)}^{{{K}^2} - K-Q}}\le T,
\label{dd_j2}
\end{equation}
\begin{equation}
d({\tilde {\cal A}_j})\le{{((\mathop {\max }\limits_{i'' \ne j,n_{i'',j}=1}{t_{i''}})n+1)}^{{{K}^2} - K-Q}}\le T.
\label{dd_j3}
\end{equation}
Therefore, we obtain that the following inequalities must hold:
\begin{equation}
D_N ({\tilde {\cal C}_j})+D_N({\tilde {\cal A}_j})={{t_j}^{{{K}^2} - K-Q}}+\mathop {\max }\limits_{i'' \ne j,n_{i'',j}=1}{({t_{i''}}^{{{K}^2} - K-Q})}\le1,
\label{d_j1}
\end{equation}
\begin{equation}
D_N ({\tilde {\cal C}_j})={{t_j}^{{{K}^2} - K-Q}}\le1,
\label{d_j2}
\end{equation}
\begin{equation}
D_N({\tilde {\cal A}_j})=\mathop {\max }\limits_{i'' \ne j,n_{i'',j}=1}{({t_{i''}}^{{{K}^2} - K-Q})}\le1.
\label{d_j3}
\end{equation}
By assumptions (\ref{T1}) and (\ref{T2}), we can see that the DoF achieved for the $j$-th receiver is:
 \begin{equation}
{d_j} = \mathop {\lim }\limits_{n \to \infty } \frac{{d({{\tilde {\cal C}}_j})}}{T} = \mathop {\lim }\limits_{n \to \infty } \frac{{\frac{{d({{\tilde {\cal C}}_j})}}{{{n^{{K^2} - K - Q}}}}}}{{\frac{T}{{{n^{{K^2} - K - Q}}}}}} = \frac{{{D_N}({{\tilde {\cal C}}_j})}}{1} = {t_j}^{{K^2} - K - Q}.
\label{d_j4}
\end{equation}
So, by (\ref{d_j1})-(\ref{d_j4}), for each $j\in\{1,...,K\}$, we will have:
\begin{equation}
{d_j}+\mathop {\max }\limits_{i'' \ne j,n_{i'',j}=1}{d_{i''}}\le1,
\end{equation}
\begin{equation}
0\le d_j\le 1,
\end{equation}
which corresponds to the region in (\ref{DoF region const active}). Thus, each DoF vector ${\bf d}=(d_1, ..., d_K)$ which satisfies the inequalities in (\ref{DoF region const active}), is achievable. For more clarity, we present the following example:

\textit{Example 1:}
Consider a $4$-user interference channel assisted by a $6$-element active IRS  realizing the following network matrix using Eq. (5):
\begin{equation*}
{\bf N} = \left[ {\begin{array}{*{20}{c}}
  1&1&1&0 \\ 
  1&1&1&0 \\ 
  1&1&1&0 \\ 
  0&0&0&1 
\end{array}} \right].
\end{equation*}
Now, we show that the DoF vector ${\bf d}=\left[0.5\quad 0.5\quad 0.5\quad 1\right]$ is achievable. To this end, for transmitters $i\in\{1,2,3\}$ we consider the following beamforming matrices:
\begin{equation*}
\scalebox{1}[1]{$\tilde{\bf V}^{[i]}=
  \left\{ {{{\tilde {\mathbf{H}}}^{[21]}}{{\tilde {\mathbf{H}}}^{[31]}}{{\tilde {\mathbf{H}}}^{[12]}}{{\tilde {\mathbf{H}}}^{[32]}}{{\tilde {\mathbf{H}}}^{[13]}}{{\tilde {\mathbf{H}}}^{[23]}}{\bf w},...,{{\tilde {\mathbf{H}}}^{[21]}}{{\tilde {\mathbf{H}}}^{[31]}}{{\tilde {\mathbf{H}}}^{[12]}}{{\tilde {\mathbf{H}}}^{[32]}}{{\tilde {\mathbf{H}}}^{[13]}}{{\left( {{{\tilde {\mathbf{H}}}^{[23]}}} \right)}^{\frac{n}{2}}}{\bf w},} \right. $}
\end{equation*}
\begin{equation*}
{\tilde {\mathbf{H}}^{[21]}}{\tilde {\mathbf{H}}^{[31]}}{\tilde {\mathbf{H}}^{[12]}}{\tilde {\mathbf{H}}^{[32]}}{\left( {{{\tilde {\mathbf{H}}}^{[13]}}} \right)^2}{\tilde {\mathbf{H}}^{[23]}}{\bf w},...,{\tilde {\mathbf{H}}^{[21]}}{\tilde {\mathbf{H}}^{[31]}}{\tilde {\mathbf{H}}^{[12]}}{\tilde {\mathbf{H}}^{[32]}}{\left( {{{\tilde {\mathbf{H}}}^{[13]}}} \right)^2}{\left( {{{\tilde {\mathbf{H}}}^{[23]}}} \right)^{\frac{n}{2}}}{\bf w},
\end{equation*}
\begin{equation*}
{\tilde {\mathbf{H}}^{[21]}}{\tilde {\mathbf{H}}^{[31]}}{\tilde {\mathbf{H}}^{[12]}}{\tilde {\mathbf{H}}^{[32]}}{\left( {{{\tilde {\mathbf{H}}}^{[13]}}} \right)^3}{\tilde {\mathbf{H}}^{[23]}}{\bf w},...,{\tilde {\mathbf{H}}^{[21]}}{\tilde {\mathbf{H}}^{[31]}}{\tilde {\mathbf{H}}^{[12]}}{\tilde {\mathbf{H}}^{[32]}}{\left( {{{\tilde {\mathbf{H}}}^{[13]}}} \right)^3}{\left( {{{\tilde {\mathbf{H}}}^{[23]}}} \right)^{\frac{n}{2}}}{\bf w},
\end{equation*}
\begin{equation*}
\vdots 
\end{equation*}
\begin{equation*}
\scalebox{.89}[1]{$\left. {{{\left( {{{\tilde {\mathbf{H}}}^{[21]}}} \right)}^n}{{\left( {{{\tilde {\mathbf{H}}}^{[31]}}} \right)}^n}{{\left( {{{\tilde {\mathbf{H}}}^{[12]}}} \right)}^n}{{\left( {{{\tilde {\mathbf{H}}}^{[32]}}} \right)}^n}{{\left( {{{\tilde {\mathbf{H}}}^{[13]}}} \right)}^n}{{\tilde {\mathbf{H}}}^{[23]}}{\bf w},...,{{\left( {{{\tilde {\mathbf{H}}}^{[21]}}} \right)}^n}{{\left( {{{\tilde {\mathbf{H}}}^{[31]}}} \right)}^n}{{\left( {{{\tilde {\mathbf{H}}}^{[12]}}} \right)}^n}{{\left( {{{\tilde {\mathbf{H}}}^{[32]}}} \right)}^n}{{\left( {{{\tilde {\mathbf{H}}}^{[13]}}} \right)}^n}{{\left( {{{\tilde {\mathbf{H}}}^{[23]}}} \right)}^{\frac{n}{2}}}}{\bf w} \right\}$},
\end{equation*}
where $n$ is an auxiliary variable, which is assumed to be even, ${\bf w}={[1\quad1\quad .... \quad1]}^T$, and $\tilde{\bf H}^{[ji]}$ is a $T\times T$ diagonal matrix, whose $t$-th diagonal element is the equivalent channel coeeficient $\tilde{H}^{[ji]}(t)$ from the $i$-th transmitter to the $j$-th receiver in the $t$-th time slot. Moreover,  notation $\{\}$ for  matrix ${\bf V}^{[i]}$ means that the order of the column vectors of this matrix is not important. The dimension of ${\bf V}^{[i]}, i\in\{1,2,3\}$, is $T\times \frac{1}{2}n^6$.
In addition, we consider:
\begin{equation*}
\scalebox{1}[1]{$\tilde{\bf V}^{[4]}=
  \left\{ {{{\tilde {\mathbf{H}}}^{[21]}}{{\tilde {\mathbf{H}}}^{[31]}}{{\tilde {\mathbf{H}}}^{[12]}}{{\tilde {\mathbf{H}}}^{[32]}}{{\tilde {\mathbf{H}}}^{[13]}}{{\tilde {\mathbf{H}}}^{[23]}}{\bf w},...,{{\tilde {\mathbf{H}}}^{[21]}}{{\tilde {\mathbf{H}}}^{[31]}}{{\tilde {\mathbf{H}}}^{[12]}}{{\tilde {\mathbf{H}}}^{[32]}}{{\tilde {\mathbf{H}}}^{[13]}}{{\left( {{{\tilde {\mathbf{H}}}^{[23]}}} \right)}^{n}}{\bf w},} \right. $}
\end{equation*}
\begin{equation*}
{\tilde {\mathbf{H}}^{[21]}}{\tilde {\mathbf{H}}^{[31]}}{\tilde {\mathbf{H}}^{[12]}}{\tilde {\mathbf{H}}^{[32]}}{\left( {{{\tilde {\mathbf{H}}}^{[13]}}} \right)^2}{\tilde {\mathbf{H}}^{[23]}}{\bf w},...,{\tilde {\mathbf{H}}^{[21]}}{\tilde {\mathbf{H}}^{[31]}}{\tilde {\mathbf{H}}^{[12]}}{\tilde {\mathbf{H}}^{[32]}}{\left( {{{\tilde {\mathbf{H}}}^{[13]}}} \right)^2}{\left( {{{\tilde {\mathbf{H}}}^{[23]}}} \right)^{n}}{\bf w},
\end{equation*}
\begin{equation*}
{\tilde {\mathbf{H}}^{[21]}}{\tilde {\mathbf{H}}^{[31]}}{\tilde {\mathbf{H}}^{[12]}}{\tilde {\mathbf{H}}^{[32]}}{\left( {{{\tilde {\mathbf{H}}}^{[13]}}} \right)^3}{\tilde {\mathbf{H}}^{[23]}}{\bf w},...,{\tilde {\mathbf{H}}^{[21]}}{\tilde {\mathbf{H}}^{[31]}}{\tilde {\mathbf{H}}^{[12]}}{\tilde {\mathbf{H}}^{[32]}}{\left( {{{\tilde {\mathbf{H}}}^{[13]}}} \right)^3}{\left( {{{\tilde {\mathbf{H}}}^{[23]}}} \right)^{n}}{\bf w},
\end{equation*}
\begin{equation*}
\vdots 
\end{equation*}
\begin{equation*}
\scalebox{.89}[1]{$\left. {{{\left( {{{\tilde {\mathbf{H}}}^{[21]}}} \right)}^n}{{\left( {{{\tilde {\mathbf{H}}}^{[31]}}} \right)}^n}{{\left( {{{\tilde {\mathbf{H}}}^{[12]}}} \right)}^n}{{\left( {{{\tilde {\mathbf{H}}}^{[32]}}} \right)}^n}{{\left( {{{\tilde {\mathbf{H}}}^{[13]}}} \right)}^n}{{\tilde {\mathbf{H}}}^{[23]}}{\bf w},...,{{\left( {{{\tilde {\mathbf{H}}}^{[21]}}} \right)}^n}{{\left( {{{\tilde {\mathbf{H}}}^{[31]}}} \right)}^n}{{\left( {{{\tilde {\mathbf{H}}}^{[12]}}} \right)}^n}{{\left( {{{\tilde {\mathbf{H}}}^{[32]}}} \right)}^n}{{\left( {{{\tilde {\mathbf{H}}}^{[13]}}} \right)}^n}{{\left( {{{\tilde {\mathbf{H}}}^{[23]}}} \right)}^{n}}} {\bf w}\right\}$}.
\end{equation*}
The dimension of ${\bf V}^{[4]}$ is $T\times n^6$.
Then, the $i$-th transmitter generates an independent symbol for each column vector of $\tilde{\bf V}^{[i]}$. The symbol vector of this stream is $\tilde{\bf x}^{[i]}$. Therefore, each transmitter sends $\tilde{\bf V}^{[i]}\tilde{\bf x}^{[i]}$, i.e., this is the precoding procedure used at each transmitter.
On the other hand, the received signals at receivers $j\in\{1,2,3\}$ can be written as follows:
\begin{equation*}
{\bold{y}^{[j]}} = \sum\limits_{i = 1}^3 {{\tilde{\bold{H}}^{[ji]}}{\bold{x}^{[i]}}}     + {\bold{z}^{[j]}},
\end{equation*}
and at the $4$-th receiver, we have:
\begin{equation*}
{\bold{y}^{[4]}} = {{\tilde{\bold{H}}^{[44]}}{\bold{x}^{[4]}}}     + {\bold{z}^{[4]}}.
\end{equation*}
Thus, for $j\in\{1,2,3\}$, the message and interference subspaces can be characterized as follows:
\begin{equation*}
\tilde{\cal C}_j={\rm span}\{\tilde{\bf H}^{[jj]}\tilde{\bf V}^{[j]}\}=
\end{equation*}
\begin{equation*}
\scalebox{1}[1]{$
{\rm span}  \left\{ {\tilde{\bf H}^{[jj]}{{\tilde {\mathbf{H}}}^{[21]}}{{\tilde {\mathbf{H}}}^{[31]}}{{\tilde {\mathbf{H}}}^{[12]}}{{\tilde {\mathbf{H}}}^{[32]}}{{\tilde {\mathbf{H}}}^{[13]}}{{\tilde {\mathbf{H}}}^{[23]}}{\bf w},...,\tilde{\bf H}^{[jj]}{{\tilde {\mathbf{H}}}^{[21]}}{{\tilde {\mathbf{H}}}^{[31]}}{{\tilde {\mathbf{H}}}^{[12]}}{{\tilde {\mathbf{H}}}^{[32]}}{{\tilde {\mathbf{H}}}^{[13]}}{{\left( {{{\tilde {\mathbf{H}}}^{[23]}}} \right)}^{\frac{n}{2}}}{\bf w},} \right. $}
\end{equation*}
\begin{equation*}
\tilde{\bf H}^{[jj]}{\tilde {\mathbf{H}}^{[21]}}{\tilde {\mathbf{H}}^{[31]}}{\tilde {\mathbf{H}}^{[12]}}{\tilde {\mathbf{H}}^{[32]}}{\left( {{{\tilde {\mathbf{H}}}^{[13]}}} \right)^2}{\tilde {\mathbf{H}}^{[23]}}{\bf w},...,\tilde{\bf H}^{[jj]}{\tilde {\mathbf{H}}^{[21]}}{\tilde {\mathbf{H}}^{[31]}}{\tilde {\mathbf{H}}^{[12]}}{\tilde {\mathbf{H}}^{[32]}}{\left( {{{\tilde {\mathbf{H}}}^{[13]}}} \right)^2}{\left( {{{\tilde {\mathbf{H}}}^{[23]}}} \right)^{\frac{n}{2}}}{\bf w},
\end{equation*}
\begin{equation*}
\tilde{\bf H}^{[jj]}{\tilde {\mathbf{H}}^{[21]}}{\tilde {\mathbf{H}}^{[31]}}{\tilde {\mathbf{H}}^{[12]}}{\tilde {\mathbf{H}}^{[32]}}{\left( {{{\tilde {\mathbf{H}}}^{[13]}}} \right)^3}{\tilde {\mathbf{H}}^{[23]}}{\bf w},...,\tilde{\bf H}^{[jj]}{\tilde {\mathbf{H}}^{[21]}}{\tilde {\mathbf{H}}^{[31]}}{\tilde {\mathbf{H}}^{[12]}}{\tilde {\mathbf{H}}^{[32]}}{\left( {{{\tilde {\mathbf{H}}}^{[13]}}} \right)^3}{\left( {{{\tilde {\mathbf{H}}}^{[23]}}} \right)^{\frac{n}{2}}}{\bf w},
\end{equation*}
\begin{equation*}
\vdots 
\end{equation*}
\begin{equation*}
\scalebox{.83}[1]{$\left. {\tilde{\bf H}^{[jj]}{{\left( {{{\tilde {\mathbf{H}}}^{[21]}}} \right)}^n}{{\left( {{{\tilde {\mathbf{H}}}^{[31]}}} \right)}^n}{{\left( {{{\tilde {\mathbf{H}}}^{[12]}}} \right)}^n}{{\left( {{{\tilde {\mathbf{H}}}^{[32]}}} \right)}^n}{{\left( {{{\tilde {\mathbf{H}}}^{[13]}}} \right)}^n}{{\tilde {\mathbf{H}}}^{[23]}}{\bf w},...,\tilde{\bf H}^{[jj]}{{\left( {{{\tilde {\mathbf{H}}}^{[21]}}} \right)}^n}{{\left( {{{\tilde {\mathbf{H}}}^{[31]}}} \right)}^n}{{\left( {{{\tilde {\mathbf{H}}}^{[12]}}} \right)}^n}{{\left( {{{\tilde {\mathbf{H}}}^{[32]}}} \right)}^n}{{\left( {{{\tilde {\mathbf{H}}}^{[13]}}} \right)}^n}{{\left( {{{\tilde {\mathbf{H}}}^{[23]}}} \right)}^{\frac{n}{2}}}}{\bf w} \right\}$},
\end{equation*}
\begin{equation*}
\scalebox{1}[1]{$\tilde{\cal A}_j={\rm span}
  \left\{ {{{\tilde {\mathbf{H}}}^{[21]}}{{\tilde {\mathbf{H}}}^{[31]}}{{\tilde {\mathbf{H}}}^{[12]}}{{\tilde {\mathbf{H}}}^{[32]}}{{\tilde {\mathbf{H}}}^{[13]}}{{\tilde {\mathbf{H}}}^{[23]}}{\bf w},...,{{\tilde {\mathbf{H}}}^{[21]}}{{\tilde {\mathbf{H}}}^{[31]}}{{\tilde {\mathbf{H}}}^{[12]}}{{\tilde {\mathbf{H}}}^{[32]}}{{\tilde {\mathbf{H}}}^{[13]}}{{\left( {{{\tilde {\mathbf{H}}}^{[23]}}} \right)}^{\frac{n}{2}+1}}{\bf w},} \right. $}
\end{equation*}
\begin{equation*}
{\tilde {\mathbf{H}}^{[21]}}{\tilde {\mathbf{H}}^{[31]}}{\tilde {\mathbf{H}}^{[12]}}{\tilde {\mathbf{H}}^{[32]}}{\left( {{{\tilde {\mathbf{H}}}^{[13]}}} \right)^2}{\tilde {\mathbf{H}}^{[23]}}{\bf w},...,{\tilde {\mathbf{H}}^{[21]}}{\tilde {\mathbf{H}}^{[31]}}{\tilde {\mathbf{H}}^{[12]}}{\tilde {\mathbf{H}}^{[32]}}{\left( {{{\tilde {\mathbf{H}}}^{[13]}}} \right)^2}{\left( {{{\tilde {\mathbf{H}}}^{[23]}}} \right)^{\frac{n}{2}+1}}{\bf w},
\end{equation*}
\begin{equation*}
{\tilde {\mathbf{H}}^{[21]}}{\tilde {\mathbf{H}}^{[31]}}{\tilde {\mathbf{H}}^{[12]}}{\tilde {\mathbf{H}}^{[32]}}{\left( {{{\tilde {\mathbf{H}}}^{[13]}}} \right)^3}{\tilde {\mathbf{H}}^{[23]}}{\bf w},...,{\tilde {\mathbf{H}}^{[21]}}{\tilde {\mathbf{H}}^{[31]}}{\tilde {\mathbf{H}}^{[12]}}{\tilde {\mathbf{H}}^{[32]}}{\left( {{{\tilde {\mathbf{H}}}^{[13]}}} \right)^3}{\left( {{{\tilde {\mathbf{H}}}^{[23]}}} \right)^{\frac{n}{2}+1}}{\bf w},
\end{equation*}
\begin{equation*}
\vdots 
\end{equation*}
\begin{equation*}
\scalebox{.75}[1]{$\left. {{{\left( {{{\tilde {\mathbf{H}}}^{[21]}}} \right)}^{n+1}}{{\left( {{{\tilde {\mathbf{H}}}^{[31]}}} \right)}^{n+1}}{{\left( {{{\tilde {\mathbf{H}}}^{[12]}}} \right)}^{n+1}}{{\left( {{{\tilde {\mathbf{H}}}^{[32]}}} \right)}^{n+1}}{{\left( {{{\tilde {\mathbf{H}}}^{[13]}}} \right)}^{n+1}}{{\tilde {\mathbf{H}}}^{[23]}}{\bf w},...,{{\left( {{{\tilde {\mathbf{H}}}^{[21]}}} \right)}^{n+1}}{{\left( {{{\tilde {\mathbf{H}}}^{[31]}}} \right)}^{n+1}}{{\left( {{{\tilde {\mathbf{H}}}^{[12]}}} \right)}^{n+1}}{{\left( {{{\tilde {\mathbf{H}}}^{[32]}}} \right)}^{n+1}}{{\left( {{{\tilde {\mathbf{H}}}^{[13]}}} \right)}^{n+1}}{{\left( {{{\tilde {\mathbf{H}}}^{[23]}}} \right)}^{\frac{n}{2}+1}}}{\bf w} \right\}$}.
\end{equation*}
Now, by the argument given in the proof of Theorem 1, if we choose $T=\frac{1}{2}n^6+{(n+1)}^5{(\frac{n}{2}+1)}$, all message and interference subspaces will be linearly independent and the messages of each receiver can be decoded by zero forcing the interference subspace at each receiver, i.e.,   ${\bf y}^{[j]}$ is multiplied by the null space of $\tilde{\cal A}_j$. 
On the other hand, we have:
\begin{equation*}
d(\tilde{\cal C}_j)=\frac{1}{2}n^6,
\end{equation*}
\begin{equation*}
d(\tilde{\cal A}_j)={(n+1)}^5{\left(\frac{n}{2}+1\right)},
\end{equation*}
where $d(\tilde{\cal A}_j)$ and $d(\tilde{\cal C}_j)$ denote the numbers of base vectors of $\tilde{\cal A}_j$ and $\tilde{\cal C}_j$, respectively. This fact indicates that each receiver $j\in\{1,2,3\}$ can receive $\frac{1}{2}n^6$ interference-free symbols in $\frac{1}{2}n^6+{(n+1)}^5{(\frac{n}{2}+1)}$ time slots, thus, each one of the receivers can achieve $\frac{\frac{1}{2}n^6}{\frac{1}{2}n^6+{(n+1)}^5{(\frac{n}{2}+1)}}$ DoF.
For the $4$-th receiver, we have:
\begin{equation*}
\tilde{\cal C}_4={\rm span}\{\tilde{\bf H}^{[44]}\tilde{\bf V}^{[4]}\}=
\end{equation*}
\begin{equation*}
\scalebox{1}[1]{$
{\rm span}  \left\{ {\tilde{\bf H}^{[44]}{{\tilde {\mathbf{H}}}^{[21]}}{{\tilde {\mathbf{H}}}^{[31]}}{{\tilde {\mathbf{H}}}^{[12]}}{{\tilde {\mathbf{H}}}^{[32]}}{{\tilde {\mathbf{H}}}^{[13]}}{{\tilde {\mathbf{H}}}^{[23]}}{\bf w},...,\tilde{\bf H}^{[44]}{{\tilde {\mathbf{H}}}^{[21]}}{{\tilde {\mathbf{H}}}^{[31]}}{{\tilde {\mathbf{H}}}^{[12]}}{{\tilde {\mathbf{H}}}^{[32]}}{{\tilde {\mathbf{H}}}^{[13]}}{{\left( {{{\tilde {\mathbf{H}}}^{[23]}}} \right)}^{n}}{\bf w},} \right. $}
\end{equation*}
\begin{equation*}
\tilde{\bf H}^{[44]}{\tilde {\mathbf{H}}^{[21]}}{\tilde {\mathbf{H}}^{[31]}}{\tilde {\mathbf{H}}^{[12]}}{\tilde {\mathbf{H}}^{[32]}}{\left( {{{\tilde {\mathbf{H}}}^{[13]}}} \right)^2}{\tilde {\mathbf{H}}^{[23]}}{\bf w},...,\tilde{\bf H}^{[44]}{\tilde {\mathbf{H}}^{[21]}}{\tilde {\mathbf{H}}^{[31]}}{\tilde {\mathbf{H}}^{[12]}}{\tilde {\mathbf{H}}^{[32]}}{\left( {{{\tilde {\mathbf{H}}}^{[13]}}} \right)^2}{\left( {{{\tilde {\mathbf{H}}}^{[23]}}}{\bf w} \right)^{n}},
\end{equation*}
\begin{equation*}
\tilde{\bf H}^{[44]}{\tilde {\mathbf{H}}^{[21]}}{\tilde {\mathbf{H}}^{[31]}}{\tilde {\mathbf{H}}^{[12]}}{\tilde {\mathbf{H}}^{[32]}}{\left( {{{\tilde {\mathbf{H}}}^{[13]}}} \right)^3}{\tilde {\mathbf{H}}^{[23]}}{\bf w},...,\tilde{\bf H}^{[44]}{\tilde {\mathbf{H}}^{[21]}}{\tilde {\mathbf{H}}^{[31]}}{\tilde {\mathbf{H}}^{[12]}}{\tilde {\mathbf{H}}^{[32]}}{\left( {{{\tilde {\mathbf{H}}}^{[13]}}} \right)^3}{\left( {{{\tilde {\mathbf{H}}}^{[23]}}} \right)^{n}}{\bf w},
\end{equation*}
\begin{equation*}
\vdots 
\end{equation*}
\begin{equation*}
\scalebox{.84}[1]{$\left. {\tilde{\bf H}^{[44]}{{\left( {{{\tilde {\mathbf{H}}}^{[21]}}} \right)}^n}{{\left( {{{\tilde {\mathbf{H}}}^{[31]}}} \right)}^n}{{\left( {{{\tilde {\mathbf{H}}}^{[12]}}} \right)}^n}{{\left( {{{\tilde {\mathbf{H}}}^{[32]}}} \right)}^n}{{\left( {{{\tilde {\mathbf{H}}}^{[13]}}} \right)}^n}{{\tilde {\mathbf{H}}}^{[23]}}{\bf w},...,\tilde{\bf H}^{[44]}{{\left( {{{\tilde {\mathbf{H}}}^{[21]}}} \right)}^n}{{\left( {{{\tilde {\mathbf{H}}}^{[31]}}} \right)}^n}{{\left( {{{\tilde {\mathbf{H}}}^{[12]}}} \right)}^n}{{\left( {{{\tilde {\mathbf{H}}}^{[32]}}} \right)}^n}{{\left( {{{\tilde {\mathbf{H}}}^{[13]}}} \right)}^n}{{\left( {{{\tilde {\mathbf{H}}}^{[23]}}} \right)}^{n}}}{\bf w} \right\}$},
\end{equation*}
which yields:
\begin{equation*}
d({\cal C}_4)=n^6.
\end{equation*}
In addition, there is no interference subspace for the $4$-th receiver, because it is interference-free due to the IRS design. Thus, the $4$-th receiver can receive $n^6$ interference-free symbols in $\frac{1}{2}n^6+{(n+1)}^5{(\frac{n}{2}+1)}$ time slots, i.e., it can achieve $\frac{n^6}{\frac{1}{2}n^6+{(n+1)}^5{(\frac{n}{2}+1)}}$ DoF.
From this, we can see that the following DoF vector is achievable:
\begin{equation*}
{\bf d}=
\end{equation*}
\begin{equation*}
\left[ {\frac{{\frac{1}{2}{n^6}}}{{\frac{1}{2}{n^6} + {{(n + 1)}^5}(\frac{n}{2} + 1)}},\frac{{\frac{1}{2}{n^6}}}{{\frac{1}{2}{n^6} + {{(n + 1)}^5}(\frac{n}{2} + 1)}},\frac{{\frac{1}{2}{n^6}}}{{\frac{1}{2}{n^6} + {{(n + 1)}^5}(\frac{n}{2} + 1)}},\frac{{{n^6}}}{{\frac{1}{2}{n^6} + {{(n + 1)}^5}(\frac{n}{2} + 1)}}} \right],
\end{equation*}
which tends to ${\bf d}=[0.5\quad 0.5\quad 0.5\quad 1]$ for sufficiently large $n$.

\textbf{\textit{B. Converse}}

To show the converse part of the proof, we remark that it has been proved in \cite[Lemma 1]{Cadambe1} that if there is a nonzero cross link between the $i$-th transmitter and the $j$-th receiver ($i\ne j$), then we have:
\begin{equation}
d_i+d_j\le 1.
\end{equation} 
The inequalities $d_i\le 1$ are obvious because we have:
\begin{equation*}
{r_i} \le \frac{1}{T}I({w^{[i]}};{{\bf y}^{[i]}}) \le \log (\rho ) + o(\log (\rho )).
\end{equation*}
Hence, the DoF region is upper bounded by the inequalities in (\ref{DoF region const active}).

\section{}
\label{appendix2}
The proof of this theorem is straightforward and based on time sharing. Assume that we have  $T$ time slots. Then, we divide these $T$ time slots into ${\left| {{{\cal N}_Q}} \right|}$  groups, such that the $i$-th group contains $a_iT$ time slots. We also set the IRS such that in the time slots corresponding to the $i$-th group, the network matrix is ${\bf N}_i$, so the DoF vector achieved in these time slots is given by ${\cal D}_{{\bf N}_i}$ according to Theorem \ref{thm DoF reg const active IRS}. Hence, the achievability of (\ref{DoF region active inner2}) is proved.

\section{}
\label{appendix3}

Consider given $i,j\in\{1,...,K\}$. 
Using (\ref{output}), we construct a new variable as follows:
\begin{equation}
{{\bf y}^{[i]}}^\prime  = {{\tilde {\tilde {\bf H}}}^{[ji]}}{\left( {{{\tilde {\bf H}}^{[ii]}}} \right)^{ - 1}}{{\bf y}^{[i]}}= {{\tilde {\tilde {\bf H}}}^{[ji]}}{\left( {{{\tilde {\bf H}}^{[ii]}}} \right)^{ - 1}}\sum\limits_{k =1 }^K {{{\tilde {\bf H}}^{[ik]}}{{\bf x}^{[k]}}}  + {{\tilde {\tilde {\bf H}}}^{[ji]}}{\left( {{{\tilde {\bf H}}^{[ii]}}} \right)^{ - 1}}{{\bf z}^{[i]}},\\
\end{equation}
where ${\tilde {\tilde {\bf H}}}^{[ji]}$ is a diagonal matrix whose $t$-th main diagonal element is ${{\tilde { H}}^{[ji]}}( t)$ if $n_{i,j}$ is $1$ in the $t$-th time slot, otherwise it is $1$.

Because the capacity region of the interference channel depends only on the noise marginals, we define
\begin{equation}
{{\bf{z}}^{[i]}}^\prime  = {{\tilde {\tilde {\bf H}}}^{[ji]}}{\left( {{{\tilde {\bf{H}}}^{[ii]}}} \right)^{ - 1}}{{\bf{z}}^{[i]}} = {\bf{\bar z}} + {{{\bf{\bar z}}}^{[i]}},
\end{equation}
\begin{equation}
{{\bf{z}}^{[i]}} = {\bf{ {\bar z}}} + {{{\bf{\bar{ \bar z}}}}^{[i]}},
\end{equation}
where
\begin{equation}
{\bf{\bar z}} \sim {\cal CN}\left( {{\bf 0},\alpha {\bf I}} \right),
\end{equation}
\begin{equation}
{{{\bf{\bar z}}}^{[i]}} \sim {\cal CN}\left( {{\bf 0},{{{\tilde {\tilde {\bf H}}}}^{[ji]}}{{\left( {{{\tilde {\bf{H}}}^{[ii]}}} \right)}^{ - 1}}{{\left( {{{\left( {{{\tilde {\bf{H}}}^{[ii]}}} \right)}^{ - 1}}} \right)}^H}{{\left( {{{{\tilde {\tilde {\bf H}}}}^{[ji]}}} \right)}^H}N_0 - \alpha {\bf I}} \right),
\end{equation}
\begin{equation}
{{{\bf{\bar{ \bar z}}}}^{[i]}} \sim {\cal CN}\left( {{\bf 0},({N_0} - \alpha ){\bf I}} \right),
\end{equation}

Similarly, we define the following variables
\begin{equation}
{{\bf{y}}^{[j]}}^\prime  = {{\tilde {\tilde {\bf H}}}^{[ij]}}{\left( {{{\tilde {\bf{H}}}^{[jj]}}} \right)^{ - 1}}{{\bf{y}}^{[j]}} = {{\tilde {\tilde {\bf H}}}^{[ij]}}{\left( {{{\tilde {\bf{H}}}^{[jj]}}} \right)^{ - 1}}\sum\limits_{k =1 }^K {{{\tilde {\bf{H}}}^{[jk]}}{{\bf{x}}^{[k]}}}  + {{\tilde {\tilde {\bf H}}}^{[ij]}}{\left( {{{\tilde {\bf{H}}}^{[jj]}}} \right)^{ - 1}}{{\bf{z}}^{[j]}}.
\end{equation}
In addition, we define
\begin{equation}
{{\bf{z}}^{[j]}}^\prime  = {{\tilde {\tilde {\bf H}}}^{[ij]}}{\left( {{{\tilde {\bf{H}}}^{[jj]}}} \right)^{ - 1}}{{\bf{z}}^{[j]}} = {\bf{ \bar z}} + {{{\bf{ {\bar z}}}}^{[j]}},
\end{equation}
\begin{equation}
{{\bf{z}}^{[j]}} = {\bf{\bar {z}}} + {{{\bf{\bar {\bar z}}}}^{[j]}},
\end{equation}
where
\begin{equation}
{{{\bf{\bar z}}}^{[j]}} \sim {\cal CN}\left( {{\bf 0},{{{\tilde {\tilde {\bf H}}}}^{[ij]}}{{\left( {{{\tilde {\bf{H}}}^{[jj]}}} \right)}^{ - 1}}{{\left( {{{\left( {{{\tilde {\bf{H}}}^{[jj]}}} \right)}^{ - 1}}} \right)}^H}{{\left( {{{{\tilde {\tilde {\bf H}}}}^{[ij]}}} \right)}^H}N_0 - \alpha {\bf I}} \right),
\end{equation}
\begin{equation}
{{{\bf{\bar{ \bar z}}}}^{[j]}} \sim {\cal CN}\left( {{\bf 0},({N_0} - \alpha ){\bf I}} \right).
\end{equation}
Here, $N_0$ is the variance of the original noise and
\begin{equation}
\alpha  = \min \left\{ {{N_0},{\lambda _{\min }}\left( {{{{\tilde {\tilde {\bf H}}}}^{[ji]}}{{\left( {{{\tilde {\bf{H}}}^{[ii]}}} \right)}^{ - 1}}{{\left( {{{\left( {{{\tilde {\bf{H}}}^{[ii]}}} \right)}^{ - 1}}} \right)}^H}{{\left( {{{{\tilde {\tilde {\bf H}}}}^{[ji]}}} \right)}^H}} \right)N_0,} \right.
\end{equation}
\begin{equation}
\left. {{\lambda _{\min }}\left( {{{{\tilde {\tilde {\bf H}}}}^{[ij]}}{{\left( {{{\tilde {\bf{H}}}^{[jj]}}} \right)}^{ - 1}}{{\left( {{{\left( {{{\tilde {\bf{H}}}^{[jj]}}} \right)}^{ - 1}}} \right)}^H}{{\left( {{{{\tilde {\tilde {\bf H}}}}^{[ij]}}} \right)}^H}} \right)N_0} \right\},
\end{equation}
where $\lambda _{\min}({\bf A})$ denotes the smallest eigenvalue of matrix $\bf A$. We also define the following random variables
\begin{equation}
{{\bf{y}}^{[i]}}^{\prime \prime } = \sum\limits_{k =1 }^K {{{\tilde {\bf{H}}}^{[ik]}}{{\bf{x}}^{[k]}}}  + {\bf{\bar z}},
\end{equation}
\begin{equation}
{{\bf{y}}^{[i]}}^{\prime \prime \prime } = {{\tilde {\tilde {\bf H}}}^{[ji]}}{\left( {{{\tilde {\bf{H}}}^{[ii]}}} \right)^{ - 1}}\sum\limits_{k =1}^K {{{\tilde {\bf{H}}}^{[ik]}}{{\bf{x}}^{[k]}}}  + {\bf{\bar z}},
\end{equation}
\begin{equation}
{{\bf{y}}^{[j]}}^{\prime \prime } = \sum\limits_{k =1 }^K {{{\tilde {\bf{H}}}^{[jk]}}{{\bf{x}}^{[k]}}}  + {\bf{\bar z}},
\end{equation}
\begin{equation}
{{\bf{y}}^{[j]}}^{\prime \prime \prime } = {{\tilde {\tilde {\bf H}}}^{[ij]}}{\left( {{{\tilde {\bf{H}}}^{[jj]}}} \right)^{ - 1}}\sum\limits_{k =1 }^K {{{\tilde {\bf{H}}}^{[jk]}}{{\bf{x}}^{[k]}}}  + {\bf{\bar z}}.
\end{equation}
By Fano's inequality, the data processing inequality, and the independence of messages $w^{[i]}$, we can  show that the following inequalities hold:
\begin{equation}
T\times{r_i} \le I\left( {{w^{[i]}};{{\bf{y}}^{[i]}}} \right)+\epsilon \le I\left( {{w^{[i]}};{{\bf{y}}^{[i]}}^{\prime \prime }} \right)+\epsilon,
\label{eqa1}
\end{equation}
\begin{equation}
T\times{r_i} \le I\left( {{w^{[i]}};{{\bf{y}}^{[i]}}^\prime } \right) +\epsilon\le I\left( {{w^{[i]}};{{\bf{y}}^{[i]}}^{\prime \prime \prime }} \right)+\epsilon \le I\left( {{w^{[i]}};{{\bf{w}}_{i' \ne i}},{{\bf{y}}^{[i]}}^{\prime \prime \prime }} \right)+\epsilon = I\left( {{w^{[i]}};{{\bf{y}}^{[i]}}^{\prime \prime \prime }\left| {{{\bf{w}}_{i' \ne i}}} \right.} \right)+\epsilon,
\label{eqa2}
\end{equation}
\begin{equation}
T\times{r_j} \le I\left( {{w^{[j]}};{{\bf{y}}^{[j]}}} \right) +\epsilon\le I\left( {{w^{[j]}};{{\bf{y}}^{[j]}}^{\prime \prime }} \right)+\epsilon,
\label{eqa3}
\end{equation}
\begin{equation}
T\times{r_j} \le I\left( {{w^{[j]}};{{\bf{y}}^{[j]}}^\prime } \right)+\epsilon \le I\left( {{w^{[j]}};{{\bf{y}}^{[j]}}^{\prime \prime \prime }} \right) +\epsilon\le I\left( {{w^{[j]}};{{\bf{w}}_{i' \ne j}},{{\bf{y}}^{[j]}}^{\prime \prime \prime }} \right)+\epsilon = I\left( {{w^{[j]}};{{\bf{y}}^{[j]}}^{\prime \prime \prime }\left| {{{\bf{w}}_{i' \ne j}}} \right.} \right)+\epsilon,
\label{eqa4}
\end{equation}
where ${\bf w}_{i'\ne i}=(w^{[1]},...,w^{[i-1]},w^{[i+1]},...,w^{[K]})$ and $\epsilon$ is a parameter, which fullfills $\mathop {\lim }\limits_{T \to \infty } \frac{\epsilon }{T} = 0$.  
By inequalities (\ref{eqa1})-(\ref{eqa4}), we can write:
\begin{equation}
T\times{r_i} \le \frac{1}{2}I\left( {{w^{[i]}};{{\bf{y}}^{[i]}}^{\prime \prime }} \right) + \frac{1}{2}I\left( {{w^{[i]}};{{\bf{y}}^{[i]}}^{\prime \prime \prime }\left| {{{\bf{w}}_{i' \ne i}}} \right.} \right)+\epsilon.
\label{converse ineq3}
\end{equation}
\begin{equation}
T\times{r_j} \le \frac{1}{2}I\left( {{w^{[j]}};{{\bf{y}}^{[j]}}^{\prime \prime }} \right) + \frac{1}{2}I\left( {{w^{[j]}};{{\bf{y}}^{[j]}}^{\prime \prime \prime }\left| {{{\bf{w}}_{i' \ne j}}} \right.} \right)+\epsilon.
\end{equation}
 We also have the  following expansions:
\begin{equation}
I\left( {{w^{[i]}};{{\bf{y}}^{[i]}}^{\prime \prime }} \right) = \sum\limits_{m = 1}^T {I\left( {{w^{[i]}};{Y^{[i]}}^{\prime \prime }({ {{t_m}}})\left| {{Y^{[i]}}^{\prime \prime }({ {{t_1}}}),...,{Y^{[i]}}^{\prime \prime }({ {{t_{m - 1}}}})} \right.} \right)}  ,
\label{converse ineq6}
\end{equation}
\begin{equation}
I\left( {{w^{[i]}};{{\bf{y}}^{[i]}}^{\prime \prime \prime }\left| {{{\bf{w}}_{i' \ne i}}} \right.} \right) = \sum\limits_{m = 1}^T {I\left( {{w^{[i]}};{Y^{[i]}}^{\prime \prime \prime }({ {{{t'}_m}}})\left| {{Y^{[i]}}^{\prime \prime \prime }({ {{{t'}_1}}}),...,{Y^{[i]}}^{\prime \prime \prime }({ {{{t'}_{m - 1}}}}),{{\bf{w}}_{i' \ne i}}} \right.} \right)}  ,
\label{converse ineq7}
\end{equation}
\begin{equation}
I\left( {{w^{[j]}};{{\bf{y}}^{[j]}}^{\prime \prime }} \right) = \sum\limits_{m = 1}^T {I\left( {{w^{[j]}};{Y^{[j]}}^{\prime \prime }({ {{{t'}_m}}})\left| {{Y^{[j]}}^{\prime \prime }({ {{{t'}_1}}}),...,{Y^{[j]}}^{\prime \prime }({ {{{t'}_{m - 1}}}})} \right.} \right)}   ,
\label{eqa5}
\end{equation}
\begin{equation}
I\left( {{w^{[j]}};{{\bf{y}}^{[j]}}^{\prime \prime \prime }\left| {{{\bf{w}}_{i' \ne j}}} \right.} \right) = \sum\limits_{m = 1}^T {I\left( {{w^{[j]}};{Y^{[j]}}^{\prime \prime \prime }({ {{t_m}}})\left| {{Y^{[j]}}^{\prime \prime \prime }({ {{t_1}}}),...,{Y^{[j]}}^{\prime \prime \prime }({ {{t_{m - 1}}}}),{{\bf{w}}_{i' \ne j}}} \right.} \right)}  ,
\label{eqa6}
\end{equation}
where $t_1,...,t_T$ are ordered in such a manner that $\exists m^{\prime}\in \mathbb{N}$, so that in the $t_m$-th time slot $(m\in\{1,2,...,m^{\prime}\})$, we have $n_{j,i}=1$ and in the $t_m$-th time slot $(m\in\{m^{\prime}+1,...,T\})$, we have $n_{j,i}=0$.
Similarly, $t^{\prime}_1,...,t^{\prime}_T$ is ordered in such a manner that $\exists m^{\prime \prime}\in \mathbb{N}$, so that in the $t^{\prime}_m$-th time slot $(m\in\{1,2,...,m^{\prime \prime}\})$, we have $n_{i,j}=1$ and in the $t^{\prime}_m$-th time slot $(m\in\{m^{\prime \prime}+1,...,T\})$, we have $n_{i,j}=0$.
 So, based on (\ref{converse ineq3})-(\ref{eqa6}), we have:
\begin{equation*}
T\times{r_i} \le \frac{1}{2}\sum\limits_{m = 1}^T {I\left( {{w^{[i]}};{Y^{[i]}}^{\prime \prime }({ {{t_m}}})\left| {{Y^{[i]}}^{\prime \prime }({ {{t_1}}}),...,{Y^{[i]}}^{\prime \prime }({ {{t_{m - 1}}}})} \right.} \right)} 
\end{equation*}
\begin{equation}
 + \frac{1}{2}\sum\limits_{m = 1}^T {I\left( {{w^{[i]}};{Y^{[i]}}^{\prime \prime \prime }({ {{{t'}_m}}})\left| {{Y^{[i]}}^{\prime \prime \prime }({ {{{t'}_1}}}),...,{Y^{[i]}}^{\prime \prime \prime }({ {{{t'}_{m - 1}}}}),{{\bf{w}}_{i' \ne i}}} \right.} \right)} +\epsilon ,
\label{converse ineq8}
\end{equation}
\begin{equation*}
T\times{r_j} \le \frac{1}{2}\sum\limits_{m = 1}^T {I\left( {{w^{[j]}};{Y^{[j]}}^{\prime \prime }({ {{{t'}_m}}})\left| {{Y^{[j]}}^{\prime \prime }({ {{{t'}_1}}}),...,{Y^{[j]}}^{\prime \prime }({ {{{t'}_{m - 1}}}})} \right.} \right)} 
\end{equation*}
\begin{equation}
 + \frac{1}{2}\sum\limits_{m = 1}^T {I\left( {{w^{[j]}};{Y^{[j]}}^{\prime \prime \prime }({ {{t_m}}})\left| {{Y^{[j]}}^{\prime \prime \prime }({ {{t_1}}}),...,{Y^{[j]}}^{\prime \prime \prime }({ {{t_{m - 1}}}}),{{\bf{w}}_{i' \ne j}}} \right.} \right)} +\epsilon .
 \label{converse ineq8ddd}
\end{equation}

Now, assume that the network matrix is ${\bf N}$ in the $t$-th time slot, then for each $i,j\in \{1,...,K\},i\ne j$, there are three possible cases:

{\bf Case 1:} If $m\in\{1,...,\min\{m^{\prime},m^{\prime \prime}\}\}$, then for the $t_m$-th and $t^{\prime}_m$-th time slots, we have $n_{i,j}{( {t^{\prime}_m})}=n_{j,i}( {t_m})=1$. Thus, in these time slots, the cross coefficient from the $i$-th transmitter to the  $j$-th receiver and the cross coefficient from the $j$-th transmitter to the $i$-th receiver will be nonzero, so we have:
\begin{equation*}
\frac{1}{2}{I\left( {{w^{[i]}};{Y^{[i]}}^{\prime \prime }({ {t_m}})\left| {{Y^{[i]}}^{\prime \prime }({ {t_1}}),...,{Y^{[i]}}^{\prime \prime }({ {{t^{\prime}_{m-1}}}})} \right.} \right)}
\end{equation*}
\begin{equation*}
 + \frac{1}{2}{I\left( {{w^{[i]}};{Y^{[i]}}^{\prime \prime \prime }({ {{{t^{\prime}_{m}}}}})\left| {{Y^{[i]}}^{\prime \prime \prime }({ {{{t^{\prime}_{1}}}}}),...,{Y^{[i]}}^{\prime \prime \prime }({ {{{t^{\prime}_{m-1}}}}}),{{\bf{w}}_{i' \ne i}}} \right.} \right)}
\end{equation*}
\begin{equation*}
+\frac{1}{2}{I\left( {{w^{[j]}};{Y^{[j]}}^{\prime \prime }({ {{{t^{\prime}_{m}}}}})\left| {{Y^{[j]}}^{\prime \prime }({ {{{t^{\prime}_{1}}}}}),...,{Y^{[j]}}^{\prime \prime }({ {{{t^{\prime}_{m-1}}}}})} \right.} \right)}
\end{equation*}
\begin{equation*}
 + \frac{1}{2}{I\left( {{w^{[j]}};{Y^{[j]}}^{\prime \prime \prime }({ {{{t_{m}}}}})\left| {{Y^{[j]}}^{\prime \prime \prime }({ {{{t_{1}}}}}),...,{Y^{[j]}}^{\prime \prime \prime }({ {{{t_{m-1}}}}}),{{\bf{w}}_{i' \ne j}}} \right.} \right)}
\end{equation*}
\begin{equation*}
\le\frac{1}{2}{I\left( {{w^{[i]}};{Y^{[i]}}^{\prime \prime }({ {{{t_{m}}}}})\left| {{Y^{[i]}}^{\prime \prime }({ {{{t_{1}}}}}),...,{Y^{[i]}}^{\prime \prime }({ {{{t_{m-1}}}}})} \right.} \right)}
\end{equation*}
\begin{equation*}
 + \frac{1}{2}{I\left( {{w^{[i]}};{Y^{[j]}}^{ \prime \prime }({ {{{t^{\prime}_{m}}}}})\left| {{Y^{[j]}}^{ \prime \prime }({ {{{t^{\prime}_{1}}}}}),...,{Y^{[j]}}^{ \prime \prime }({ {{{t^{\prime}_{m-1}}}}}),{{\bf{w}}_{i' \ne i}}} \right.} \right)}
\end{equation*}
\begin{equation*}
+\frac{1}{2}{I\left( {{w^{[j]}};{Y^{[j]}}^{\prime \prime }({ {{{t^{\prime}_{m}}}}})\left| {{Y^{[j]}}^{\prime \prime }({ {{{t^{\prime}_{1}}}}}),...,{Y^{[j]}}^{\prime \prime }({ {{{t^{\prime}_{m-1}}}}})} \right.} \right)}
\end{equation*}
\begin{equation}
 + \frac{1}{2}{I\left( {{w^{[j]}};{Y^{[i]}}^{ \prime \prime }({ {{{t_{m}}}}})\left| {{Y^{[i]}}^{ \prime \prime }({ {{{t_{1}}}}}),...,{Y^{[i]}}^{ \prime \prime }({ {{{t_{m-1}}}}}),{{\bf{w}}_{i' \ne j}}} \right.} \right)}
 \label{converse ineq9}
\end{equation}
\begin{equation*}
\le\frac{1}{2}{I\left( {{{\bf w}_{i'\ne j}};{Y^{[i]}}^{\prime \prime }({ {{{t_{m}}}}})\left| {{Y^{[i]}}^{\prime \prime }({ {{{t_{1}}}}}),...,{Y^{[i]}}^{\prime \prime }({ {{{t_{m-1}}}}})} \right.} \right)}
\end{equation*}
\begin{equation*}
 + \frac{1}{2}{I\left( {{w^{[i]}};{Y^{[j]}}^{ \prime \prime }({ {{{t^{\prime}_{m}}}}})\left| {{Y^{[j]}}^{ \prime \prime }({ {{{t^{\prime}_{1}}}}}),...,{Y^{[j]}}^{ \prime \prime }({ {{{t^{\prime}_{m-1}}}}}),{{\bf{w}}_{i' \ne i}}} \right.} \right)}
\end{equation*}
\begin{equation*}
 + \frac{1}{2}{I\left( {{{\bf w}_{i'\ne i}};{Y^{[j]}}^{ \prime \prime }({ {{{t^{\prime}_{m}}}}})\left| {{Y^{[j]}}^{ \prime \prime }({ {{{t^{\prime}_{1}}}}}),...,{Y^{[j]}}^{ \prime \prime }({ {{{t^{\prime}_{m-1}}}}})} \right.} \right)}
\end{equation*}
\begin{equation*}
 + \frac{1}{2}{I\left( {{w^{[j]}};{Y^{[i]}}^{ \prime \prime }({ {{{t_{m}}}}})\left| {{Y^{[i]}}^{ \prime \prime }({ {{{t_{1}}}}}),...,{Y^{[i]}}^{ \prime \prime }({ {{{t_{m-1}}}}}),{{\bf{w}}_{i' \ne j}}} \right.} \right)}
\end{equation*}
\begin{equation*}
 = \frac{1}{2}I\left( {{w^{[1]}},...,{w^{[K]}};{Y^{[i]}}^{\prime \prime }({ {{{t_{m}}}}})\left| {{Y^{[i]}}^{\prime \prime }({ {{{t_{1}}}}}),...,{Y^{[i]}}{{^{\prime \prime }}}({ {{{t_{m-1}}}}})} \right.} \right)
\end{equation*}
\begin{equation*}
 + \frac{1}{2}I\left( {{w^{[1]}},...,{w^{[K]}};{Y^{[j]}}^{\prime \prime }({ {{{t^{\prime}_{m}}}}})\left| {{Y^{[j]}}^{\prime \prime }({ {{{t^{\prime}_{1}}}}}),...,{Y^{[j]}}^{\prime \prime }({ {{{t^{\prime}_{m-1}}}}})} \right.} \right) \le \log (\rho ) + o(\log (\rho ))
\end{equation*}
\begin{equation}
 = \left(2 - \frac{{{n_{j,i}( {t_m})} + {n_{i,j}}( {t^{\prime}_m})}}{2}\right)\log (\rho ) + o(\log (\rho )),
\label{converse ineq10}
\end{equation}
where ${n_{j,i}( {t_m})}$ is the element in the $j$-th row and the $i$-th column of the network matrix in the $t_m$-th time slot.
The inequality in (\ref{converse ineq9}) follows from:
\begin{equation*}
{I\left( {{w^{[i]}};{Y^{[i]}}^{\prime \prime \prime }({ {{{t^{\prime}_{m}}}}})\left| {{Y^{[i]}}^{\prime \prime \prime }({ {{{t^{\prime}_{1}}}}}),...,{Y^{[i]}}^{\prime \prime \prime }({ {{{t^{\prime}_{m-1}}}}}),{{\bf{w}}_{i' \ne i}}} \right.} \right)}
\end{equation*}
\begin{equation*}
=H\left( {{w^{[i]}}\left| {{Y^{[i]}}^{\prime \prime \prime }({ {{{t^{\prime}_{1}}}}}),...,{Y^{[i]}}^{\prime \prime \prime }({ {{{t^{\prime}_{m-1}}}}}),{{\bf{w}}_{i' \ne i}}} \right.} \right) - H\left( {{w^{[i]}}\left| {{Y^{[i]}}^{\prime \prime \prime }({ {{{t^{\prime}_{1}}}}}),...,{Y^{[i]}}^{\prime \prime \prime }({ {{{t^{\prime}_{m}}}}}),{{\bf{w}}_{i' \ne i}}} \right.} \right)
\end{equation*}
\begin{equation*}
 \scalebox{.92}[1]{$= H\left({w^{[i]}}\left| {{Y^{[i]}}^{\prime \prime \prime }({ {{{t^{\prime}_{1}}}}}),...,{Y^{[i]}}^{\prime \prime \prime }({ {{{t^{\prime}_{m-1}}}}}),{{\tilde {\tilde H}}^{[ji]}}({ {{t^{\prime}_{1}}}}){X^{[i]}}({ {{{t^{\prime}_{1}}}}}) + \bar Z({ {{t^{\prime}_{1}}}}),...,{{\tilde{ \tilde H}}^{[ji]}}({ {{{t^{\prime}_{m-1}}}}}){X^{[i]}}({ {{{t^{\prime}_{m-1}}}}}) + \bar Z({ {{{t^{\prime}_{m-1}}}}}),{{\bf{w}}_{i' \ne i}}} \right.\right)$}
\end{equation*}
\begin{equation}
 \scalebox{.92}[1]{$- H\left({w^{[i]}}\left| {{Y^{[i]}}^{\prime \prime \prime }({ {{{t^{\prime}_{1}}}}}),...,{Y^{[i]}}^{\prime \prime \prime }({ {{{t^{\prime}_{m}}}}}),{{\tilde {\tilde H}}^{[ji]}}({ {{t^{\prime}_{1}}}}){X^{[i]}}({ {{{t^{\prime}_{1}}}}}) + \bar Z({ {{t^{\prime}_{1}}}}),...,{{\tilde{ \tilde H}}^{[ji]}}({ {{{t^{\prime}_{m}}}}}){X^{[i]}}({ {{{t^{\prime}_{m}}}}}) + \bar Z({ {{{t^{\prime}_{m}}}}}),{{\bf{w}}_{i' \ne i}}} \right.\right)$}
\label{converse ineq11}
\end{equation}
\begin{equation*}
 \scalebox{.92}[1]{$= H\left({w^{[i]}}\left| {{Y^{[j]}}^{ \prime \prime }({ {t^{\prime}_{1}}}),...,{Y^{[j]}}^{ \prime \prime }({ {{t^{\prime}_{m-1}}}}),{{\tilde {\tilde H}}^{[ji]}}({ {t^{\prime}_{1}}}){X^{[i]}}({ {t^{\prime}_{1}}}) + \bar Z({ {t^{\prime}_{1}}}),...,{{\tilde{ \tilde H}}^{[ji]}}({ {{t^{\prime}_{m-1}}}}){X^{[i]}}({ {{t^{\prime}_{m-1}}}}) + \bar Z({ {{t^{\prime}_{m-1}}}}),{{\bf{w}}_{i' \ne i}}} \right.\right)$}
\end{equation*}
\begin{equation}
 \scalebox{.92}[1]{$- H\left({w^{[i]}}\left| {{Y^{[j]}}^{ \prime \prime }({ {t^{\prime}_{1}}}),...,{Y^{[j]}}^{ \prime \prime }({ {{t^{\prime}_{m}} }}),{{\tilde {\tilde H}}^{[ji]}}({ {t^{\prime}_{1}}}){X^{[i]}}({ {t^{\prime}_{1}}}) + \bar Z({ {t^{\prime}_{1}}}),...,{{\tilde{ \tilde H}}^{[ji]}}({ {{t^{\prime}_{m}}}}){X^{[i]}}({ {{t^{\prime}_{m}} }}) + \bar Z({ {{t^{\prime}_{m}} }}),{{\bf{w}}_{i' \ne i}}} \right.\right)$}
\label{converse ineq11-2}
\end{equation}
\begin{equation*}
 \scalebox{.92}[1]{$= H\left({w^{[i]}}\left| {{Y^{[j]}}^{ \prime \prime }({ {t^{\prime}_{1}}}),...,{Y^{[j]}}^{ \prime \prime }({ {{t^{\prime}_{m-1}}}}),{{\tilde {\tilde H}}^{[ji]}}({ {t^{\prime}_{1}}}){X^{[i]}}({ {t^{\prime}_{1}}}) + \bar Z({ {t^{\prime}_{1}}}),...,{{\tilde{ \tilde H}}^{[ji]}}({ {{t^{\prime}_{m-1}}}}){X^{[i]}}({ {{t^{\prime}_{m-1}}}}) + \bar Z({ {{t^{\prime}_{m-1}}}}),{{\bf{w}}_{i' \ne i}}} \right.\right)$}
\end{equation*}
\begin{equation}
 \scalebox{.92}[1]{$- H\left({w^{[i]}}\left| {{Y^{[j]}}^{ \prime \prime }({ {t^{\prime}_{1}}}),...,{Y^{[j]}}^{ \prime \prime }({ {{t^{\prime}_{m}} }}),{{\tilde {\tilde H}}^{[ji]}}({ {t^{\prime}_{1}}}){X^{[i]}}({ {t^{\prime}_{1}}}) + \bar Z({ {t^{\prime}_{1}}}),...,{{\tilde{ \tilde H}}^{[ji]}}({ {{t^{\prime}_{m-1}}}}){X^{[i]}}({ {{t^{\prime}_{m-1}} }}) + \bar Z({ {{t^{\prime}_{m-1}} }}),{{\bf{w}}_{i' \ne i}}} \right.\right)$}
\label{converse ineq12}
\end{equation}
\begin{equation*}
 \scalebox{.85}[1]{$=I\left( {{w^{[i]}};{Y^{[j]}}^{\prime \prime }({ {t_m^\prime }})\left| {{Y^{[j]}}^{\prime \prime }({ {t_1^\prime }}),...,{Y^{[j]}}^{\prime \prime }({ {t_{m - 1}^\prime }}),{{\tilde{ \tilde H}}^{[ji]}}({ {t_1^\prime }}){X^{[i]}}({ {t_1^\prime }}) + \bar Z({ {t_1^\prime }}),...,{{\tilde {\tilde H}}^{[ji]}}({ {t_{m - 1}^\prime }}){X^{[i]}}({ {t_{m - 1}^\prime }}) + \bar Z({ {t_{m - 1}^\prime }}),{{\bf{w}}_{i' \ne i}}} \right.} \right)$}
\end{equation*}
\begin{equation}
 \scalebox{.85}[1]{$ \le I\left( {{w^{[i]}},{{\tilde {\tilde H}}^{[ji]}}({ {t_1^\prime }}){X^{[i]}}({ {t_1^\prime }}) + \bar Z({ {t_1^\prime }}),...,{{\tilde {\tilde H}}^{[ji]}}({ {t_{m - 1}^\prime }}){X^{[i]}}({ {t_{m - 1}^\prime }}) + \bar Z({ {t_{m - 1}^\prime }});{Y^{[j]}}^{\prime \prime }({ {t_m^\prime }})\left| {{Y^{[j]}}^{\prime \prime }({ {t_1^\prime }}),...,{Y^{[j]}}^{\prime \prime }({ {t_{m - 1}^\prime }}),{{\bf{w}}_{i' \ne i}}} \right.} \right)$}
\label{converse ineq12-2}
\end{equation}
\begin{equation*}
 = I\left( {{w^{[i]}};{Y^{[j]}}^{\prime \prime }({ {t_m^\prime }})\left| {{Y^{[j]}}^{\prime \prime }({ {t_1^\prime }}),...,{Y^{[j]}}^{\prime \prime }({ {t_{m - 1}^\prime }}),{{\bf{w}}_{i' \ne i}}} \right.} \right)
\end{equation*}
\begin{equation*}
 \scalebox{.8}[1]{$ + I\left( {{{\tilde {\tilde H}}^{[ji]}}({ {t_1^\prime }}){X^{[i]}}({ {t_1^\prime }}) + \bar Z({ {t_1^\prime }}),...,{{\tilde {\tilde H}}^{[ji]}}({ {t_{m - 1}^\prime }}){X^{[i]}}({ {t_{m - 1}^\prime }}) + \bar Z({ {t_{m - 1}^\prime }});{Y^{[j]}}^{\prime \prime }({ {t_m^\prime }})\left| {{Y^{[j]}}^{\prime \prime }({ {t_1^\prime }}),...,{Y^{[j]}}^{\prime \prime }({ {t_{m - 1}^\prime }}),} \right.{w^{[1]}},...,{w^{[K]}}} \right)$}
\end{equation*}
\begin{equation}
 = I\left( {{w^{[i]}};{Y^{[j]}}^{\prime \prime }({ {t_m^\prime }})\left| {{Y^{[j]}}^{\prime \prime }({ {t_1^\prime }}),...,{Y^{[j]}}^{\prime \prime }({ {t_{m - 1}^\prime }}),{{\bf{w}}_{i' \ne i}}} \right.} \right).
\label{converse ineq13}
\end{equation}
Since conditioned on ${{Y^{[i]}}^{\prime \prime \prime }({ {t_1^\prime }}),...,{Y^{[i]}}^{\prime \prime \prime }({ {t_{m }^\prime }})}$ and ${{\bf{w}}_{i' \ne i}}$, the terms ${{\tilde{\tilde H}}^{[ji]}}({ {t_1^\prime }}){X^{[i]}}({ {t_1^\prime }}) + \bar Z({ {t_1^\prime }})$ $,...,{{\tilde {\tilde H}}^{[ji]}}({ {t_{m}^\prime }})$ ${X^{[i]}}({ {t_{m }^\prime }}) + \bar Z({ {t_{m }^\prime }})$ become deterministic,  we can add them to the condition of the entropy in (\ref{converse ineq11}) without changing its value, i.e., $H(A|B,C)=H(A|B)$, if conditioned on $B$,  random variable $C$ becomes deterministic. 
Conditioned on ${{\tilde{\tilde H}}^{[ji]}}({ {t_1^\prime }}){X^{[i]}}({ {t_1^\prime }}) + \bar Z({ {t_1^\prime }}),...,{{\tilde {\tilde H}}^{[ji]}}({ {t_{m }^\prime }}){X^{[i]}}({ {t_{m }^\prime }}) + \bar Z({ {t_{m }^\prime }})$ and ${{\bf{w}}_{i' \ne i}}$, both ${{Y^{[i]}}^{\prime \prime \prime }({ {t_1^\prime }}),...,{Y^{[i]}}^{\prime \prime \prime }({ {t_{m }^\prime }})}$ and ${{Y^{[j]}}^{ \prime \prime }({ {t_1^\prime }}),...,{Y^{[j]}}^{ \prime \prime }({ {t_{m}^\prime }})}$ become deterministic quantities,  hence, we can remove the latter and add the former to the condition of the entropy in (\ref{converse ineq11-2}). 
Inequality (\ref{converse ineq12}) follows from the fact  that we have assumed that the cross coefficient from the  $i$-th transmitter to the  $j$-th receiver is nonzero ($n_{i,j}( {t^{\prime}_m})=1$), so conditioned on ${Y^{[j]}}^{ \prime \prime }({ {t_{m}^\prime }})$ and ${{\bf{w}}_{i' \ne i}}$,  the term ${{\tilde {\tilde H}}^{[ji]}}({ {t_{m }^\prime }}){X^{[i]}}({ {t_{m }^\prime }}) + \bar Z({ {t_{m }^\prime }})$ becomes deterministic and we can omit it from the condition. 
Inequality (\ref{converse ineq12-2}) follows from the inequality $I(A;B|C)\le I(A,C;B)$.
 Eq. (\ref{converse ineq13}) follows from the fact that conditioned on ${ {{Y^{[j]}}^{\prime \prime }({ {t_1^\prime }}),...,{Y^{[j]}}^{\prime \prime }({ {t_{m - 1}^\prime }}),} {w^{[1]}},...,{w^{[K]}}}$, random variables ${{\tilde {\tilde H}}^{[ji]}}({ {t_1^\prime }}){X^{[i]}}({ {t_1^\prime }}) + \bar Z({ {t_1^\prime }}),...,{{\tilde {\tilde H}}^{[ji]}}({ {t_{m - 1}^\prime }})$ ${X^{[i]}}({ {t_{m - 1}^\prime }}) + \bar Z({ {t_{m - 1}^\prime }})$  become deterministic.

Similar considerations hold for ${I\left( {{w^{[j]}};{Y^{[j]}}^{\prime \prime \prime }({ {{{t_{m}}}}})\left| {{Y^{[j]}}^{\prime\prime  \prime }({ {{{t_{1}}}}}),...,{Y^{[j]}}^{ \prime \prime\prime }({ {{{t_{m-1}}}}}),{{\bf{w}}_{i' \ne j}}} \right.} \right)}$.

{\bf Case 2:} For the case $m\in\{\min\{m',m''\}+1,\max\{m',m''\}\}$, we distinguish two sub-cases: 

\textbf{Sub-case 1:} If $m^{\prime}> m^{\prime\prime}$, then, for $m\in\{m^{\prime \prime}+1,...,m^{\prime}\}$,  $n_{j,i}( {t_m})=1,n_{i,j}( {t^{\prime}_m})=0$ holds. Then, only the cross coefficient from the $i$-th transmitter to the  $j$-th receiver  will be nonzero, so we have:
\begin{equation*}
\frac{1}{2}{I\left( {{w^{[i]}};{Y^{[i]}}^{\prime \prime }({ {t_m}})\left| {{Y^{[i]}}^{\prime \prime }({ {t_1}}),...,{Y^{[i]}}^{\prime \prime }({ {{t^{\prime}_{m-1}}}})} \right.} \right)}
\end{equation*}
\begin{equation*}
 + \frac{1}{2}{I\left( {{w^{[i]}};{Y^{[i]}}^{\prime \prime \prime }({ {{{t^{\prime}_{m}}}}})\left| {{Y^{[i]}}^{\prime \prime \prime }({ {{{t^{\prime}_{1}}}}}),...,{Y^{[i]}}^{\prime \prime \prime }({ {{{t^{\prime}_{m-1}}}}}),{{\bf{w}}_{i' \ne i}}} \right.} \right)}
\end{equation*}
\begin{equation*}
+\frac{1}{2}{I\left( {{w^{[j]}};{Y^{[j]}}^{\prime \prime }({ {{{t^{\prime}_{m}}}}})\left| {{Y^{[j]}}^{\prime \prime }({ {{{t^{\prime}_{1}}}}}),...,{Y^{[j]}}^{\prime \prime }({ {{{t^{\prime}_{m-1}}}}})} \right.} \right)}
\end{equation*}
\begin{equation*}
 + \frac{1}{2}{I\left( {{w^{[j]}};{Y^{[j]}}^{\prime \prime \prime }({ {{{t_{m}}}}})\left| {{Y^{[j]}}^{\prime \prime \prime }({ {{{t_{1}}}}}),...,{Y^{[j]}}^{\prime \prime \prime }({ {{{t_{m-1}}}}}),{{\bf{w}}_{i' \ne j}}} \right.} \right)}
\end{equation*}
\begin{equation*}
\le\frac{1}{2}{I\left( {{w^{[i]}};{Y^{[i]}}^{\prime \prime }({ {{{t_{m}}}}})\left| {{Y^{[i]}}^{\prime \prime }({ {{{t_{1}}}}}),...,{Y^{[i]}}^{\prime \prime }({ {{{t_{m-1}}}}})} \right.} \right)}
\end{equation*}
\begin{equation*}
 + \frac{1}{2}{I\left( {{w^{[i]}};{Y^{[i]}}^{\prime \prime \prime }({ {{{t^{\prime}_{m}}}}})\left| {{Y^{[i]}}^{\prime \prime \prime }({ {{{t^{\prime}_{1}}}}}),...,{Y^{[i]}}^{\prime \prime \prime }({ {{{t^{\prime}_{m-1}}}}}),{{\bf{w}}_{i' \ne i}}} \right.} \right)}
\end{equation*}
\begin{equation*}
+\frac{1}{2}{I\left( {{w^{[j]}};{Y^{[j]}}^{\prime \prime }({ {{{t^{\prime}_{m}}}}})\left| {{Y^{[j]}}^{\prime \prime }({ {{{t^{\prime}_{1}}}}}),...,{Y^{[j]}}^{\prime \prime }({ {{{t^{\prime}_{m-1}}}}})} \right.} \right)}
\end{equation*}
\begin{equation}
 + \frac{1}{2}{I\left( {{w^{[j]}};{Y^{[i]}}^{ \prime \prime }({ {{{t_{m}}}}})\left| {{Y^{[i]}}^{ \prime \prime }({ {{{t_{1}}}}}),...,{Y^{[i]}}^{ \prime \prime }({ {{{t_{m-1}}}}}),{{\bf{w}}_{i' \ne j}}} \right.} \right)}
 \label{converse ineq15}
\end{equation}
\begin{equation*}
\le\frac{1}{2}{I\left( {{{\bf w}_{i'\ne j}};{Y^{[i]}}^{\prime \prime }({ {{{t_{m}}}}})\left| {{Y^{[i]}}^{\prime \prime }({ {{{t_{1}}}}}),...,{Y^{[i]}}^{\prime \prime }({ {{{t_{m-1}}}}})} \right.} \right)}
\end{equation*}
\begin{equation*}
 + \frac{1}{2}{I\left( {{w^{[i]}};{Y^{[i]}}^{\prime \prime \prime }({ {{{t^{\prime}_{m}}}}})\left| {{Y^{[i]}}^{\prime \prime \prime }({ {{{t^{\prime}_{1}}}}}),...,{Y^{[i]}}^{\prime \prime \prime }({ {{{t^{\prime}_{m-1}}}}}),{{\bf{w}}_{i' \ne i}}} \right.} \right)}
\end{equation*}
\begin{equation*}
+\frac{1}{2}{I\left( {{w^{[j]}};{Y^{[j]}}^{\prime \prime }({ {{{t^{\prime}_{m}}}}})\left| {{Y^{[j]}}^{\prime \prime }({ {{{t^{\prime}_{1}}}}}),...,{Y^{[j]}}^{\prime \prime }({ {{{t^{\prime}_{m-1}}}}})} \right.} \right)}
\end{equation*}
\begin{equation*}
 + \frac{1}{2}{I\left( {{w^{[j]}};{Y^{[i]}}^{ \prime \prime }({ {{{t_{m}}}}})\left| {{Y^{[i]}}^{ \prime \prime }({ {{{t_{1}}}}}),...,{Y^{[i]}}^{ \prime \prime }({ {{{t_{m-1}}}}})},{{\bf{w}}_{i' \ne j}} \right.} \right)}
\end{equation*}
\begin{equation*}
 = \frac{1}{2}I\left( {{w^{[1]}},...,{w^{[K]}};{Y^{[i]}}^{\prime \prime }({ {{{t_{m}}}}})\left| {{Y^{[i]}}^{\prime \prime }({ {{{t_{1}}}}}),...,{Y^{[i]}}{{^{\prime \prime }}}({ {{{t_{m-1}}}}})} \right.} \right)
\end{equation*}
\begin{equation*}
 + \frac{1}{2}{I\left( {{w^{[i]}};{Y^{[i]}}^{\prime \prime \prime }({ {{{t^{\prime}_{m}}}}})\left| {{Y^{[i]}}^{\prime \prime \prime }({ {{{t^{\prime}_{1}}}}}),...,{Y^{[i]}}^{\prime \prime \prime }({ {{{t^{\prime}_{m-1}}}}}),{{\bf{w}}_{i' \ne i}}} \right.} \right)}
\end{equation*}
\begin{equation*}
+\frac{1}{2}{I\left( {{w^{[j]}};{Y^{[j]}}^{\prime \prime }({ {{{t^{\prime}_{m}}}}})\left| {{Y^{[j]}}^{\prime \prime }({ {{{t^{\prime}_{1}}}}}),...,{Y^{[j]}}^{\prime \prime }({ {{{t^{\prime}_{m-1}}}}})} \right.} \right)}
\end{equation*}
\begin{equation}
 \le \frac{3}{2} \log (\rho ) + o(\log (\rho ))= (2 - \frac{{{n_{j,i}( {t_m})} + {n_{i,j}}( {t^{\prime}_m})}}{2})\log (\rho ) + o(\log (\rho )),
 \label{boundbb}
\end{equation}
where inequality (\ref{converse ineq15}) follows from the same arguments as those provided for (\ref{converse ineq11})-(\ref{converse ineq13}). 

\textbf{Sub-case 2:}
If we have $m^{\prime}< m^{\prime\prime}$, then, for $m\in\{m^{\prime }+1,...,m^{\prime\prime}\}$, we have $n_{j,i}( {t_m})=0,n_{i,j}( {t^{\prime}_m})=1$, and by a similar argument as the one presented for Sub-case 1, we obtain the bound in (\ref{boundbb}).

{\bf Case 3:} For $m\in\{\max\{m^{\prime},m^{\prime\prime}\}+1,...,T\}$, we have $n_{j,i}( {t_m})=0,n_{i,j}( {t^{\prime}_m})=0$. Thus, there is no cross link and we have:
\begin{equation*}
\frac{1}{2}{I\left( {{w^{[i]}};{Y^{[i]}}^{\prime \prime }({ {t_m}})\left| {{Y^{[i]}}^{\prime \prime }({ {t_1}}),...,{Y^{[i]}}^{\prime \prime }({ {{t^{\prime}_{m-1}}}})} \right.} \right)}
\end{equation*}
\begin{equation*}
 + \frac{1}{2}{I\left( {{w^{[i]}};{Y^{[i]}}^{\prime \prime \prime }({ {{{t^{\prime}_{m}}}}})\left| {{Y^{[i]}}^{\prime \prime \prime }({ {{{t^{\prime}_{1}}}}}),...,{Y^{[i]}}^{\prime \prime \prime }({ {{{t^{\prime}_{m-1}}}}}),{{\bf{w}}_{i' \ne i}}} \right.} \right)}
\end{equation*}
\begin{equation*}
+\frac{1}{2}{I\left( {{w^{[j]}};{Y^{[j]}}^{\prime \prime }({ {{{t^{\prime}_{m}}}}})\left| {{Y^{[j]}}^{\prime \prime }({ {{{t^{\prime}_{1}}}}}),...,{Y^{[j]}}^{\prime \prime }({ {{{t^{\prime}_{m-1}}}}})} \right.} \right)}
\end{equation*}
\begin{equation*}
 + \frac{1}{2}{I\left( {{w^{[j]}};{Y^{[j]}}^{\prime \prime \prime }({ {{{t_{m}}}}})\left| {{Y^{[j]}}^{\prime \prime \prime }({ {{{t_{1}}}}}),...,{Y^{[j]}}^{\prime \prime \prime }({ {{{t_{m-1}}}}}),{{\bf{w}}_{i' \ne j}}} \right.} \right)}
\end{equation*}
\begin{equation}
 \le 2\log (\rho ) + o(\log (\rho ))= (2 - \frac{{{n_{j,i}( {t_m})} + {n_{i,j}}( {t^{\prime}_m})}}{2})\log (\rho ) + o(\log (\rho )).
 \label{boundbb2}
\end{equation}

As we can see from (\ref{converse ineq10}), (\ref{boundbb}), and (\ref{boundbb2}), all three cases yield the same upper bound expression. Combining these results with (\ref{converse ineq8}) and (\ref{converse ineq8ddd}) yields the following upper bound for $r_i+r_j$:
\begin{equation*}
{r_i} + {r_j} \le \frac{1}{T}\sum\limits_{m = 1}^T {(2 - \frac{{{n_{j,i}}({ {{t_m}}}) + {n_{i,j}}({ {t_m^\prime }})}}{2})\log (\rho )}  + o(\log (\rho ))
\end{equation*}
\begin{equation}
 = \frac{1}{T}\sum\limits_{t = 1}^T {(2 - \frac{{{n_{j,i}}({ t}) + {n_{i,j}}({ t})}}{2})\log (\rho )}  + o(\log (\rho )).
\label{expansion123}
\end{equation}
 If we set $a_m$ in (\ref{equation_a_m}) equal to the fraction of time slots in which the  network matrix is ${\bf N}_m$, then we can see that the DoF region must be a subset of the set in (\ref{DoF region active outer2}), and the proof is complete.

\section{}
\label{appendix4}
By (\ref{found cancel}) and Theorems \ref{thm DoF reg const active IRS} and \ref{active IRS inner DoF reg}, we can design the IRS element coefficients such that the network matrix satisfies the following conditions almost surely:
\begin{equation*}
\begin{cases}
{{n_{i,j}} = 0,i \in \{ 1,...,W\} ,j \in \{ 1,...,K\} ,i \ne j},\\
{{n_{i,j}} = 0,i \in \{ W + 1,...,K\} ,j \in \{ 1,...,W\} },\\
{{n_{i,j}} = 1,i,j \in \{ W + 1,...,K\} },\\
{{n_{i,i}} = 1,i \in \{ 1,...,W\} }.
\end{cases}
\end{equation*}
Then, by Theorems \ref{thm DoF reg const active IRS} and \ref{active IRS inner DoF reg},  the following DoF vector is achievable:
\begin{equation*}
\begin{cases}
{{d_i} = 1,i \in \{ 1,...,W\} },\\
{{d_i} = \frac{1}{2},i \in \{ W + 1,...,K\} },
\end{cases}
\end{equation*}
which implies that the sum DoF of $\frac{K+W}{2}$ is achievable.

\section{}
\label{appendix5}
Based on Theorem \ref{active IRS outer DoF reg}, for every vector ${\bf d}\in {\cal D}_{active-out}$, there exists a vector ${\bf a}\in {\cal A}$, where $\cal A$  is given in (\ref{A_eq}), for which  we have:
\begin{equation*}
\sum\limits_{n = 1}^K {{d_n}}  \le \frac{1}{{2(K - 1)}}\sum\limits_{i \ne j} {\sum\limits_{m = 1}^{\left| {{{\cal N}_Q}} \right|} {{a_m}{{\tilde d}^{[ij]}}_{{{\bf{N}}_m}}} }  = \frac{1}{{2(K - 1)}}\sum\limits_{m = 1}^{\left| {{{\cal N}_Q}} \right|} {\sum\limits_{i \ne j} {{a_m}{{\tilde d}^{[ij]}}_{{{\bf{N}}_m}}} } 
\end{equation*}
\begin{equation*}
 \le \frac{1}{{2(K - 1)}}\mathop {\max }\limits_{m \in \{ 1,...,\left| {{{\cal N}_Q}} \right|\} } \sum\limits_{i \ne j} {{{\tilde d}^{[ij]}}_{{{\bf{N}}_m}}}  = K - \mathop {\min }\limits_{{\bf{N}} \in {{\cal N}_Q}} \frac{1}{{2(K - 1)}}\sum\limits_{i \ne j} {\frac{{{n_{i,j}} + {n_{j,i}}}}{2}} 
\label{sum up N}
\end{equation*}
\begin{equation}
 = K - \mathop {\min }\limits_{{\bf N} \in {{\cal N}_Q}} \sum\limits_{i \ne j} {\frac{{{n_{i,j}}}}{{2(K - 1)}}}  = K - \frac{{K(K - 1) - Q}}{{2(K - 1)}} = \frac{K}{2} + \frac{Q}{{2(K - 1)}}.
\end{equation}
 
\section{}
\label{appendix6}
Let $X$ be a random variable with possible events ${\cal E}_{{Q_i}},i\in\{1,...,2^{\left| {{{\cal N}_Q}} \right|-1}\}$, and let $X^T$ be $T$ i.i.d. realizations of $X$. We define $\pi ({{\cal E}_{{Q_i}}}\left| {{X^T}} \right.)$ as the fraction of time slots, in which event ${\cal E}_{Q_i}$  occurs in $T$ i.i.d. realizations of $X$. By the law of large numbers,  for each event ${\cal E}_{Q_i}$ and for each $\delta>0$, there exists a sequence $\epsilon (\delta,T)$ such that:
\begin{equation}
\Pr \left\{ {\left| {\pi ({{\cal E}_{{Q_i}}}\left| {{X^T}} \right.) - \Pr \{ {{\cal E}_{{Q_i}}}\} } \right| > \delta } \right\} < \epsilon (\delta,T),
\label{L.N.N}
\end{equation}
where $\forall \delta>0\to\mathop {\lim }\limits_{T \to \infty } \epsilon (\delta,T) = 0$. 
We note that inequality (\ref{expansion123}) derived in the proof of Theorem \ref{active IRS outer  DoF reg} is valid for both active and passive IRSs. The only difference is that for passive IRSs the elements of the network matrix, i.e., $n_{i,j}$, in (\ref{expansion123}) are constrained since only certain network matrices are realizable for each realization of channel coefficients. 
Recall that ${\cal E}_{Q_i}$ characterizes the set of channel realizations, where only the network matrices in ${\cal N}_{Q_i}$ are realizable. Therefore, the outer bound of the DoF region can be represented as (\ref{DoF region passive outer2}) because in at most $\Pr\{{\cal E}_{Q_i}\}+\delta$ time slots,  ${\cal E}_{Q_i}$ occurs for each $i\in\{1,...,2^{\left| {{{\cal N}_Q}} \right|-1}\}$  with a probability higher than $1-\epsilon$ for a sufficiently large $T$ (by (\ref{L.N.N})). This completes the proofs.

\section{}
\label{appendix7}
The proof of this theorem is similar to that of Theorem \ref{passive outer DoF reg}.
Let $X$ be a random variable with possible events ${\cal F}_{{Q_i}},i\in\{1,...,2^{\left| {{{\cal N}_Q}} \right|-1}\}$, and let $X^T$ be $T$ i.i.d. realizations of $X$.  By the law of large numbers, for each event ${\cal F}_{{Q_i}}$ and for each $\delta>0$, there exists a sequence $\epsilon (\delta,T)$ such that:
\begin{equation}
\Pr \left\{ {\left| {\pi ({{\cal F}_{{Q_i}}}\left| {{X^T}} \right.) - \Pr \{ {{\cal F}_{{Q_i}}}\} } \right| > \delta } \right\} < \epsilon (\delta,T),
\label{L.N.N2}
\end{equation}
where $\mathop {\lim }\limits_{T \to \infty } \epsilon (\delta,T) = 0,\forall \delta>0$. 

\begin{lemma}
In time slots, in which ${\cal F}_{{Q_i}}$  occurs, DoF region (\ref{DoF region passive inner1}) is achievable.
\label{lemma4}
\end{lemma}

\begin{IEEEproof}
The proof of this lemma follows the same steps as the proof of Theorems \ref{thm DoF reg const active IRS} and \ref{active IRS inner DoF reg}, exept the fact that we have to use the pseudo inverse instead of the regular inverse in (\ref{Interference cancelation}). However, this does not change the validity of the argument of the proof.
\end{IEEEproof}

By Lemma \ref{lemma4} and inequality (\ref{L.N.N2}), for a sufficiently large $T$, with a probability higher than $1-\epsilon$ each event ${\cal F}_{{Q_i}}$ occurs in at least $T(\Pr\{{\cal F}_{{Q_i}}\}-\delta)$ time slots. Thus, region (\ref{DoF region passive inner2})  is achievable with a probability higher than $1-\epsilon$.  This completes the proof.

\section{}

\label{appendix10}
We prove that $\mathop {\lim }\limits_{Q \to \infty } \Pr \{ {{\cal F}_{{Q_i}}}\}  = 0$ holds for $\forall i\neq 1$. For each $i\ne 1$, there exists at least one network matrix ${\bf N}^*_i$ such that ${\bf{N}}_i^* \in {{\cal N}_Q},{\bf{N}}_i^* \notin {{\tilde {\cal N}}_{{Q_i}}}$.
For future reference, we rewrite (\ref{Interference cancelation passive IRS}) for ${\bf N}_{i}^{*}$ in matrix form ${\bold{H}}_{{\bf N}^*_i}{\bold{\tau}}_{{\bf N}^*_i}={\bold h}_{{\bf N}^*_i}$. 
Then, we have the following Lemma:

\begin{lemma}
If we assume $Q>K(K-1)$
and define ${\bf L}={{\bf{H}}_{{\bf N}^*_i}}{\bf{H}}_{{\bf N}^*_i}^H$, then we obtain:
\begin{equation}
\Pr \left\{ {\frac{1}{Q}\left| {{l_{n,n}} - \sum\limits_{u = 1}^Q {E\left\{ {{{\left| {H_{{\rm{TI}}}^{[u{i_n}]}(t)H_{{\rm{IR}}}^{[{j_n}u]}(t)} \right|}^2}} \right\}} } \right| > \delta } \right\} < {\varepsilon _1}(Q),
\label{HHH1}
\end{equation}
\begin{equation}
\Pr \left\{ {\frac{1}{Q}\left| {{l_{n,m}}} \right| > \delta } \right\} < {\varepsilon _2}(Q),n\ne m
\label{HHH2}
\end{equation}
where $\mathop {\lim }\limits_{Q \to \infty } {\varepsilon _1}(Q) = \mathop {\lim }\limits_{Q \to \infty } {\varepsilon _2}(Q) = 0$, and $\left[ {\begin{array}{*{20}{c}}
{H_{{\rm{TI}}}^{[1{i_n}]}(t)H_{{\rm{IR}}}^{[{j_n}1]}(t)}& \cdots &{H_{{\rm{TI}}}^{[Q{i_n}]}(t)H_{{\rm{IR}}}^{[{j_n}Q]}(t)}
\end{array}} \right]$ is the $n$-th row of ${\bold{H}}_{{\bf N}^*_i}$.
\label{lemma5}
\end{lemma}

\begin{IEEEproof}
We have:
\begin{equation*}
\Pr \left\{ {\frac{1}{Q}\left| {{l_{n,n}} - \sum\limits_{u = 1}^Q {E\left\{ {{{\left| {H_{{\rm{TI}}}^{[u{i_n}]}(t)H_{{\rm{IR}}}^{[{j_n}u]}(t)} \right|}^2}} \right\}} } \right| > \delta } \right\}
\end{equation*}
\begin{equation*}
 = \Pr \left\{ {\frac{1}{{{Q^2}}}{{\left( {{l_{n,n}} - \sum\limits_{u = 1}^Q {E\left\{ {{{\left| {H_{{\rm{TI}}}^{[u{i_n}]}(t)H_{{\rm{IR}}}^{[{j_n}u]}(t)} \right|}^2}} \right\}} } \right)}^2} > {\delta ^2}} \right\}
\end{equation*}
\begin{equation}
 \le \frac{{E\left\{ {{{\left( {{l_{n,n}} - \sum\limits_{u = 1}^Q {E\left\{ {{{\left| {H_{{\rm{TI}}}^{[u{i_n}]}(t)H_{{\rm{IR}}}^{[{j_n}u]}(t)} \right|}^2}} \right\}} } \right)}^2}} \right\}}}{{{Q^2}{\delta ^2}}}
\label{HHH3}
\end{equation}
\begin{equation}
 = \frac{{QE\left\{ {{{\left( {{{\left| {H_{{\rm{TI}}}^{[u{i_n}]}(t)H_{{\rm{IR}}}^{[{j_n}u]}(t)} \right|}^2} - E\left\{ {{{\left| {H_{{\rm{TI}}}^{[u{i_n}]}(t)H_{{\rm{IR}}}^{[{j_n}u]}(t)} \right|}^2}} \right\}} \right)}^2}} \right\}}}{{{Q^2}{\delta ^2}}} = {\varepsilon _1}(Q),
\label{HHH4}
\end{equation}

\begin{equation}
\Pr \left\{ {\frac{1}{Q}\left| {{l_{n,m}}} \right| > \delta } \right\} = \Pr \left\{ {\frac{1}{{{Q^2}}}{{\left| {{l_{n,m}}} \right|}^2} > {\delta ^2}} \right\} \le \frac{{E\left\{ {{{\left| {{l_{n,m}}} \right|}^2}} \right\}}}{{{Q^2}{\delta ^2}}}
\label{HHH5}
\end{equation}
\begin{equation}
 = \frac{{QE\left\{ {{{\left| {H_{{\rm{TI}}}^{[u{i_n}]}(t)H_{{\rm{IR}}}^{[{j_n}u]}(t)} \right|}^2}{{\left| {H_{{\rm{TI}}}^{[u{i_m}]}(t)H_{{\rm{IR}}}^{[{j_m}u]}(t)} \right|}^2}} \right\}}}{{{Q^2}{\delta ^2}}} = {\varepsilon _2}(Q),
\label{HHH6}
\end{equation}
where (\ref{HHH3}) and (\ref{HHH5}) follow from the Markov inequality and (\ref{HHH4}) and (\ref{HHH6}) follow from the independent and identical distributions  of the channel coefficients for all values of $u$, i.e., (\ref{prop1})-(\ref{prop4}). Note that due to (\ref{prop5}) and (\ref{prop6}), the expectations in (\ref{HHH4}) and (\ref{HHH6}) are bounded.
\end{IEEEproof}
By Lemma \ref{lemma5}, if we define events ${\cal G}_i$ and ${\cal G}_i^c $ as follows:
\begin{equation}
{{\cal G}_i} = \bigcap\limits_n {\left\{ {\left\{ {\frac{1}{Q}\left| {{l_{n,n}} - \sum\limits_{u = 1}^Q {E\left\{ {{{\left| {H_{{\text{TI}}}^{[u{i_n}]}(t)H_{{\text{IR}}}^{[{j_n}u]}(t)} \right|}^2}} \right\}} } \right| \leqslant \delta } \right\}\bigcap {\left\{ {\bigcap\limits_{m \ne n} {\left\{ {\frac{1}{Q}\left| {{l_{n,m}}} \right| \leqslant \delta } \right\}} } \right\}} } \right\}} 
\end{equation}
\begin{equation}
{\cal G}_i^c = \bigcup\limits_n {\left\{ {\left\{ {\frac{1}{Q}\left| {{l_{n,n}} - \sum\limits_{u = 1}^Q {E\left\{ {{{\left| {H_{{\text{TI}}}^{[u{i_n}]}(t)H_{{\text{IR}}}^{[{j_n}u]}(t)} \right|}^2}} \right\}} } \right| > \delta } \right\}\bigcup {\left\{ {\bigcup\limits_{m \ne n} {\left\{ {\frac{1}{Q}\left| {{l_{n,m}}} \right| > \delta } \right\}} } \right\}} } \right\}} ,
\end{equation}
then, by the union bound, we have:
\begin{equation}
\Pr \left\{ {{\cal G}_i^c} \right\} < {\varepsilon _3}(Q),
\label{G3i}
\end{equation}
where $\mathop {\lim }\limits_{Q \to \infty } {\varepsilon _3}(Q) = 0$. Note that if we fix $(i_n,j_n)$, then the expectation ${E\left\{ {{{\left| {H_{{\rm{TI}}}^{[u{i_n}]}(t)H_{{\rm{IR}}}^{[{j_n}u]}(t)} \right|}^2}} \right\}}$ is idependent of $u$. Now, we state the following useful lemma.
\begin{lemma}
Conditioned on event ${\cal G}_i$, if we consider
\begin{equation*}
{\left( {\frac{1}{Q}{{\bf{H}}_{{\bf{N}}_i^*}}{\bf{H}}_{{\bf{N}}_i^*}^H} \right)^{ - 1}}
\end{equation*}
\begin{equation}
\scalebox{.8}[1]{$= \frac{1}{{1 + {a_{0,0}}}}{\left[ {\begin{array}{*{20}{c}}
{{{\left( {E\left\{ {{{\left| {H_{{\rm{TI}}}^{[u{i_1}]}(t)H_{{\rm{IR}}}^{[{j_1}u]}(t)} \right|}^2}} \right\}} \right)}^{ - 1}} + {a_{1,1}}}&{{a_{1,2}}}& \cdots \\
{{a_{2,1}}}&{{{\left( {E\left\{ {{{\left| {H_{{\rm{TI}}}^{[u{i_2}]}(t)H_{{\rm{IR}}}^{[{j_2}u]}(t)} \right|}^2}} \right\}} \right)}^{ - 1}} + {a_{2,2}}}& \cdots \\
 \vdots & \vdots & \ddots 
\end{array}} \right]_{\left| {{M_{{\bf{N}}_i^*}}} \right| \times \left| {{M_{{\bf{N}}_i^*}}} \right|}} = \frac{1}{{1 + {a_{0,0}}}}({\bf E} + {\bf{A}})$},
\label{A}
\end{equation}
where
\begin{equation*}
{\bf E} = {\left[ {\begin{array}{*{20}{c}}
{{{\left( {E\left\{ {{{\left| {H_{{\rm{TI}}}^{[u{i_1}]}(t)H_{{\rm{IR}}}^{[{j_1}u]}(t)} \right|}^2}} \right\}} \right)}^{ - 1}}}&0& \cdots \\
0&{{{\left( {E\left\{ {{{\left| {H_{{\rm{TI}}}^{[u{i_2}]}(t)H_{{\rm{IR}}}^{[{j_2}u]}(t)} \right|}^2}} \right\}} \right)}^{ - 1}}}& \cdots \\
 \vdots & \vdots & \ddots 
\end{array}} \right]_{\left| {{M_{{\bf{N}}_i^*}}} \right| \times \left| {{M_{{\bf{N}}_i^*}}} \right|}}.
\end{equation*}
Then, we have $\left| {{a_{i,j}}} \right| \le \left| {{P^{[ji]}}(\delta )} \right|$, where $P^{[ji]}(\delta)$ are polynomials with zero constant term (i.e., $\mathop {\lim }\limits_{\delta  \to 0} {P^{[ji]}}(\delta ) = 0$), $\forall i,j$ (also for $i=j=0$). 

\label{lemma6}
\end{lemma}

\begin{IEEEproof}
From linear algebra, we know that the inverse of matrix $\bf D$ has the following form:
\begin{equation}
{{\bf D}^{ - 1}} = \frac{1}{{\det ({\bf D})}}{{\bf C}^T},
\label{A-1}
\end{equation}
where $\bf C$ is the cofactor matrix, whose elements have the following form:
\begin{equation*}
c_{i,j}={(-1)}^{(i+j)}m_{i,j},
\end{equation*}
where $m_{i,j}$ is the determinant of the submatrix generated by eliminating the $i$-th row and $j$-th column of $\bf D$. Then, we can see that if  event ${\cal G}_i$ occurs, by defining
\begin{equation*}
{\bf{L^{'}}} = \frac{1}{{Q}}{{\bf{H}}_{{\bf{N}}_i^*}}{\bf{H}}_{{\bf{N}}_i^*}^H,
\end{equation*}
we will have:
\begin{equation}
\left| {{l^{'}_{n,n}} - E\left\{ {{{\left| {H_{{\rm{TI}}}^{[u{i_n}]}(t)H_{{\rm{IR}}}^{[{j_n}u]}(t)} \right|}^2}} \right\}} \right| \le \delta ,
\label{L'1}
\end{equation}
\begin{equation}
\left| {{l^{'}_{n,m}}} \right| \le \delta ,\forall m \ne n.
\label{L'2}
\end{equation}
Substituting (\ref{L'2}) into (\ref{A-1})  leads to (\ref{A}) and concludes Lemma \ref{lemma6}.
\end{IEEEproof}

Now,  for $\Pr\{{\cal F}_{{Q_i}}\},i\ne1$, we have:
\begin{equation*}
\Pr \left\{ {{{\cal F}_{{Q_i}}}} \right\} = \Pr \left\{ {{{\cal F}_{{Q_i}}}\bigcap {{{\cal G}_i}} } \right\} + \Pr \left\{ {{{\cal F}_{{Q_i}}}\bigcap {{\cal G}_i^c} } \right\}
\end{equation*}
\begin{equation}
 \le \Pr \left\{ {{{\cal F}_{{Q_i}}}\bigcap {{{\cal G}_i}} } \right\} + \Pr \left\{ {{\cal G}_i^c} \right\} \le \Pr \left\{ {{{\cal F}_{{Q_i}}}\bigcap {{{\cal G}_i}} } \right\} + {\varepsilon _3}(Q),
\label{pupp}
\end{equation}
where ${\cal G}_i^c$ is the complement of ${\cal G}_i$ and the last inequality in (\ref{pupp}) follows from (\ref{G3i}). In addition, we have:
\begin{equation}
\Pr \left\{ {{{\cal F}_{{Q_i}}}\bigcap {{{\cal G}_i}} } \right\} = \Pr \left\{ {{{\cal F}_{{Q_i}}}\left| {{{\cal G}_i}} \right.} \right\}\Pr \left\{ {{{\cal G}_i}} \right\}.
\label{chain rule}
\end{equation}
By Lemma \ref{lemma6}, we can  see that conditioned on ${\cal G}_i$, we have:
\begin{equation*}
{{{\tau }}_{{\bf{N}}_i^*}} = {\bf{H}}_{{\bf{N}}_i^*}^H{\left( {{{\bf{H}}_{{\bf{N}}_i^*}}{\bf{H}}_{{\bf{N}}_i^*}^H} \right)^{ - 1}}{{\bf{h}}_{{\bf{N}}_i^*}}=\frac{1}{Q}{\bf{H}}_{{\bf{N}}_i^*}^H{\left( {\frac{1}{Q}{{\bf{H}}_{{\bf{N}}_i^*}}{\bf{H}}_{{\bf{N}}_i^*}^H} \right)^{ - 1}}{{\bf{h}}_{{\bf{N}}_i^*}}
\end{equation*}
\begin{equation*}
= \frac{1}{{Q\gamma (1 + {a_{0,0}})}}{\bf{H}}_{{\bf{N}}_i^*}^H{\bf E} {{\bf{h}}_{{\bf{N}}_i^*}} + \frac{1}{{Q\gamma (1 + {a_{0,0}})}}{\bf{H}}_{{\bf{N}}_i^*}^H{\bf{A}}{{\bf{h}}_{{\bf{N}}_i^*}}.
\end{equation*}
where 
\begin{equation*}
\gamma  = E\left\{ {{{\left| {{H_{\rm TI}^{[{u}i']}}({ t}){H_{\rm IR}^{[j'{u}]}}({ t})} \right|}^2}} \right\},
\end{equation*}
and $\bf A$ is given by (\ref{A}). If we define $\delta ' = \mathop {\max }\limits_{i,j} \left| {{P^{[ji]}}(\delta )} \right|$, we can see that:
\begin{equation}
\left| {\tau _{{\bf{N}}_i^*}^{[u]}} \right| \le \left| {\frac{1}{{Q\gamma (1 - \delta ')}}{{\left[ {{\bf{H}}_{{\bf{N}}_i^*}^H{\bf E}{{\bf{h}}_{{\bf{N}}_i^*}}} \right]}_u}} \right| + \left| {\frac{{\delta '}}{{Q\gamma (1 - \delta ')}}{{\left[ {\left| {{{\bf{H}}^H_{{\bf{N}}_i^*}}} \right|{\bf R}\left| {{{\bf{h}}_{{\bf{N}}_i^*}}} \right|} \right]}_u}} \right|,
\end{equation}
where 
\begin{equation*}
{\bf R} = {\left[ {\begin{array}{*{20}{c}}
1&1& \cdots \\
1&1& \cdots \\
 \vdots & \vdots & \ddots 
\end{array}} \right]_{\left| {{{\cal M}_{{\bf{N}}_i^*}}} \right| \times \left| {{{\cal M}_{{\bf{N}}_i^*}}} \right|}},
\end{equation*}
and the matrix ${\left| {{{\bf{H}}^H_{{\bf{N}}_i^*}}} \right|}$  corresponds to matrix ${{{\bf{H}}^H_{{\bf{N}}_i^*}}}$ with all its elements replaced by their absolute values. ${\left| {{{\bf{h}}_{{\bf{N}}_i^*}}} \right|}$ is constructed in a similar manner. In addition, ${{{\left[ {{\bf{H}}_{{\bf{N}}_i^*}^H{\bf E}{{\bf{h}}_{{\bf{N}}_i^*}}} \right]}_u}}$ and ${{{\left[ {\left| {{{\bf{H}}^H_{{\bf{N}}_i^*}}} \right|{\bf R}\left| {{{\bf{h}}_{{\bf{N}}_i^*}}} \right|} \right]}_u}}$ denote the $u$-th elements of  vectors ${{\bf{H}}_{{\bf{N}}_i^*}^H{\bf E}{{\bf{h}}_{{\bf{N}}_i^*}}}$ and ${\left| {{{\bf{H}}^H_{{\bf{N}}_i^*}}} \right|{\bf R}\left| {{{\bf{h}}_{{\bf{N}}_i^*}}} \right|}$, respectively. In order to derive an upper bound on $\Pr \left\{ {{{\cal F}_{{Q_i}}}\left| {{{\cal G}_i}} \right.} \right\}$, we define events ${\cal J}_i$ and ${\cal L}_i$ as follows: 
\begin{equation}
{{\cal J}_i} = \bigcup\limits_{u = 1}^Q {\left\{ {\left| {\frac{1}{{Q\gamma (1 - \delta ')}}{{\left[ {{\bf{H}}_{{\bf{N}}_i^*}^H{\bf E}{{\bf{h}}_{{\bf{N}}_i^*}}} \right]}_u}} \right| + \left| {\frac{{\delta '}}{{Q\gamma (1 - \delta ')}}{{\left[ {\left| {{{\bf{H}}^H_{{\bf{N}}_i^*}}} \right|{\bf R}\left| {{{\bf{h}}_{{\bf{N}}_i^*}}} \right|} \right]}_u}} \right| > 1} \right\}}  ,
\end{equation}
\begin{equation}
{{\cal L}_i} = \bigcup\limits_{u = 1}^Q {\left\{ {\left| {\tau _{{\bf{N}}_i^*}^{[u]}} \right| > 1} \right\}} .
\end{equation}
We have ${{\cal F}_{{Q_i}}} \subseteq {{\cal L}_i} \subseteq {{\cal J}_i}$, so we conclude:
\begin{equation}
\Pr \left\{ {{{\cal F}_{{Q_i}}}\left| {{{\cal G}_i}} \right.} \right\} \le \Pr \left\{ {{{\cal L}_i}\left| {{{\cal G}_i}} \right.} \right\} \le \Pr \left\{ {{{\cal J}_i}\left| {{{\cal G}_i}} \right.} \right\},
\label{chain2}
\end{equation}
and by (\ref{chain rule}) and (\ref{chain2}), we have:
\begin{equation*}
\Pr \left\{ {{{\cal F}_{{Q_i}}}\bigcap {{{\cal G}_i}} } \right\} \le \Pr \left\{ {{{\cal J}_i}\left| {{{\cal G}_i}} \right.} \right\}\Pr \left\{ {{{\cal G}_i}} \right\} = \Pr \left\{ {{{\cal J}_i}\bigcap {{{\cal G}_i}} } \right\} \le \Pr \left\{ {{{\cal J}_i}} \right\}
\end{equation*}
\begin{equation*}
\Pr \left\{ {{{\cal J}_i}} \right\} \le Q\Pr \left\{ {\left| {\frac{1}{{Q\gamma (1 - \delta ')}}{{\left[ {{\bf{H}}_{{\bf{N}}_i^*}^H{\bf E}{{\bf{h}}_{{\bf{N}}_i^*}}} \right]}_u}} \right| + \left| {\frac{{\delta '}}{{Q\gamma (1 - \delta ')}}{{\left[ {\left| {{{\bf{H}}^H_{{\bf{N}}_i^*}}} \right|{\bf R}\left| {{{\bf{h}}_{{\bf{N}}_i^*}}} \right|} \right]}_u}} \right| > 1} \right\}
\end{equation*}
\begin{equation*}
= Q\Pr \left\{ {{{\left\{ {\left| {{{\left[ {{\bf{H}}_{{\bf{N}}_i^*}^H{\bf E}{{\bf{h}}_{{\bf{N}}_i^*}}} \right]}_u}} \right| + \left| {\delta '{{\left[ {\left| {{{\bf{H}}^H_{{\bf{N}}_i^*}}} \right|{\bf R}\left| {{{\bf{h}}_{{\bf{N}}_i^*}}} \right|} \right]}_u}} \right|} \right\}}^2} > {Q^2}{\gamma ^2}{{(1 - \delta ')}^2}} \right\}
\end{equation*}
\begin{equation}
 \le Q\frac{{E\left\{ {{{\left\{ {\left| {{{\left[ {{\bf{H}}_{{\bf{N}}_i^*}^H{\bf E}{{\bf{h}}_{{\bf{N}}_i^*}}} \right]}_u}} \right| + \left| {\delta '{{\left[ {\left| {{{\bf{H}}^H_{{\bf{N}}_i^*}}} \right|{\bf R}\left| {{{\bf{h}}_{{\bf{N}}_i^*}}} \right|} \right]}_u}} \right|} \right\}}^2}} \right\}}}{{{Q^2}{\gamma ^2}{{(1 - \delta ')}^2}}},
\label{Markov1}
\end{equation}
where (\ref{Markov1}) follows from Markov's inequality. Thus, (\ref{pupp}) and (\ref{Markov1}) show that by choosing a sufficiently small  $\delta$ and a sufficiently large $Q$, $\Pr\{{\cal F}_{{Q_i}}\}$ can be made arbitrarily small for $i\ne1$.

\section{}
\label{appendix8}
From Theorem \ref{passive outer DoF reg}, we observe that, for every $\delta>0$ and for large enough $T$, with probability more than $1-\epsilon$, we have ${\bf d}\in {\cal D}_{out}(\delta)$. So, with probability more than $1-\epsilon$, there exist vectors ${\bf a}^{[1]}\in {\cal A}_1,...,{\bf a}^{[{2^{\left| {{{\cal N}_Q}} \right| - 1}}]}\in {\cal A}_{{2^{\left| {{{\cal N}_Q}} \right| - 1}}}$, for which  we have:
\begin{equation*}
\sum\limits_{k = 1}^K {{d_k}}  \le \frac{1}{{2(K - 1)}}\sum\limits_{i \ne j} {\sum\limits_{l = 1}^{{2^{\left| {{{\cal N}_Q}} \right| - 1}}} {\left( {\Pr \{ {{\cal E}_{{Q_l}}}\}  + \delta } \right)\sum\limits_{m = 1}^{\left| {{{\tilde {\cal N}}_{{Q_l}}}} \right|} {a_m^{[l]}{{\tilde d}^{[ij]}}_{{\bf{N}}_m^{[l]}}} } } 
\end{equation*}
\begin{equation*}
\le \frac{1}{{2(K - 1)}}\sum\limits_{l = 1}^{{2^{\left| {{{\cal N}_Q}} \right| - 1}}} {\left( {\Pr \{ {{\cal E}_{{Q_l}}}\}  + \delta } \right)} \left( {\sum\limits_{i \ne j} {\sum\limits_{m = 1}^{\left| {{{\tilde {\cal N}}_{{Q_l}}}} \right|} {a_m^{[l]}{{\tilde d}^{[ij]}}_{{\bf{N}}_m^{[l]}}} } } \right)
\end{equation*}
\begin{equation*}
 \le \frac{1}{{2(K - 1)}}\sum\limits_{l = 1}^{{2^{\left| {{{\cal N}_Q}} \right| - 1}}} {\left( {\Pr \{ {{\cal E}_{{Q_l}}}\}  + \delta } \right)} \mathop {\max }\limits_{{\bf{N}} \in {{\tilde {\cal N}}_{{Q_l}}}} \left( {\sum\limits_{i \ne j} {{{\tilde d}^{[ij]}}_{\bf{N}}} } \right)
\end{equation*}
\begin{equation*}
 \le \sum\limits_{l = 1}^{{2^{\left| {{{\cal N}_Q}} \right| - 1}}} {\left( {\Pr \{ {{\cal E}_{{Q_l}}}\}  + \delta } \right)\left( {K - \mathop {\min }\limits_{{\bf{N}} \in {{\tilde {\cal N}}_{{Q_l}}}} \sum\limits_{i \ne j} {\frac{{{n_{i,j}}}}{{2(K - 1)}}} } \right)} 
\end{equation*}
\begin{equation*}
 = \sum\limits_{l = 1}^{{2^{\left| {{{\cal N}_Q}} \right| - 1}}} {\left( {\Pr \{ {{\cal E}_{{Q_l}}}\}  + \delta } \right)\left( {K - \mathop {\min }\limits_{{\bf{N}} \in {{\tilde {\cal N}}_{{Q_l}}}} \frac{{K(K - 1) - \left| {{{\cal M}_{\bf{N}}}} \right|}}{{2(K - 1)}}} \right)} 
\end{equation*}
\begin{equation}
 = \sum\limits_{l = 1}^{{2^{\left| {{{\cal N}_Q}} \right| - 1}}} {\left( {\Pr \{ {{\cal E}_{{Q_l}}}\}  + \delta } \right)\mathop {\max }\limits_{{\bf{N}} \in {{\tilde {\cal N}}_{{Q_l}}}} \left( {\frac{K}{2} + \frac{{\left| {{{\cal M}_{\bf{N}}}} \right|}}{{2(K - 1)}}} \right)} .
\end{equation}

\section{}
\label{appendix9}
From Theorem \ref{passive inner DoF reg}, we observe that, for every $\delta>0$ and for sufficiently large $T$, with probability higher than $1-\epsilon$, in at least $T(\Pr\{{\cal F}_{{Q_i}}\}-\delta)$ time slots, ${\cal F}_{{Q_i}}$ occurs. So, in these slots, if we choose the network matrix such that ${\bf N}\in {\tilde {\cal N}}^{W^*}_{Q_i}$, where $W^*=\mathop {\max }\limits_{{\tilde {\cal N}}_{{Q_i}}^W \ne \Phi }  {{{ W}}}$, we can show that $\frac{K+W^*}{2}$ sum DoF are achievable, see the proof of Theorem \ref{active IRS sum in}. Thus,  by time sharing, the total ${\sum\limits_{i = 1}^{{2^{\left| {{{\cal N}_Q}} \right| - 1}}} {\left( {\Pr \{ {{\cal F}_{{Q_i}}}\}  - \delta } \right)\mathop {\max }\limits_{{\tilde {\cal N}}_{{Q_i}}^W \ne \Phi } \left( {\frac{{K + W}}{2}} \right)} }$ DoFs are achievable with probability higher than $1-\epsilon$.

\section{}
\label{appendix11}
Define $\pi ({ x}\left| {{{\Lambda}^T}} \right.)$ as the fraction of time slots, in which event $x\in\{1,0\}$ occurs in $T$ realizations of random variable $\Lambda ( t)$ ($t\in \{1,...,T\}$). Then, by the law of large numbers,  for each ${ x}\in {\{0,1\}}$ and for each $0<\delta$, there exists a sequence $\epsilon (\delta,T)$ such that:
\begin{equation*}
\Pr \left\{ {\left| {\pi ({ x}\left| {{{\Lambda}^T}} \right.) - \Pr \left\{ {\Lambda ({ t}) = {x}} \right\}} \right| > \delta } \right\} < \epsilon (\delta,T),
\end{equation*}
where $\forall \delta  > 0 \to \mathop {\lim }\limits_{T \to \infty } \epsilon (\delta ,T) = 0$. Thus, for sufficiently large  $T$, with probability higher than $1-\epsilon$, in at least $T\times\left( {\Pr \left\{ {\Lambda ({ t}) = 1} \right\} - \delta } \right)$  time slots,  the interference can be cancelled for all users by at least one subset of $\varepsilon$-relaxed passive lossless IRS elements ${\cal N}_{k'}$. Therefore,  the total $T\times\left( {\Pr \left\{ {\Lambda ({ t}) = 1} \right\} - \delta } \right) {K} $ DoFs are achievable in these slots. 
In addition, the interference cancellation equations in (\ref{intercancel11}) cannot be satisfied by any $K^2$ elements of the $\varepsilon$-relaxed passive lossless IRS in at least $T\times\left( {\Pr \left\{ {\Lambda ({ t}) = 0} \right\} - \delta } \right)$  time slots with a probability higher than $1-\epsilon$, assuming $T$ is sufficiently large.
 Thus, in these time slots, a total of $T\times\left( {\Pr \left\{ {\Lambda ({ t}) = 0} \right\} - \delta } \right)\left( {\frac{K}{2}} \right)$ DoFs are achievable. 
This completes the proof.

\section{}
\label{appendix12}
We assume that $Q=mK^2$, where $m\in \mathbb{N}$. 
Then, we prove the following expression:
\begin{equation}
\mathop {\lim }\limits_{m \to \infty } \Pr \{ \Lambda ({ t}) = 0\}  = 0.
\label{lambda limit}
\end{equation}
Considering the construction of $\Lambda( t)$ in (\ref{Lambda construction}), we have:
\begin{equation*}
\Pr \{ \Lambda ({ t}) = 0\}  = \Pr \left\{ {\forall k' \in \left\{ {1,...,\left( {\begin{array}{*{20}{c}}
Q\\
{K^2}
\end{array}} \right)} \right\}:\exists u \in {{\cal N}_{k'}}:1 < \left| {\tau _{k'}^{[u]}({ t})} \right|or\left| {\tau _{k'}^{[u]}({ t})} \right| < 1 - \varepsilon } \right\}
\end{equation*}
\begin{equation}
 = \Pr \left\{ {\bigcap\limits_{k' \in \left\{ {1,...,\left( {\begin{array}{*{20}{c}}
Q\\
{K^2}
\end{array}} \right)} \right\}} {\left\{ {\exists u \in {{\cal N}_{k'}}:1 < \left| {\tau _{k'}^{[u]}({ t})} \right|or\left| {\tau _{k'}^{[u]}({ t})} \right| < 1 - \varepsilon } \right\}} } \right\}.
\label{intersection}
\end{equation}
Without loss of generality, we assume that:
\begin{equation*}
{{\cal N}_{k'}} = \{ 1 + K^2(k' - 1),2 + K^2(k' - 1),..., K^2k'\} ,k' \in \{ 1,...,m\}.
\end{equation*}
Therefore, by reducing the number of intersections in (\ref{intersection}), we obtain:
\begin{equation}
\Pr \{ \Lambda ({ t}) = 0\}  \le \Pr \left\{ {\bigcap\limits_{k' \in \left\{ {1,...,m} \right\}} {\left\{ {\exists u \in {{\cal N}_{k'}}:1 < \left| {\tau _{k'}^{[u]}({ t})} \right|or\left| {\tau _{k'}^{[u]}({ t})} \right| < 1 - \varepsilon } \right\}} } \right\}.
\label{upperbound}
\end{equation}
To analyze the upper bound in (\ref{upperbound}), we define the following event:
\begin{equation*}
{{\cal T}_{k'}} = \left\{ {\exists u \in {{\cal N}_{k'}}:1 < \left| {\tau _{k'}^{[u]}( t)} \right|or\left| {\tau _{k'}^{[u]}( t)} \right| < 1 - \varepsilon } \right\}.
\end{equation*}
Thus, the right hand of (\ref{upperbound}) is equal to:
\begin{equation*}
\Pr \left\{ {\bigcap\limits_{k' \in \left\{ {1,...,m} \right\}} {{{\cal T}_{k'}}} } \right\} = 
\end{equation*}
\begin{equation}
\int\scalebox{.75}[1]{$ {\Pr \left\{ {\bigcap\limits_{k' \in \left\{ {1,...,m} \right\}} {{{\cal T}_{k'}}} \left| {{H^{[11]}}({ t})...{H^{[KK]}}({ t})} \right.} \right\}\prod\limits_{i,j} {\left( {f_{H_r^{[ji]}}^r\left( {H_r^{[ji]}({ t})} \right)f_{H_i^{[ji]}}^i\left( {H_i^{[ji]}({ t})} \right)} \right)} \prod\limits_{i,j} {\left( {dH_r^{[ji]}({ t})dH_i^{[ji]}({ t})} \right)} }  $},
\label{conditioned}
\end{equation}
where $H_r^{[ji]}( t)$ and $H_i^{[ji]}( t)$ are the  real and imaginary parts of $H^{[ji]}( t)$, respectively.
Conditioned on \scalebox{.9}[1]{${H^{[11]}}({ t}),...,{H^{[KK]}}$}, the events ${\cal T}_{k'},k'\in\{1,...,m\}$, will be independent. Thus, (\ref{conditioned}) can be rewritten as:
\begin{equation}
\int \scalebox{.75}[1]{${\left( {\prod\limits_{k' \in \left\{ {1,...,m} \right\}} {\Pr \left\{ {{{\cal T}_{k'}}\left| {{H^{[11]}}({ t})...{H^{[KK]}}({ t})} \right.} \right\}} } \right)\prod\limits_{i,j} {\left( {f_{H_r^{[ji]}}^r\left( {H_r^{[ji]}({ t})} \right)f_{H_i^{[ji]}}^i\left( {H_i^{[ji]}({ t})} \right)} \right)} \prod\limits_{i,j} {\left( {dH_r^{[ji]}({ t})dH_i^{[ji]}({ t})} \right)} }  $}.
\label{conditioned2}
\end{equation}
Based on (\ref{prop1})-(\ref{prop4}), the events ${\cal T}_{k'},\forall k'\in\{1,...,M\}$, conditioned on ${{H^{[11]}}({ t}),...,{H^{[KK]}}({ t})}$ have the same probability. Thus, we define:
\begin{equation}
h\left( {{H^{[11]}}({ t}),...,{H^{[KK]}}({ t})} \right) = \Pr \left\{ {{{\cal T}_{k'}}\left| {{H^{[11]}}({ t})...{H^{[KK]}}({ t})} \right.} \right\}.
\label{h_def}
\end{equation}
Then, (\ref{conditioned2}) can be rewritten as follows:
\begin{equation}
\int {{h^m}\left( {{H^{[11]}}({ t}),...,{H^{[KK]}}({ t})} \right)\prod\limits_{i,j} {\left( {f_{H_r^{[ji]}}^r\left( {H_r^{[ji]}({ t})} \right)f_{H_i^{[ji]}}^i\left( {H_i^{[ji]}({ t})} \right)} \right)} \prod\limits_{i,j} {\left( {dH_r^{[ji]}({ t})dH_i^{[ji]}({ t})} \right)} } .
\label{upper bound}
\end{equation}
Now, we rewrite (\ref{intercancel11}) in matrix form  ${\bold{H}}_{k'}{{\tau}_{k'}}=\bold h$ , where ${\bold{H}}_{k'}$ is a matrix with elements ${H_{\rm TI}^{[{u}i]}}({t}){H_{\rm IR}^{[j{u}]}}({t}),$ $u\in {\cal N}_{k'}$, ${\bold{\tau}}_{k'}$ is a column vector with elements ${\tau ^{[u]}}({t}),u\in {\cal N}_{k'}$, and ${\bold h}_{\bf N}$ is a column vector with elements $- {H^{[ji]}}({t}),i\ne j,$ and $0$ (the index $k'$ indicates that the operating subset of the IRS elements is ${\cal N}_{k'}$).
To bound  (\ref{upper bound}), we prove the following Lemma:

\begin{lemma}
For the space of channel coefficients ${{H^{[11]}}({ t}),...,{H^{[KK]}}({ t})}$, except for a subset of this space that has zero measure, there exists an invertible matrix ${\bf H}_{k'}$ for each $k' \in \left\{ {1,...,m} \right\}$ that satisfies $1-\varepsilon<|\tau_{k'}^{[u]}( t)|<1,\forall u\in {\cal N}_{k'}$.
\label{H0}
\end{lemma}
\begin{IEEEproof}
Without loss of generality, we arrange the elements of ${\bf H}_{k'}$ in the following manner:
\begin{equation}
{\left[ {{{\bf H}_{k'}}} \right]_{K(i - 1) + j,(\mathop {\bmod }\limits_{{K^2}} u) + 1}} = {H_{\rm TI}^{[{u}i]}}({ t}){H_{\rm IR}^{[j{u}]}}({ t}),i,j \in \{ 1,...,K\} ,u \in {{\cal N}_{k'}},
\end{equation}
where $\mathop {\bmod }\limits_{{K^2}} u$  is an integer $p\in\{0,1,...,K^2-1\}$, for which we have $u=ZK^2+p,Z\in\mathbb{Z}$. 
If the channel coefficients have the following feasible values:
\begin{equation*}
{H_{\rm TI}^{[{u}i]}}({ t}) = \frac{{\left[ {\sum\limits_{z = 1}^{{K^2}} {\exp \left\{ {\frac{{ - (z - 1)(\mathop {\bmod }\limits_{{K^2}} u)}}{{{K^2}}}2\pi \sqrt { - 1} } \right\}{{\left[ {\bf h} \right]}_z}} } \right]}}{{(1 - \frac{\varepsilon }{2})K^2}}\exp \left\{ {\frac{{K(i - 1)(\mathop {\bmod }\limits_{{K^2}} u)}}{{{K^2}}}2\pi \sqrt { - 1} } \right\},
\end{equation*}
\begin{equation*}
{H_{\rm IR}^{[j{u}]}}({ t}) = \exp \left\{ {\frac{{(j - 1)(\mathop {\bmod }\limits_{{K^2}} u)}}{{{K^2}}}2\pi \sqrt { - 1} } \right\},j \in \{ 1,...,K\} ,u \in {{\cal N}_{k'}},
\end{equation*}
where $\sqrt { - 1} $ is the  imaginary unit, for which we have $(\sqrt { - 1})^2=-1$. Then, we will have:
\begin{equation}
{\left[ {{{\bf H}_{k'}}} \right]_{v,(\mathop {\bmod }\limits_{{K^2}} u) + 1}} = \frac{{\left[ {\sum\limits_{z = 1}^{{K^2}} {\exp \left\{ {\frac{{ - (z - 1)(\mathop {\bmod }\limits_{{K^2}} u)}}{{{K^2}}}2\pi \sqrt { - 1} } \right\}{{\left[ {\bf h} \right]}_z}} } \right]}}{{(1 - \frac{\varepsilon }{2})K^2}}\exp \left\{ {\frac{{(v - 1)(\mathop {\bmod }\limits_{{K^2}} u)}}{{{K^2}}}2\pi \sqrt { - 1} } \right\},
\end{equation}
and its inverse matrix
\begin{equation}
{\left[ {{\bf H}_{k'}^{ - 1}} \right]_{(\mathop {\bmod }\limits_{{K^2}} u) + 1,v}} = \frac{{1 - \frac{\varepsilon }{2}}}{{\left[ {\sum\limits_{z = 1}^{{K^2}} {\exp \left\{ {\frac{{ - (z - 1)(\mathop {\bmod }\limits_{{K^2}} u)}}{{{K^2}}}2\pi \sqrt { - 1} } \right\}{{\left[ {\bf h} \right]}_z}} } \right]}}\exp \left\{ {\frac{{ - (v - 1)(\mathop {\bmod }\limits_{{K^2}} u)}}{{{K^2}}}2\pi \sqrt { - 1} } \right\}.
\end{equation}
Then, for each $\forall u\in {\cal N}_{k'}$, we can see that:
\begin{equation*}
\tau _{k'}^{[u]}({ t}) = {\left[ {{\bf{H}}_{k'}^{ - 1}{\bf{h}}} \right]_u}
\end{equation*}
\begin{equation*}
= \frac{{1 - \frac{\varepsilon }{2}}}{{\left[ {\sum\limits_{z = 1}^{{K^2}} {\exp \left\{ {\frac{{ - (z - 1)(\mathop {\bmod }\limits_{{K^2}} u)}}{{{K^2}}}2\pi \sqrt { - 1} } \right\}{{\left[{\bf h} \right]}_z}} } \right]}} \times \left[ {\sum\limits_{v = 1}^{{K^2}} {\exp \left\{ {\frac{{ - (v - 1)(\mathop {\bmod }\limits_{{K^2}} u)}}{{{K^2}}}2\pi \sqrt { - 1} } \right\}{{\left[ {\bf h} \right]}_v}} } \right]
\end{equation*}
\begin{equation}
=1-\frac{\varepsilon}{2}.
\end{equation}
Therefore,  exept for the  set 
\begin{equation}
{\cal B}_0 = \bigcup\limits_{u \in {{\cal N}_{k'}}} {\left\{ {\sum\limits_{z = 1}^{{K^2}} {\exp \left\{ {\frac{{ - (z - 1)(\mathop {\bmod }\limits_{{K^2}} u)}}{{{K^2}}}2\pi \sqrt { - 1} } \right\}{{\left[ {\bf h} \right]}_z}} }=0 \right\}} 
\label{B_set}
\end{equation}
 with zero measure in the space of channel coefficients ${{H^{[11]}}({ t}),...,{H^{[KK]}}({ t})}$, the statement of the Lemma is correct.

\end{IEEEproof}

\begin{lemma}
If ${\bf h}\notin {\cal B}_0$ (${\cal B}_0$ is given by (\ref{B_set})), then we have: 
\begin{equation}
{h}\left( {{H^{[11]}}({ t}),...,{H^{[KK]}}({ t})} \right)<1,
\label{strict ineq}
\end{equation}
where $h(\cdot)$ is defined in (\ref{h_def}).
\label{lemma50}
\end{lemma}

\begin{IEEEproof}
For given \scalebox{.92}[1]{${H^{[11]}}({ t}),...,{H^{[KK]}}$}, vector $\bold h$ is fixed and by  Lemma \ref{H0}, if ${\bold h}\notin {\cal B}_0$, there exits a matrix ${\bf H}_{k',0}$ for which we have $\forall u\in {\cal N}_{k'},1-\varepsilon<|\tau_{k'}^{[u]}( t)|<1$ and we call its channel coefficients generators  $ h^{[ui]}_{\rm TI_0}( t)$ and $h_{\rm IR_0}^{[j{u'}]}( t)$.   Now, we consider an $\epsilon$-neighbourhood (${\cal D}_{\epsilon}$) of $ h_{\rm TI_0}^{[ui]}( t)$ and $ h_{\rm IR_0}^{[j{u'}]}( t)$ in the space of channel coefficients $H_{\rm TI}^{[ui]}( t)$ and $H_{\rm IR}^{[j{u'}]}( t)$,$i,j\in\{1,...,K\},u,u'\in {\cal N}_{k'}$. By the continuity of ${\left( {{{\mathbf{H}}_{k'}}} \right)^{ - 1}}{\mathbf{h}}$ arround $ h_{\rm TI_0}^{[ui]}( t)$ and $ h_{\rm IR_0}^{[j{u'}]}( t)$ (because the determinant of ${\bold H}_{k',0}$ is nonzero), we can choose a sufficiently small $\epsilon$ such that for all  $H_{\rm TI}^{[ui]}( t)$ and $H_{\rm IR}^{[j{u'}]}( t)$,$i,j\in\{1,...,K\},u,u'\in {\cal N}_{k'}$ in ${\cal D}_{\epsilon}$, we have $\forall u \in {{\cal N}_{k'}}, 1-\varepsilon\le\left| {\tau _{11}^{[u]}} \right| \le1$. Thus, we have:
\begin{equation*}
h\left( {{H^{[11]}}({ t}),...,{H^{[KK]}}({ t})} \right) \le
\end{equation*}
\begin{equation*}
\scalebox{.72}[1]{$1 - \int\limits_{{{\cal D}_\varepsilon }} {\prod\limits_{i,u} {\left( {f_H^r\left( {H_{{\rm TI}_r}^{[{u}i]}({ t})} \right)f_H^i\left( {H_{{\rm TR}_i}^{[{u}i]}({ t})} \right)} \right)\prod\limits_{u,j} {\left( {f_H^r\left( {H_{{\rm IR}_r}^{[j{u}]}({ t})} \right)f_H^i\left( {H_{{\rm IR}_i}^{[j{u}]}({ t})} \right)} \right)} } \prod\limits_{i,u} {\left( {dH_{{\rm TI}_r}^{[{u}i]}({ t})dH_{{\rm TI}_i}^{[{u}i]}({ t})} \right)} \prod\limits_{u,j} {\left( {dH_{{\rm IR}_r}^{[j{u}]}({ t})dH_{{\rm IR}_i}^{[j{u}]}({ t})} \right)} } $},
\end{equation*}
and by assumption (\ref{axiom1}), we obtain (\ref{strict ineq}).
\end{IEEEproof}

Now, we can bound (\ref{upper bound}).
To this end, we define
\begin{equation*}
\begin{cases}
\scalebox{1}[1]{${y_m} = {h^m}\left( {{H^{[11]}}({ t}),...,{H^{[KK]}}({ t})} \right)\prod\limits_{i,j} {\left( {f_{H_r^{[ji]}}^r\left( {H_r^{[ji]}({ t})} \right)f_{H_i^{[ji]}}^i\left( {H_i^{[ji]}({ t})} \right)} \right)} $} ,\\
g = \prod\limits_{i,j} {\left( {f_{H_r^{[ji]}}^r\left( {H_r^{[ji]}({ t})} \right)f_{H_i^{[ji]}}^i\left( {H_i^{[ji]}({ t})} \right)} \right)},\\
d\vec H = \prod\limits_{i,j} {\left( {dH_r^{[ji]}(t)dH_i^{[ji]}(t)} \right)} .
\end{cases}
\end{equation*}
Next, we use the Dominated Convergence Theorem, which states the conditions for a sequence of functions $f_n(x)$ to yield $\mathop {\lim }\limits_{n \to \infty } \int {{f_n}} (x)dx = \int {\mathop {\lim }\limits_{n \to \infty } {f_n}} (x)dx$, see \cite[Theorem 1.13]{Stein} for details. 
We can see that $y_{m}\le g$, $\mathop {\lim }\limits_{m \to \infty } {y_m} = 0$ for ${\bf h}\notin {\cal B}_0$ by Lemma \ref{lemma50},  and $\int\limits_{{\cal B}_0^c} {\left| g \right|d\vec H }  = 1$. Thus, by the Dominated Convergence Theorem \cite[Theorem 1.13]{Stein}, we obtain:
\begin{equation*}
\mathop {\lim }\limits_{m \to \infty } \int {{y_m}d\vec H }  = \mathop {\lim }\limits_{m \to \infty } \int\limits_{{\cal B}_0^c} {{y_m}d\vec H }  + \int\limits_{{{\cal B}_0}} {{y_m}d\vec H }  = \mathop {\lim }\limits_{m \to \infty } \int\limits_{{\cal B}_0^c} {{y_md\vec H }}  = \int\limits_{{\cal B}_0^c} {\mathop {\lim }\limits_{m \to \infty } {y_md\vec H }}  = 0.
\end{equation*}
Therefore, we have $\mathop {\lim }\limits_{Q \to \infty } \Pr \left\{ {\Lambda ({t}) = 0} \right\} = 0$, as stated in (\ref{lossless convergence1}).

\section{}
\label{appendix14}

By the law of large numbers, we can see that there exists a number $T^{\prime}$, such that for $T>T^{\prime}$, in at least $T{\left( {\Pr \left\{ {\bigcap\limits_{k = 1}^K {\left\{ {\left\{ {{\rm{SINR}}_k^r \ge {\rho ^{1 - \varepsilon }}} \right\}\bigcap {\left\{ {{\rm{SINR}}_k^i \ge {\rho ^{1 - \varepsilon }}} \right\}} } \right\}} } \right\} - \delta } \right)}$ time slots and with probability higher than $1-\epsilon$, we have $\textrm{SINR}_k^r\ge \rho^{1-\varepsilon},\textrm{SINR}_k^i\ge \rho^{1-\varepsilon},\forall k\in\{1,...,K\}$, and in these time slots, the $\rho$-limited sum DoF is at least $K(1-\varepsilon)$, because we have:
\begin{equation*}
\scalebox{1}[1]{${\rm{SINR}}_k=\frac{{{{\left| {{\rm{Re}}\left\{ {{H^{[kk]}}(t) + \sum\limits_{u = 1}^Q {{H_{\rm IR}^{[k{u}]}}(t){\tau ^{[u]}}(t){H_{\rm TI}^{[{u}k]}}(t)} } \right\}} \right|}^2}\frac{\rho }{2} + {{\left| {{\mathop{\rm Im}\nolimits} \left\{ {{H^{[kk]}}(t) + \sum\limits_{u = 1}^Q {{H_{\rm IR}^{[k{u}]}}(t){\tau ^{[u]}}(t){H_{\rm TI}^{[{u}k]}}(t)} } \right\}} \right|}^2}\frac{\rho }{2}}}{{\sum\limits_{i = 1,i \ne k}^K {{{\left| {{\rm{Re}}\left\{ {{H^{[ki]}}(t) + \sum\limits_{u = 1}^Q {{H_{\rm IR}^{[k{u}]}}(t){\tau ^{[u]}}(t){H_{\rm TI}^{[{u}i]}}(t)} } \right\}} \right|}^2}\frac{\rho }{2}}  + \sum\limits_{i = 1,i \ne k}^K {{{\left| {{\rm{Re}}\left\{ {{H^{[ki]}}(t) + \sum\limits_{u = 1}^Q {{H_{\rm IR}^{[k{u}]}}(t){\tau ^{[u]}}(t){H_{\rm TI}^{[{u}i]}}(t)} } \right\}} \right|}^2}\frac{\rho }{2}}  + \frac{{{N_0}}}{2}}} \ge $}
\end{equation*}
\begin{equation*}
\scalebox{1}[1]{$\frac{{{{\left| {{\rm{Re}}\left\{ {{H^{[kk]}}(t) + \sum\limits_{u = 1}^Q {{H_{\rm IR}^{[k{u}]}}(t){\tau ^{[u]}}(t){H_{\rm TI}^{[{u}k]}}(t)} } \right\}} \right|}^2}\frac{\rho }{2} + {{\left| {{\mathop{\rm Im}\nolimits} \left\{ {{H^{[kk]}}(t) + \sum\limits_{u = 1}^Q {{H_{\rm IR}^{[k{u}]}}(t){\tau ^{[u]}}(t){H_{\rm TI}^{[{u}k]}}(t)} } \right\}} \right|}^2}\frac{\rho }{2}}}{{\sum\limits_{i = 1,i \ne k}^K {{{\left| {{\rm{Re}}\left\{ {{H^{[ki]}}(t) + \sum\limits_{u = 1}^Q {{H_{\rm IR}^{[k{u}]}}(t){\tau ^{[u]}}(t){H_{\rm TI}^{[{u}i]}}(t)} } \right\}} \right|}^2}\frac{\rho }{2}}  + \sum\limits_{i = 1,i \ne k}^K {{{\left| {{\rm{Re}}\left\{ {{H^{[ki]}}(t) + \sum\limits_{u = 1}^Q {{H_{\rm IR}^{[k{u}]}}(t){\tau ^{[u]}}(t){H_{\rm TI}^{[{u}i]}}(t)} } \right\}} \right|}^2}\frac{\rho }{2}}  + {N_0}}} \ge {\rho ^{1 - \varepsilon }}.$}
\end{equation*}
Therefore,  the rate $\log(1+\rho^{1-\varepsilon})\ge\log(\rho^{1-\varepsilon})=(1-\varepsilon)\log(\rho)$, is achievable for each user.

\section{}
\label{appendix13}
 Let us consider $Q=nK$ and define set ${\cal U}_k=\{n(k-1)+1,...,nk\}$ for each $k\in \{1,...,K\}$. Then, for each $k\in\{1,...,K\}$, we set:
\begin{equation}
{\phi ^{[u]}}({ t}) =  - \angle {H_{\rm TI}^{[{u}k]}}({ t}) - \angle {H_{\rm IR}^{[k{u}]}}({ t})+\frac{\pi}{4},u \in {{\cal U}_k},
\label{phase}
\end{equation}
 where ${\phi ^{[u]}}({ t})$ is the phase of ${\tau ^{[u]}}({ t})$.  We prove theorem for $\textrm{SINR}_k^r$, as the procedure of the proof for $\textrm{SINR}_k^i$ is similar. First, we analyze the numerator of $\textrm{SINR}_k^r$ in (\ref{SINR_k_r}). Thus, we have:
\begin{equation*}
\scalebox{.85}[1]{${H^{[kk]}}({ t}) + \sum\limits_{u = 1}^Q {{H_{\rm IR}^{[k{u}]}}({ t}){\tau ^{[u]}}({ t}){H_{\rm TI}^{[{u}k]}}({ t})} = {H^{[kk]}}({ t}) + \sum\limits_{u \in {{\cal U}_k}} {{H_{\rm IR}^{[k{u}]}}({ t}){\tau ^{[u]}}({ t}){H_{\rm TI}^{[{u}k]}}({ t})}  + \sum\limits_{u = 1,u \notin {{\cal U}_k}}^Q {{H_{\rm IR}^{[k{u}]}}({ t}){\tau ^{[u]}}({ t}){H_{\rm TI}^{[{u}k]}}({ t})} $}
\end{equation*}
\begin{equation*}
\quad\quad\quad\quad\quad\quad\quad\quad\quad\quad\quad\quad\quad\quad\quad\scalebox{.78}[1]{$ = {H^{[kk]}}(t) + \sum\limits_{u \in {{\cal U}_k}} {\left| {{H_{\rm IR}^{[k{u}]}}(t){H_{\rm TI}^{[{u}k]}}(t)} \right|} \exp \left\{ {\frac{\pi }{4}\sqrt { - 1} } \right\} + \sum\limits_{u = 1,u \notin {{\cal U}_k}}^Q {{H_{\rm IR}^{[k{u}]}}(t){\tau ^{[u]}}(t){H_{\rm TI}^{[{u}k]}}(t)} $}.
\end{equation*}
Hence, we obtain:
\begin{equation*}
{\mathop{\rm Re}\nolimits} \left\{ {{H^{[kk]}}({ t}) + \sum\limits_{u = 1}^Q {{H_{\rm IR}^{[k{u}]}}({ t}){\tau ^{[u]}}({ t}){H_{\rm TI}^{[{u}k]}}({ t})} } \right\}
\end{equation*}
\begin{equation}
 = \frac{{\sqrt 2 }}{2}\sum\limits_{u \in {{\cal U}_k}} {\left| {{H_{\rm IR}^{[k{u}]}}(t){H_{\rm TI}^{[{u}k]}}(t)} \right|}  + {\rm{Re}}\left\{ {{H^{[kk]}}(t) + \sum\limits_{u = 1,u \notin {{\cal U}_k}}^Q {{H_{\rm IR}^{[k{u}]}}(t){\tau ^{[u]}}(t){H_{\rm TI}^{[{u}k]}}(t)} } \right\}.
\end{equation}
Now, by Markov's inequality, we obtain:
\begin{equation*}
\Pr \left\{ {\frac{{\sqrt 2 }}{2}\left| {\sum\limits_{u \in {{\cal U}_k}} {\left[ {\left| {{H_{\rm IR}^{[k{u}]}}(t){H_{\rm TI}^{[{u}k]}}(t)} \right| - E\left\{ {\left| {{H_{\rm IR}^{[k{u}]}}(t){H_{\rm TI}^{[{u}k]}}(t)} \right|} \right\}} \right]} } \right| \ge n\delta } \right\}
\end{equation*}
\begin{equation}
 \le \frac{{E\left\{ {{{\left\{ {\left| {{H_{\rm IR}^{[k{u}]}}({ t}){H_{\rm TI}^{[{u}k]}}({ t})} \right| - E\left\{ {\left| {{H_{\rm IR}^{[k{u}]}}({ t}){H_{\rm TI}^{[{u}k]}}({ t})} \right|} \right\}} \right\}}^2}} \right\}}}{{2n{\delta ^2}}}=\varepsilon_1(n),
\label{Markov SINR1}
\end{equation}
\begin{equation*}
\Pr \left\{ {\left| {{\mathop{\rm Re}\nolimits} \left\{ {{H^{[kk]}}({ t}) + \sum\limits_{u = 1,u \notin {{\cal U}_k}}^Q {{H_{\rm IR}^{[k{u}]}}({ t}){\tau ^{[u]}}({ t}){H_{\rm TI}^{[{u}k]}}({ t})} } \right\}} \right| \ge n\delta } \right\}
\end{equation*}
\begin{equation*}
 \le \frac{{E\left\{ {{{\left\{ {{\mathop{\rm Re}\nolimits} \left\{ {{H^{[kk]}}({ t}) + \sum\limits_{u = 1,u \notin {{\cal U}_k}}^Q {{H_{\rm IR}^{[k{u}]}}({ t}){\tau ^{[u]}}({ t}){H_{\rm TI}^{[{u}k]}}({ t})} } \right\}} \right\}}^2}} \right\}}}{{{n^2}{\delta ^2}}}
\end{equation*}
\begin{equation}
 = \frac{{E\left\{ {{{\left\{ {{\mathop{\rm Re}\nolimits} \left\{ {{H^{[kk]}}({ t})} \right\}} \right\}}^2}} \right\} + n(K - 1)E\left\{ {{{\left\{ {{\mathop{\rm Re}\nolimits} \left\{ {{H_{\rm IR}^{[k{{u'}}]}}({ t}){\tau ^{[u']}}({ t}){H_{\rm TI}^{[{{u'}}k]}}({ t})} \right\}} \right\}}^2}} \right\}}}{{{n^2}{\delta ^2}}}=\varepsilon_2(n),
\label{Markov SINR2}
\end{equation}
where $u' \notin {{\cal U}_k}$. Note that inequality (\ref{Markov SINR2}) follows from:
\begin{equation*}
E\left\{ {{H_{\rm IR}^{[k{{u'}}]}}({ t}){\tau ^{[u']}}({ t}){H_{\rm TI}^{[{{u'}}k]}}({ t})} \right\} = 0,u' \notin {{\cal U}_k},
\end{equation*}
which holds due to (\ref{phase}). Exploiting the following inequality for two random variables $X$ and $Y$
\begin{equation}
\Pr \left\{ {\left| {X + Y} \right| \ge 2a} \right\} \le \Pr \left\{ {\left| X \right| \ge a\bigcup {\left| Y \right| \ge a} } \right\} \le \Pr \left\{ {\left| X \right| \ge a} \right\} + \Pr \left\{ {\left| Y \right| \ge a} \right\},
\end{equation}
we obtain:
\begin{equation*}
\Pr \left\{ {\left| {{\mathop{\rm Re}\nolimits} \left\{ {{H^{[kk]}}({ t}) + \sum\limits_{u = 1}^Q {{H_{\rm IR}^{[k{u}]}}({ t}){\tau ^{[u]}}({ t}){H_{\rm TI}^{[{u}k]}}({ t})} } \right\} - n\frac{{\sqrt 2 }}{2}E\left\{ {\left| {{H_{\rm IR}^{[k{u}]}}({ t}){H_{\rm TI}^{[{u}k]}}({ t})} \right|} \right\}} \right| \ge 2n\delta } \right\}
\end{equation*}
\begin{equation}
\le \varepsilon_1(n)+\varepsilon_2(n).
\label{union0}
\end{equation}

For the denominator of  ${\textrm{SINR}}^r_k$, from (\ref{phase}), we have:
\begin{equation*}
E\left\{ {{H^{[ki]}}({ t}) + \sum\limits_{u = 1}^Q {{H_{\rm IR}^{[k{u}]}}({ t}){\tau ^{[u]}}({ t}){H_{\rm TI}^{[{u}i]}}({ t})} } \right\} = 0,i \ne k,
\end{equation*} 
and by the same argument, we obtain:
\begin{equation*}
\Pr \left\{ {\left| {{\mathop{\rm Re}\nolimits} \left\{ {{H^{[ki]}}({ t}) + \sum\limits_{u = 1}^Q {{H_{\rm IR}^{[k{u}]}}({ t}){\tau ^{[u]}}({ t}){H_{\rm TI}^{[{u}i]}}({ t})} } \right\}} \right| \ge n\delta } \right\}
\end{equation*}
\begin{equation}
\le\frac{{E\left\{ {{{\left\{ {{\mathop{\rm Re}\nolimits} \left\{ {{H^{[ki]}}({ t})} \right\}} \right\}}^2}} \right\} + KnE\left\{ {{{\left\{ {{\mathop{\rm Re}\nolimits} \left\{ {{H_{\rm IR}^{[k{u}]}}({ t}){\tau ^{[u]}}({ t}){H_{\rm TI}^{[{u}i]}}({ t})} \right\}} \right\}}^2}} \right\}}}{{{n^2}{\delta ^2}}} = {\varepsilon _3}(n),
\end{equation}

Now, we can write:
\begin{equation*}
\Pr \left\{ {{\textrm{SINR}}_k^r < m} \right\} \le \Pr \left\{ {{{\left| {{\mathop{\rm Re}\nolimits} \left\{ {{H^{[kk]}}(t) + \sum\limits_{u = 1}^Q {{H_{\rm IR}^{[k{u}]}}(t){\tau ^{[u]}}(t){H_{\rm TI}^{[{u}k]}}(t)} } \right\}} \right|}^2}\frac{\rho }{2} < m\left( {(K - 1)\frac{\rho }{2}{n^2}{\delta ^2} + \frac{{{N_0}}}{2}} \right)} \right.
\end{equation*}
\begin{equation*}
\left. {\bigcup {\sum\limits_{i = 1,i \ne k}^K {{{\left| {{\mathop{\rm Re}\nolimits} \left\{ {{H^{[ki]}}(t) + \sum\limits_{u = 1}^Q {{H_{\rm IR}^{[k{u}]}}(t){\tau ^{[u]}}(t){H_{\rm TI}^{[{u}i]}}(t)} } \right\}} \right|}^2}\frac{\rho }{2}}  + \frac{{{N_0}}}{2} > (K - 1)\frac{\rho }{2}{n^2}{\delta ^2} + \frac{{{N_0}}}{2}} } \right\}
\end{equation*}
\begin{equation*}
 \le \Pr \left\{ {{{\left| {{\mathop{\rm Re}\nolimits} \left\{ {{H^{[kk]}}(t) + \sum\limits_{u = 1}^Q {{H_{\rm IR}^{[k{u}]}}(t){\tau ^{[u]}}(t){H_{\rm TI}^{[{u}k]}}(t)} } \right\}} \right|}^2}\frac{\rho }{2} < m\left( {(K - 1)\frac{\rho }{2}{n^2}{\delta ^2} + \frac{{{N_0}}}{2}} \right)} \right\}
\end{equation*}
\begin{equation}
 + \Pr \left\{ {\sum\limits_{i = 1,i \ne k}^K {{{\left| {{\mathop{\rm Re}\nolimits} \left\{ {{H^{[ki]}}(t) + \sum\limits_{u = 1}^Q {{H_{\rm IR}^{[k{u}]}}(t){\tau ^{[u]}}(t){H_{\rm TI}^{[{u}i]}}(t)} } \right\}} \right|}^2}\frac{\rho }{2}}  + \frac{{{N_0}}}{2} > (K - 1)\frac{\rho }{2}{n^2}{\delta ^2} + \frac{{{N_0}}}{2}} \right\}.
\label{union1}
\end{equation}
For the second term in (\ref{union1}), we have:
\begin{equation*}
 + \Pr \left\{ {\sum\limits_{i = 1,i \ne k}^K {{{\left| {{\mathop{\rm Re}\nolimits} \left\{ {{H^{[ki]}}(t) + \sum\limits_{u = 1}^Q {{H_{\rm IR}^{[k{u}]}}(t){\tau ^{[u]}}(t){H_{\rm TI}^{[{u}i]}}(t)} } \right\}} \right|}^2}\frac{\rho }{2}}  + \frac{{{N_0}}}{2} > (K - 1)\frac{\rho }{2}{n^2}{\delta ^2} + \frac{{{N_0}}}{2}} \right\}.
\end{equation*}
\begin{equation*}
 = \Pr \left\{ {\sum\limits_{i = 1,i \ne k}^K {{{\left| {{\mathop{\rm Re}\nolimits} \left\{ {{H^{[ki]}}(t) + \sum\limits_{u = 1}^Q {{H_{\rm IR}^{[k{u}]}}(t){\tau ^{[u]}}(t){H_{\rm TI}^{[{u}i]}}(t)} } \right\}} \right|}^2}}  > (K - 1){n^2}{\delta ^2}} \right\}
\end{equation*}
\begin{equation*}
 \le \Pr \left\{ {\bigcup\limits_{i = 1,i \ne Q}^K {\left\{ {{{\left| {{\mathop{\rm Re}\nolimits} \left\{ {{H^{[ki]}}(t) + \sum\limits_{u = 1}^Q {{H_{\rm IR}^{[k{u}]}}(t){\tau ^{[u]}}(t){H_{\rm TI}^{[{u}i]}}(t)} } \right\}} \right|}^2} > {n^2}{\delta ^2}} \right\}} } \right\}
\end{equation*}
\begin{equation*}
 \le \sum\limits_{i = 1,i \ne k}^K {\Pr \left\{ {{{\left| {{\mathop{\rm Re}\nolimits} \left\{ {{H^{[ki]}}(t) + \sum\limits_{u = 1}^Q {{H_{\rm IR}^{[k{u}]}}(t){\tau ^{[u]}}(t){H_{\rm TI}^{[{u}i]}}(t)} } \right\}} \right|}^2} > {n^2}{\delta ^2}} \right\}} 
\end{equation*}
\begin{equation}
 = (K - 1){\varepsilon _3}(n).
\label{union5}
\end{equation}

For the first term in (\ref{union1}), if we choose a sufficiently small $\delta$ (we must have $\delta  < \frac{{\sqrt 2 }}{4}E\left\{ {\left| {H_{{\text{IR}}}^{[ku]}(t)H_{{\text{TI}}}^{[uk]}(t)} \right|} \right\}$),  there exists a number $n'$, such that for $\forall n>n'$, we can ensure that
\begin{equation*}
m\left( {(K - 1)\frac{\rho }{2}{n^2}{\delta ^2} + \frac{{{N_0}}}{2}} \right) < \frac{\rho }{2}{n^2}{\left( {\frac{{\sqrt 2 }}{2}E\left\{ {\left| {{H_{\rm IR}^{[k{u}]}}(t){H_{\rm TI}^{[{u}k]}}(t)} \right|} \right\} - 2\delta } \right)^2},
\end{equation*}
so, for  $n>n'$,  we have:
\begin{equation*}
\Pr \left\{ {{{\left| {{\mathop{\rm Re}\nolimits} \left\{ {{H^{[kk]}}(t) + \sum\limits_{u = 1}^Q {{H_{\rm IR}^{[k{u}]}}(t){\tau ^{[u]}}(t){H_{\rm TI}^{[{u}k]}}(t)} } \right\}} \right|}^2}\frac{\rho }{2} < m\left( {(K - 1)\frac{\rho }{2}{n^2}{\delta ^2} + \frac{{{N_0}}}{2}} \right)} \right\}
\end{equation*}
\begin{equation*}
\le \Pr \left\{ {{{\left| {{\mathop{\rm Re}\nolimits} \left\{ {{H^{[kk]}}({ t}) + \sum\limits_{u = 1}^Q {{H_{\rm IR}^{[k{u}]}}({ t}){\tau ^{[u]}}({ t}){H_{\rm TI}^{[{u}k]}}({ t})} } \right\}} \right|}^2} < {n^2}{{(\frac{{\sqrt 2 }}{2}E\left\{ {\left| {{H_{\rm IR}^{[k{u}]}}({ t}){H_{\rm TI}^{[{u}k]}}({ t})} \right|} \right\} - 2\delta )}^2}} \right\}
\end{equation*}
\begin{equation*}
\le \Pr \left\{ {{\mathop{\rm Re}\nolimits} \left\{ {{H^{[kk]}}({ t}) + \sum\limits_{u = 1}^Q {{H_{\rm IR}^{[k{u}]}}({ t}){\tau ^{[u]}}({ t}){H_{\rm TI}^{[{u}k]}}({ t})} } \right\} \le n(\frac{{\sqrt 2 }}{2}E\left\{ {\left| {{H_{\rm IR}^{[k{u}]}}({ t}){H_{\rm TI}^{[{u}k]}}({ t})} \right|} \right\} - 2\delta )} \right\}
\end{equation*}
\begin{equation*}
\le \Pr \left\{ {\left| {{\rm{Re}}\left\{ {{H^{[kk]}}({ t}) + \sum\limits_{u = 1}^Q {{H_{\rm IR}^{[k{u}]}}({ t}){\tau ^{[u]}}({ t}){H_{\rm TI}^{[{u}k]}}({ t})} } \right\} - n\frac{{\sqrt 2 }}{2}E\left\{ {\left| {{H_{\rm IR}^{[k{u}]}}({ t}){H_{\rm TI}^{[{u}k]}}({ t})} \right|} \right\}} \right| \ge 2n\delta } \right\}
\end{equation*}
\begin{equation}
 \le {\varepsilon _1}(n) + {\varepsilon _2}(n).
\label{union6}
\end{equation}
Thus, based on (\ref{union5}) and (\ref{union6}), we can find a number $n''$ (we must have $n''>n'$) such that for $\forall n>n''$, we have $\Pr\{\textrm{SINR}_k^r<m\}<\frac{\epsilon}{2}$. By a similar procedure, we can find a number $n'''$, such that for $n>n'''$, we have  $\Pr\{\textrm{SINR}_k^i<m\}<\frac{\epsilon}{2}$, and exploiting the union bound, we can prove inequality (\ref{prob SINR}).

\end{appendices}

\end{document}